\documentclass[12pt]{article}
\usepackage{amssymb}
\usepackage{cite}
\usepackage[linktocpage=true]{hyperref} % If `linktocpage=true' omitted,
\input xy
\xyoption{all}

\newenvironment{single_itemize}{
\begin{itemize}
\setlength{\itemsep}{1pt}
\setlength{\parskip}{0pt}
\setlength{\parsep}{0pt}}{\end{itemize}
}

\begin{document}

%\vspace*{0.25in}

\begin{center}

{\large\bf General aspects of heterotic string compactifications}\\
{\large\bf on 
stacks and gerbes}

\vspace*{0.2in}

Lara B. Anderson$^1$, Bei Jia$^2$, Ryan Manion$^3$, Burt Ovrut$^4$, 
Eric Sharpe$^2$

\vspace*{0.2in}

\begin{tabular}{cc}
{ \begin{tabular}{l}
$^1$ Center for the \\
$\: \:$ Fundamental Laws of Nature\\
Jefferson Laboratory\\
Harvard University\\
17 Oxford Street\\
Cambridge, MA  02138\\
\end{tabular}
} &
{ \begin{tabular}{l}
$^2$ Department of Physics\\
Robeson Hall, 0435\\
Virginia Tech \\
Blacksburg, VA  24061\\
\end{tabular} }
\\
{\begin{tabular}{l}
$^3$ Department of Mathematics\\
David Rittenhouse Laboratory\\
209 South 33rd Street\\
University of Pennsylvania\\
Philadelphia, PA  19104-6395
\end{tabular} }
& 
{\begin{tabular}{l}
$^4$ Department of Physics\\
David Rittenhouse Laboratory\\
209 South 33rd Street\\
University of Pennsylvania\\
Philadelphia, PA  19104-6395
\end{tabular}
}
\end{tabular}

{\tt lara@physics.harvard.edu}, {\tt beijia@vt.edu}, {\tt rymanion@gmail.com},
{\tt ovrut@elcapitan.hep.upenn.edu}, 
{\tt ersharpe@vt.edu}

$\,$

\end{center}

In this paper we work out some basic results 
concerning heterotic
string compactifications on stacks and, in particular, gerbes.  
A heterotic string
compactification on a gerbe can be understood as, simultaneously, 
both a compactification on a space with a restriction on nonperturbative
sectors, and also, a gauge theory in which a subgroup of the gauge group
acts trivially on the massless matter.  Gerbes admit more bundles than
corresponding spaces, which suggests they are potentially a rich
playground for heterotic string compactifications.  After we give a general
characterization of heterotic strings on stacks, we specialize to
gerbes, and consider three different classes of `building blocks' of
gerbe compactifications.  We argue that heterotic string
compactifications on one class is equivalent to
compactification of the same heterotic string on a disjoint union of 
spaces, compactification on another class is dual to compactifications of other
heterotic strings on spaces, and compactification on the 
third class is not perturbatively consistent,
so that we do not in fact recover a broad array of new heterotic
compactifications, just combinations of existing ones.
In appendices we explain how to compute massless spectra of 
heterotic string compactifications on stacks, derive some new necessary
conditions for a heterotic string on a stack or orbifold to be well-defined,
and also review some basic
properties of bundles on gerbes.

\begin{flushleft}
July 2013
\end{flushleft}

\newpage

\tableofcontents

\newpage

\section{Introduction}

The compactification of heterotic superstrings on smooth Calabi-Yau
threefolds has led to realistic $N=1$ supersymmetric particle physics in
four-dimensions. For the $E_{8} \times E_{8}$ heterotic string, the generic
structure of such vacua was presented in
\cite{Donagi:2004qk,Donagi:2004ia,Donagi:2004su,Donagi:2004ub}. Building
upon these results, many phenomenologically relevant low-energy theories
with MSSM-like matter spectra have been constructed, see for example
\cite{Bouchard:2005ag,Braun:2005ux,Braun:2005bw,Braun:2005zv,
Anderson:2009mh,Anderson:2011ns,Anderson:2012yf,Braun:2011ni} for
constructions and related work. However, the limitation of these vacua to
equivariant vector bundles over smooth Calabi-Yau manifolds seems overly
restrictive, and it is of considerable interest to try to construct
heterotic vacua over more general backgrounds.

The purpose of this paper is to outline basic results and general issues
in making sense of heterotic string compactifications on stacks, generalized
spaces admitting metrics, spinors, and all the other items needed to make
sense of a string compactification.  This essentially completes a program
started many years ago to understand the basics of string compactifications
on stacks, see {\it e.g.} 
\cite{kps,nr,msx,glsm,summ,cdhps,karp1,karp2,ps5,me-tex,me-qts}.  
The original hope of this program was to find new SCFT's, new
string compactifications, arising from these generalized spaces.
Although that has not proven to be the case, much has been learned about
the structure of string compactifications, as we shall review.

One of the most physically 
interesting kinds of stacks are known as gerbes. 
The worldsheet theory of
a string compactification on a gerbe can be understood in two 
more or less\footnote{
Mathematically,
the second description, as a gauge theory in which a finite subgroup
acts trivially, implies the first, together with a small amount of 
additional information, a certain trivialization, which we have suppressed
from the description of the first, so we should be slightly
careful in claiming that they are precisely the same. 
}
equivalent
ways:
\begin{itemize}
\item as a sigma model on a space, but with a (combinatorial\footnote{
Meaning, only instantons with degrees
satisfying certain divisibility properties are included.
})
restriction on allowed
nonperturbative sectors, or
\item as a gauge theory in which a (finite)
subgroup of the gauge group acts trivially
on the massless matter.
\end{itemize}

Viewed from the first perspective, it is clear that there is a potential
problem with cluster decomposition in these theories.  For (2,2) SCFT's,
this issue was addressed in \cite{summ}, where it was argued that the
SCFT is equivalent to that on a disjoint union of spaces with variable
$B$ fields, a result listed there as the `decomposition conjecture.'
A sigma model on a disjoint union also violates cluster decomposition,
but in an extremely mild fashion, easily understood.
This duality has since proven crucial for understanding physics issues in
many GLSM's, see {\it e.g.} \cite{cdhps,hori2,ed-nick-me,ncgw,hkm,enstx},
and also has been used to make predictions for Gromov-Witten invariants
of gerbes, predictions which have been checked in {\it e.g.}
\cite{ajt1,ajt2,ajt3,t1,gt1,xt1}.

Viewed from the second perspective, there are analogous issues concerning
whether and how physics can see a trivially-acting finite group.
This was addressed in \cite{nr,msx,glsm}, and will be reviewed later
in this paper.  Massless spectra of (2,2) SCFT's are
computed\footnote{
The papers \cite{nr,msx,glsm} contain consistency checks of this
computation.  Ultimately, demanding modular invariance forces 
the spectrum to contain multiple dimension zero
operators.
} to contain multiple dimension zero operators, another sign of
cluster decomposition issues.  These multiple dimension zero operators are
(discrete Fourier transforms of) identity operators counting the number
of components in the corresponding disjoint union of spaces \cite{summ}.

These ideas have also been recently been applied to four-dimensional
supergravity theories\footnote{
Another thrust of the same papers is a modern discussion of Fayet-Iliopoulos
parameters in supergravity -- it is argued that they can exist and are
quantized.  See {\it e.g.} \cite{dienes-thomas} for an excellent
discussion of old lore on the subject, which is circumvented in the works
above.
} \cite{nati0,git-sugrav,banks-seib,sugrav-g}.  
For example, gerbes admit line bundles
with fractional Chern classes, so the Bagger-Witten \cite{bw1}
quantization
condition on cohomology classes of K\"ahler forms is modified when the
supergravity moduli space admits a gerbe structure.
More generally, a general introduction to four-dimensional supergravities
whose moduli spaces are stacks (generic in Calabi-Yau compactification) is
in \cite{sugrav-g}.
Furthermore, it was shown in \cite{js1}[appendix B] that four-dimensional
supergravity anomalies have a natural description in terms of stacks.
See for example \cite{bgcmru,bgcmu} for other applications.

This paper is concerned with heterotic string compactifications on
stacks and, in particular, gerbes.  As the introduction above alludes,
there are many more bundles on gerbes than on corresponding spaces,
which naively suggests that there could be a rich new landscape of
(0,2) SCFT's and heterotic string compactifications obtainable from
heterotic compactifications on gerbes.
Our results break into three fundamental building blocks or classes:
\begin{itemize}
\item For heterotic compactifications on gerbes in which the gauge
bundle is a pullback from the base (equivalently, when the group
that acts trivially on the base, also acts trivially on the bundle),
the heterotic string compactification is consistent, and is equivalent
to a compactification on a disjoint union of spaces. 
Compactifications of this form are discussed in section~\ref{sect:het-decomp}. 
\item For heterotic compactifications on ${\mathbb Z}_2$ gerbes in which
the ${\mathbb Z}_2$ acts nontrivially on a rank 8 bundle, 
these compactifications
do not decompose, and (we conjecture) are T-dual to ordinary heterotic
compactifications (on spaces)
with a different left-moving GSO.  In other
words, a Spin$(32)/{\mathbb Z}_2$ compactification on such a gerbe is
equivalent to an $E_8 \times E_8$ compactification on a space.
Compactifications of this form are described in section~\ref{sect:het-gsomods}.
\item We conjecture when the bundle is
nontrivial over the gerbe, but not rank 8 or the gerbe is not
${\mathbb Z}_2$, a perturbative
heterotic string compactification is not consistent.  
That said, we do provide some seemingly consistent (0,2) SCFT's defined
by gerbes and bundles of this form, but unfortunately
they do not seem to be useful
for heterotic string compactification.
Compactifications of this form are discussed in 
section~\ref{sect:type3:twisted}.
\end{itemize}
In addition, it is also possible to build examples displaying
combinations of these classes, which are discussed in 
section~\ref{sect:combos}.

In appendix~\ref{app:spectra} we describe how to compute massless spectra
in heterotic string compactifications on general stacks. 
Along the way, we derive some new necessary conditions for 
well-definedness of a SCFT associated to a heterotic string on a stack,
generalizing old statements that ``$c_1 \equiv 0$ mod 2'' for a consistent
heterotic compactification.
Appendix~\ref{app:linebundles} describes in some depth line bundles on
gerbes over projective spaces, as a good prototype for other bundles
on more general gerbes.  Appendix~\ref{app:chern-reps} discusses
how Chern classes and characters are defined for stacks, and in particular,
discusses $c^{\rm rep}$ and ${\rm ch}^{\rm rep}$, versions of 
Chern classes and characters which encode information about twisted
sectors and which play a vital role in index theory.  
Appendix~\ref{app:canonical-roots} contains a short discussion of
roots of canonical bundles on gerbes, a technical matter that sometimes
arises in computations.

One of the original motivations of this work was the hope that
the third class above would yield new consistent heterotic string
compactifications and new consistent (0,2) SCFT's.
Although it seems there are new consistent (0,2) SCFT's, we will argue that
they do not seem to define new consistent supersymmetric heterotic string 
compactifications.

In hindsight, we can understand that result as follows\footnote{
We would like to thank J.~Gray for pointing this out to us.
}.  In an ineffective
orbifold (one in which part of the orbifold group acts trivially
on the space), the twisted sectors contain massless states whose wavefunctions
have support over the entire space.  This would seem to imply that there
are `extra' ten-dimensional massless states, but this would be a 
contradiction, since the ten-dimensional supergravity theory is known and
fixed.  Furthermore, so long as we work at low energies and close to
large-radius limits, a ten-dimensional supergravity analysis should
be applicable.  

In type II strings, this conundrum was implicity solved by the
decomposition conjecture \cite{summ}:  strings on gerbes are the same as
strings on disjoint unions of spaces.  The `extra' states are there, but
simply fill out copies of the supergravity theory.  

In heterotic strings, we will see a mix of several solutions:  in some cases,
an analogue of the decomposition conjecture exists; in other cases,
the theory is dual to a compactification on a manifold; in yet other cases,
the compactification does not seem to be consistent.

\section{Generalities}    \label{sect:generalreview}

\subsection{Strings on stacks}

Stacks are a form of `generalized spaces,' admitting smooth structures,
metrics, bundles, and other structures needed to define sigma models.
In particular, stacks are defined by the incoming maps from other spaces,
making them a natural setting for defining sigma models.

Stacks have been discussed as target `spaces' for nonlinear sigma
models in a number of references, 
including\footnote{
In addition to the references above on the physics of nonlinear sigma
models, there is also an extensive discussion of Gromov-Witten invariants
of stacks in the math literature, see for example
\cite{cr,agv,cclt,mann} for a few representative examples.}
 \cite{kps,nr,msx,glsm,summ,cdhps,karp1,karp2,ps5,me-tex,me-qts} 
for two-dimensional (2,2)
supersymmetric and \cite{git-sugrav,sugrav-g} 
for four-dimensional ${\cal N}=1$ supersymmetric
sigma models.  References for physicists on the mathematics of stacks
are, unfortunately, somewhat harder to locate.  In the
mathematics literature, standard references on algebraic
stacks include 
\cite{vistoli,gomez,lmb} and good references on topological
stacks include 
\cite{bx,heinloth1,metzler1,noohi1,noohi2,noohi3,heinloth2,bss1}.  
In addition,
we have striven to write our own papers to be reasonably self-contained
(see for example \cite{msx} for more information, oriented towards physicsts).

We can make more concrete sense of strings on stacks as follows.
Every\footnote{With minor caveats, as discussed in {\it e.g.}
\cite{msx}.} smooth (Deligne-Mumford) stack $\mathfrak{X}$
has a presentation of the
form of a global quotient $[X/G]$, where $X$ is a smooth manifold
and $G$ is a group which need
neither be finite nor act effectively.  To such a presentation,
we associate a $G$-gauged nonlinear sigma model on $X$.

Now, such presentations are not unique:  a given stack can have many
presentations of this\footnote{
In addition, stacks can have presentations of other forms.  However,
realizing other types of presentations 
in physics would require a significant generalization
of Faddeev-Popov and Batalin-Vilkovisky gauge-fixing procedures, which we
do not claim to understand, so we do
not claim that physics can be associated to all presentations.
} form.  In two dimensional (2,2) theories, 
it is believed, and has been extensively checked, that renormalization group
flow `washes out' such presentation dependence, so universality classes
depend only upon the stack, not any particular presentation.
Thus, one can meaningfully associate a two-dimensional CFT to a stack,
not merely a presentation thereof.
In four dimensions, by contrast, this is not believed to be the case.
For example, although gauge couplings are dynamically generated in two
dimensions, they are not dynamically generated in four dimensions,
and the stack does not determine a gauge coupling.  Thus, in four
dimensions we can not uniquely associate physics to stacks,
though we can certainly do the converse, and use stacks to understand
some parts of the physics of four-dimensional gauge theories,
as in \cite{sugrav-g}.

This paper is concerned with issues around perturbative heterotic
strings on stacks, {\it i.e.} (0,2) SCFT's.  
In principle, a perturbative heterotic string will be defined by
a Calabi-Yau stack $\mathfrak{X}$ together with a gauge bundle 
${\cal E}$ over the stack,
satisfying certain anomaly cancellation conditions.
We understand (0,2) SCFT's in the same fashion as above:
we pick a presentation of the stack of the form $[X/G]$.
Given such a presentation, the gauge bundle is then a $G$-equivariant
bundle ${\cal E}$ over $X$.  To this data, we associate a $G$-gauged
heterotic sigma model on $X$ with gauge bundle ${\cal E}$.
As before, there can be multiple presentations
of a stack with different UV physics, so we conjecture that
renormalization group flow washes out such presentation-dependence,
and only associate universality classes of renormalization group flow
to stacks.

Not every $(X,{\cal E},G)$ will define a consistent heterotic string
theory; for example, the data above must satisfy anomaly cancellation.
One part of anomaly cancellation is clear:  before gauging,
the heterotic sigma model on $X$ with bundle ${\cal E}$ must be
anomaly-free, meaning that ${\rm ch}_2(TX) = {\rm ch}_2({\cal E})$.

Demanding that the gauge theory be anomaly-free can impose further
constraints.  One well-known example is level-matching.
As discussed in {\it e.g.} \cite{freedvafa}, for orbifolds, 
level-matching is believed
to be equivalent to matching of second
Chern classes in equivariant cohomology.
(In particular, equivariant Chern classes can be defined intrinsically
on the stack, they are independent of the choice of presentation and
descend to well-defined objects on the stack.)
Equivariant cohomology can be defined on stacks, and in fact forms
the `naive' cohomology theory of a stack.  (See appendix~\ref{app:chern-reps}
for more subtle notions.)

However, level-matching (in the form described in \cite{freedvafa})
is not sufficient to guarantee that a given
theory is consistent \cite{dienespriv,klst}, and we shall see explicit
examples later in section~\ref{sect:type3:twisted}.  In 
appendix~\ref{app:spectra:fockconstraints}, 
we discuss another set of consistency
conditions that arise, essentially a generalization of the statement
that ``$c_1 \equiv 0$ mod 2.''  Specifically, these conditions state that
on each component $\alpha$ of the inertia stack,
the $\langle \alpha \rangle$-equivariant line bundle 
\begin{displaymath}
K_{\alpha} \otimes \det {\cal E}^{\alpha}_0
\end{displaymath}
admit a square root.  
We defer further discussion of this condition to
appendix~\ref{app:spectra:fockconstraints}.

One of the original goals of this project was to find a suitable generalization
of anomaly cancellation, a set of sufficient conditions,
valid for arbitrary stacks,
that would guarantee that the resulting
$G$-gauged heterotic sigma model is consistent, but we have been unable
to do this.  Instead, we only have the necessary
conditions above.  We leave the problem of finding sufficient
conditions for future work.

The most interesting examples of heterotic strings on stacks are the
special case of strings on gerbes.  In previous work \cite{summ}, it was
argued that (2,2) supersymmetric strings on gerbes are equivalent to
strings on disjoint unions of spaces.  For the heterotic string,
we shall argue that such a decomposition only exists in general if the
gauge bundle is a pullback from a bundle on the base space.
More general, `twisted,' bundles exist, and at least sometimes can 
appear in heterotic compactifications.  In fact, it was one of the original
goals of this work to construct new (0,2) SCFT's by using twisted bundles,
though as we shall argue later, that does not seem to be the case.

In any event, most of this paper will focus on the special case of heterotic
strings on gerbes, so in the remainder of this section we shall review some
pertinent facts.

\subsection{Review of gerbes}  \label{sect:rev-gerbes}

So far we have realized heterotic strings on stacks as gauged nonlinear
sigma models.  The special case of gerbes is realized when a subgroup
of the gauge group acts trivially on the target space.
In this case, even though part of the gauge group acts trivially
on the target, it need not act trivially on the gauge bundle, and this
will be responsible for `twisted' bundles. 

For purposes of disambiguation, let us distinguish our usage of
the term from other appearances in the literature.  
In some papers, 
gerbes are used formally to describe characteristic classes of
$B$ fields, just as principal bundles can be used to describe
characteristic classes of gauge fields, and sometimes they are used
in that sense to help characterize nontrivial $B$ fields.

However, our usage in this paper is different.  We are not using the
term `gerbe' to describe characteristic classes; instead, we are
thinking of gerbes as analogues of spaces on which strings propagate, just as 
strings can propagate on the
total space of a principal bundle.

Let us now turn to reviewing gerbes.  We review some basics here,
see \cite{summ} for another pertinent general description.
In general, to specify a $G$-gerbe
over a space $X$, given an open cover $\{U_i\}$ of $X$, one
specifies $g_{ijk} \in G$
on triple overlaps and $\varphi_{ij} \in \mbox{Aut}(G)$ on
double overlaps, obeying the constraints
\begin{equation}    \label{firstg}
\varphi_{jk}\circ\varphi_{ij}\:=\:\mbox{Ad}(g_{ijk})\circ\varphi_{ik}
\end{equation}
on triple overlaps and
\begin{equation}\label{secondg}
g_{jk\ell}\,g_{ij\ell}\:=\:\varphi_{k\ell}(g_{ijk})\,g_{ik\ell}.
\end{equation}
on quadruple overlaps.
If we let Out$(G)$ denote the quotient of the
group of all automorphisms of $G$
by inner automorphisms, then the $\varphi_{ij}$ above descend to
define a principal Out$(G)$ bundle.  If that bundle is trivializable,
then we say the gerbe is banded.  In this case, the gerbe is
effectively specified just by the $g_{ijk}$'s, which define a characteristic
class in
$H^2(X,Z(G))$.  (For example, these were the gerbes described
in \cite{hitchin}.)  The more general case, in which the Out$(G)$ bundle is
nontrivial, is known simply as non-banded.

In terms of stacks, a stack $[X/G]$ will be a ($K$-)gerbe if a 
nontrivial subgroup (denoted $K$) of $G$ acts trivially on $X$, by which we mean
$g \cdot x = x$ for all $x \in X$ and all $g \in K \subseteq G$. 
(This is known as an non-effective group action.) 
Although quotient spaces cannot detect trivial group actions,
quotient stacks can, and moreover, so too can the physics\footnote{
Historically, this was one of several confusing points in understanding
whether strings could be consistently defined on stacks.}
of gauge theories.  Although such trivial group actions are invisible
perturbatively, they show up nonperturbatively, as has been discussed
extensively in {\it e.g.} \cite{nr,msx,glsm,summ}.

As the physics of strings on gerbes will be important in this paper,
let us briefly review how nonperturbative physics can detect
trivial group actions.

One short answer is that working with a gauge theory containing a
non-effective group action is equivalent to restricting the allowed
nonperturbative sectors\footnote{
Restricting the allowed instanton sectors ordinarily breaks cluster
decomposition, and understanding how this can be consistent was,
historically, another confusing issue that had to be straightened out
to make sense of strings on stacks.  Briefly, the answer is that the
theory decomposes into a union of theories on ordinary spaces,
see {\it e.g.} \cite{summ,cdhps,sugrav-g} for discussions in two and
four-dimensional theories.  We will return to this in 
section~\ref{sect:het-gsomods}.
}.  For example, consider the ${\mathbb P}^n$ model,
described as a supersymmetric $U(1)$ gauge theory with $n+1$
chiral superfields of charge $1$, but let us instead give the fields
charge $k$ instead of charge $1$.  Mathematically, this means that a
${\mathbb Z}_k$ subgroup of $U(1)$ acts trivially on the chiral superfields,
and describes the weighted projective stack ${\mathbb P}^n_{[k,k,\cdots,k]}$,
which is a ${\mathbb Z}_k$ gerbe on ${\mathbb P}^n$.  
Physically, it is straightforward
to see that the instantons in this GLSM are the same as the instantons
of degree divisible by $k$ in the original ${\mathbb P}^n$ model.
As a practical matter, this means that the $U(1)_A$ symmetry is broken to
${\mathbb Z}_{2k(n+1)}$ rather than ${\mathbb Z}_{2(n+1)}$, for example,
and also changes correlation functions and quantum cohomology rings.

More globally, if the worldsheet is compact, then the proper definition of
the `charge' of a field is in terms of what bundle it couples to.
Changing the bundle changes the allowed zero modes, hence changes anomalies
and correlation functions \cite{nr}.  For a noncompact worldsheet, an analogous
result can be obtained in two dimensions utilizing theta angles.  We distinguish
`gerbe' cases from `non-gerbe' cases by adding massive minimally-charged
fields.  The existence of such fields can be sensed, even if their
masses are above the cutoff, by examining the periodicity of the theta
angle.  Since the theta angle acts as an electric field in two dimensions,
if we build a capacitor, then by making the plate separation large, one can
excite arbitarily-massive field configurations, hence theta angle periodicity
measures existence of massive minimally-charged fields 
\cite{nr,nati0,banks-seib}.  In four dimensions, there are analogous
methods, involving for example Reissner-Nordstrom
black holes and Hawking radiation \cite{sugrav-g}.

A simple example in toroidal orbifolds may help clarify the discussion.
Consider the orbifold $[X/D_4]$, where $D_4$ is an eight-element group
with a ${\mathbb Z}_2$ center, such that 
$D_4/{\mathbb Z}_2 = {\mathbb Z}_2 \times
{\mathbb Z}_2$.  Assume the central ${\mathbb Z}_2$ acts trivially on $X$.
From the general analysis above, one would expect that
$[X/D_4] \neq [X/{\mathbb Z}_2 \times {\mathbb Z}_2]$, {\it i.e.} that
physics `sees' the trivially-acting ${\mathbb Z}_2$, and that is exactly
what happens.

Label the elements of $D_4$ by
\begin{displaymath}
\{1, z, a, b, az, bz, ab, ba=abz \},
\end{displaymath}
where $z$ generates the ${\mathbb Z}_2$ center, so that the coset
$D_4/{\mathbb Z}_2$ is given by the images of $1$, $a$, $b$, $ab$,
which in the (${\mathbb Z}_2 \times {\mathbb Z}_2$) coset we shall denote
$\{ 1, \overline{a}, \overline{b}, \overline{ab} \}$.

The (string)
one-loop partition function of $[X/D_4]$ is obtained by summing over
twisted sectors defined by all commuting pairs in $D_4$.
For example, there are no $(a, ab)$, $(b, ab)$, $(a,b)$ twisted sectors,
as those pairs do not commute in $D_4$.
Now, if we compare the ${\mathbb Z}_2 \times {\mathbb Z}_2$ partition function,
although individual twisted sector contributions match (as the
${\mathbb Z}_2$ acts trivially), the total number is different.
For example, the ${\mathbb Z}_2 \times {\mathbb Z}_2$
contains contributions from $(\overline{a}, \overline{ab})$,
$(\overline{b},\overline{ab})$ and $(\overline{a},\overline{b})$
twisted sectors, but there are no corresponding
$(a,ab)$, $(b,ab)$, $(a,ab)$ contributions in the $D_4$ partition
function.  Thus, we see the one-loop partition functions of the
$D_4$ and ${\mathbb Z}_2 \times {\mathbb Z}_2$ partition functions are very
different, despite the fact that the theories differ by a trivially-acting
gauged ${\mathbb Z}_2$.  

In fact, in the example above, one can show that the partition function
of the $D_4$ orbifold is the same as the partition function of a disjoint
union of two ${\mathbb Z}_2 \times {\mathbb Z}_2$ orbifolds, one with and
the other without discrete torsion.  The one-loop partition function of
a disjoint union is the sum of the partition functions of the components,
and discrete torsion adds a sign to the $(\overline{a}, \overline{ab})$,
$(\overline{b},\overline{ab})$ and $(\overline{a},\overline{b})$ sectors,
so they cancel out of the partition function for the disjoint union.
This is a simple example of the `decomposition conjecture' we review
in section~\ref{sect:decomp-22review}.

\subsection{Notions of twisting}  \label{sect:twisting}

Now that we have outlined gerbes and demonstrated their physical
meaningfulness, let us turn to possible bundles over gerbes.
A gerbe was defined by a trivial group action on the base space;
however, that same group action can be nontrivial on the bundle.
The resulting bundle is then interpreted as some sort of twisted
bundle, in some sense, as we shall review here.

There are various notions of twisted bundles in the literature.
One notion, discussed for example in \cite{cks},
is of a twisted bundle in which the twisting refers to the fact
that the transition functions do not quite close on triple overlaps:
instead of
\begin{displaymath}
g_{\alpha \beta } g_{\beta \gamma} g_{\gamma \alpha} \: = \: 1
\end{displaymath}
the transition functions obey
\begin{equation}  \label{cocyc1}
g_{\alpha \beta } g_{\beta \gamma} g_{\gamma \alpha} \: = \:
h_{\alpha \beta \gamma} I
\end{equation}
for some cocycle $h_{\alpha \beta \gamma}$.
At the level of the gauge field, such a twisting means that across
coordinate patches, the gauge field receives an affine translation
in addition to a gauge transformation.
Such twisted bundles appear physically on D-branes.
After all, under a gauge transformation of the $B$ field, of the form
\begin{displaymath}
B \: \mapsto \: B \: + \: d \Lambda,
\end{displaymath} 
the Chan-Paton gauge field must necessarily transform as
\begin{displaymath}
A \: \mapsto \: A \: - \: \Lambda
\end{displaymath}
in order to preserve gauge-invariance on the open string worldsheet,
and such affine translations correspond, in terms of transition functions,
to the modified overlap condition equation~(\ref{cocyc1}).
However, although such twistings are possible for D-branes,
no such twisting is ordinarily possible in heterotic strings,
because the heterotic gauge field never picks up affine translations
across coordinate patches -- the heterotic gauge field and the
heterotic $B$ field are related in a very different fashion than in
D-branes.

A second notion of twisting appears when discussing gerbes.
Consider the weighted projective stack ${\mathbb P}^N_{[k,\cdots,k]}$,
a ${\mathbb Z}_k$ gerbe on ${\mathbb P}^N$,
described physically by an analogue of the supersymmetric ${\mathbb P}^N$
model in which chiral superfields have charge $k$ instead of $1$,
as discussed earlier.
Now, the total space of a line bundle ${\cal O}(-n) \rightarrow
{\mathbb P}^N$ can be described as a quotient of $N+1$ fields
$\phi_i$ and one field $p$ of charges $1$, $-n$, respectively.
Consider instead a quotient of the fields above in which the
$\phi_i$ have charge $k$ (and so describe ${\mathbb P}^N_{[k,\cdots,k]}$),
and the field $p$ has charge $-1$.
This quotient is the total space of a line bundle on the
gerbe sometimes denoted ${\cal O}(-1/k)$.
(We will discuss line bundles on gerbes in more detail in 
appendix~\ref{app:linebundles}.)

We can understand this second notion of twisting in much greater generality,
as follows.  First, for 
any stack $\mathfrak{X}$ presented as $\mathfrak{X} = [X/G]$ for
some space $X$ and group $G$, a vector bundle (sheaf) on $\mathfrak{X}$ is
the same as a $G$-equivariant vector bundle (sheaf) on $X$.
Now, suppose that
$G$ is an extension
\begin{displaymath}
1 \: \longrightarrow \: K \: \longrightarrow \: G \: \longrightarrow \:
H \: \longrightarrow \: 1,
\end{displaymath}
where $K$ acts trivially on $X$, and $G/K \cong H$ acts effectively.
In this case, $\mathfrak{X} = [X/G]$ is a $K$-gerbe.
A vector bundle on $\mathfrak{X}$ is a $G$-equivariant vector bundle on $X$,
and as such, the $K$ action is defined by a representation of $K$ on the
fibers of that vector bundle.  This is the more general picture of the second
notion of twisting.  Any bundle on the gerbe that is not a pullback from
the base, has a nontrivial action of $K$.

These two notions of twisting are not unrelated.
Mathematically, it is a standard result that
the category of sheaves on a gerbe decomposes into different sectors
containing
twisted sheaves on the underlying space, twisted by flat $B$ fields.
Moreover, this decomposition is complete:  there are no nonzero Ext groups
between sheaves in different sectors on the same gerbe.  This fact was one
of the inspirations for the `decomposition conjecture' presented in
\cite{summ}, which said that conformal field theories describing strings
on gerbes should factorize in the same way, that the CFT's are the same
as CFT's on disjoint unions of spaces.  The resulting factorization of
D-branes reflects the mathematical result above on factorization of sheaves
on gerbes.

For completeness, let us discuss this decomposition for
the special case of 
${\cal O}(1/k) \rightarrow {\mathbb P}^N_{[k,\cdots,k]}$.
To be twisted in the first
sense we discussed, one can show that the rank of the twisted bundle
must be divisible by the order of the twisting cocycle's cohomology
class.  Here, since
${\cal O}(1/k)$ has rank one, the order of the cocycle must be one.
Indeed, the twistings of ${\cal O}(1/k)$ appearing involve cocycles
with trivial cohomology, so there is no rank restriction.

\section{Class I:  Gauge bundle a pullback from the base}
\label{sect:het-decomp}

We have classified heterotic string compactifications on gerbes into
three fundamental classes or 
`building blocks,' from which more general
compactifications can be built.  In this and the next two sections,
we will examine properties of those classes.

The first class we consider involves the special case that the 
gauge bundle is a pullback from the base.
This is equivalent to the
statement that the subgroup $G$ of the gauge group that acts trivially on the
base, also acts trivially on the fibers of the gauge bundle.

In this case, we will argue that, at least for banded gerbes,
the heterotic (0,2) SCFT factorizes -- it is equivalent to
a heterotic string on a disjoint union of spaces with bundles, following
essentially the same mechanism as in (2,2) strings.

\subsection{Review of (2,2) decomposition conjecture}
\label{sect:decomp-22review}

As was reviewed earlier in section~\ref{sect:rev-gerbes},
gauge theories in which a subgroup of the gauge group acts trivially
on massless matter break cluster decomposition.   
However, it was argued in \cite{summ} that such theories are equivalent
to tensor products / disjoint unions of cluster-decomposition-obeying
theories.  For example, a gauged nonlinear sigma model of this form
is equivalent to a nonlinear sigma model on a disjoint union of ordinary
spaces.  The latter violates cluster decomposition, but does so in
an obviously trivial fashion, and so there is no essential difficulty
with the quantum field theory.

For (2,2) supersymmetric gauged nonlinear sigma models in two dimensions,
this was encapsulated in \cite{summ} in the ``decomposition conjecture,''
which we shall generalize to heterotic strings.
To make this paper self-contained, we take a moment here to review
the statement of the decomposition conjecture.

Suppose we have a $K$-gerbe over $[X/H]$, defined by the quotient
$[X/G]$ where
\begin{displaymath}
1 \: \longrightarrow \: K \: \longrightarrow \: G \: \longrightarrow \:
H \: \longrightarrow \: 1.
\end{displaymath}
Let $\hat{K}$ denote the set of irreducible representations of $K$.
There is a natural action of $H$ on $\hat{K}$, defined as follows:
given $h \in H$ and $\rho \in \hat{K}$, pick a lift $\tilde{h} \in G$ of $h$,
and define $h \cdot \rho$ by,
\begin{displaymath}
(h \cdot \rho)(g) \: \equiv \: \rho(\tilde{h}^{-1} g \tilde{h} )
\end{displaymath}
for all $g \in K$.  If $K$ is abelian, this is well-defined.  If $K$
is not abelian, then it can be shown
(see \cite{summ}[section 4]) that there exists an operator
intertwining the representations $h \cdot \rho$ defined by any two lifts,
hence $h \cdot \rho$ is well-defined in $\hat{K}$.

Then, the decomposition conjecture for (2,2) theories states that a string
on the gerbe $[X/G]$ is the same as a string on the disjoint union of
spaces $[ (X \times \hat{K} )/H ]$, together with a flat $B$ field defined
in \cite{summ}[section 4].

In the special case that the gerbe $[X/G]$ is banded,
the description
above simplifies.  In this case, the $H$ action on $\hat{K}$ is trivial,
and so the decomposition conjecture reduces to the statement that a string
on the gerbe $[X/G]$ is the same as a string on a disjoint union of
$| \hat{K} |$ copies of $[X/H]$, in which each copy comes with a flat
$B$ field determined by acting on the characteristic class of the gerbe
with the irreducible representation corresponding to that copy:
\begin{displaymath}
\rho \in \hat{K}: \: H^2([X/H], Z(G)) \: \longrightarrow \:
H^2([X/H], U(1) ).
\end{displaymath}

Extensive evidence was presented in \cite{summ} for this conjecture,
ranging from computations of orbifold spectra and partition functions to
GLSM results and quantum cohomology computations.
Other results have appeared since.  For reasons of brevity, we only list
two below:
\begin{itemize}
\item This conjecture makes a prediction for Gromov-Witten invariants of
stacks, namely that the Gromov-Witten invariants of gerbes are equivalent
to Gromov-Witten invariants of disjoint unions of spaces.
This was checked in the mathematics literature in
{\it e.g.} \cite{ajt1,ajt2,ajt3,t1,gt1,xt1}.
\item This conjecture plays an important role in understanding certain
GLSM's.  Specifically, it was used in \cite{cdhps} to understand
Landau-Ginzburg points of complete intersections of quadrics,
resolving some old unanswered questions, and also providing examples
of GLSM's that realize geometry in a different way than as a critical
locus of a superpotential, that contain non-birational phases, and in
some cases, that RG flow to `noncommutative resolutions' of singular
spaces, providing the first physical realizations of those mathematical
theories in CFT.  The results of \cite{cdhps} have since been
checked in {\it e.g.} \cite{hori2,ed-nick-me} and further examples
discussed in \cite{hori2,hkm,enstx}.  The same methods have also been
applied to make
predictions for Gromov-Witten invariants of noncommutative resolutions
in \cite{ncgw}. 
\end{itemize}
See also the $D_4$ orbifold discussed in section~\ref{sect:rev-gerbes} 
for another example.

The result may seem obscure, but there is a simple physical reason for it,
namely that in the path integral,
summing over the elements of the disjoint union, together
with variable $B$ fields, is equivalent to inserting a projection operator
that enforces the requirement that only instantons of certain degrees
contribute to the theory.
Schematically, for a nonlinear sigma model, we can describe
the insertion of a projection operator in the form
\begin{displaymath}
\int [D \phi] e^{-S} \left( \sum_{k=0}^{n-1} e^{i k \int \phi^* \omega}
\right)
\: = \:
\sum_{k=0}^{n-1} \int [D \phi] \exp\left( - S + i k \int \phi^* \omega
\right),
\end{displaymath}
where $\omega$ is the K\"ahler form on the target space.
The left-hand side is the partition function with a projector onto
nonperturbative states of certain degrees; the right-hand side is a 
partition function for a disjoint union of $n$ copies of the
original target space, each with a 
rotated $B$ field, rotated by an amount $k \omega$.
Nonbanded gerbes merely represent a more complicated variation.

\subsection{Heterotic decomposition conjecture}

In this section we will describe the heterotic analogue of the
decomposition conjecture, for banded gerbes.
Briefly, given a (0,2) SCFT defined by a banded gerbe $\mathfrak{X}$
over a space (or orbifold) $X$
and bundle ${\cal E} \rightarrow \mathfrak{X}$, such that 
${\cal E}$ is a pullback of a bundle on $X$, then this (0,2) SCFT
is the same as a (0,2) SCFT on a disjoint union of copies of $X$.

Now, let us define terms more precisely.
Suppose we have a $K$-gerbe over $[X/H]$, defined by the quotient
$\mathfrak{X} = [X/G]$ where
\begin{displaymath}
1 \: \longrightarrow \: K \: \longrightarrow \: G \: \longrightarrow \:
H \: \longrightarrow \: 1.
\end{displaymath}
Suppose we also have a holomorphic vector bundle ${\cal E}$ over
$[X/G]$ ({\it i.e.} a $G$-equivariant bundle on $X$), defining a consistent
(0,2) SCFT.

We assume that ${\cal E}$ is a pullback of a bundle
${\cal E}'$ on $[X/H]$.  This can be understood in several equivalent
ways, for example:
\begin{itemize}
\item $K$ acts trivially on both $X$ and ${\cal E}$,
\item ${\cal E}$ is in the weight-zero part of the decomposition of
sheaves on $[X/G]$,
\end{itemize}
which imply that the $G$-equivariant structure on ${\cal E}$ (as a bundle
on $X$) descends to an $H$-equivariant structure.

The heterotic decomposition conjecture for (0,2) theories is that,
in these circumstances, if the gerbe is banded,
a heterotic string on $([X/G], {\cal E})$
is the same as a heterotic string on the disjoint union
\begin{displaymath}
\amalg_{\hat{K}} [X/H]
\end{displaymath}
with varying $B$ fields and gauge bundle ${\cal E}'$ on each copy of
$[X/H]$.

As a consistency check, in the special case that ${\cal E}=T\mathfrak{X}$ 
({\it i.e.} $TX$ with
its natural $G$-equivariant structure), then ${\cal E}' = TX$ with its natural
$H$-equivariant structure, and this reduces to the (2,2) decomoposition
conjecture (for banded gerbes).

Other examples are easy to construct.  For example, if we take
an anomaly-free heterotic (0,2) SCFT defined by a bundle ${\cal E}$
on a space $X$, and take a global orbifold of $X$ by a finite group
that acts trivially on both $X$ and ${\cal E}$, it is trivial to see
that the twisted sector states will all be copies of the untwisted sector
states, in agreement with the prediction of the decomposition conjecture
above that this (0,2) SCFT should be the same as that for a disjoint union of
copies of $(X, {\cal E})$.

Another set of examples is provided by (0,2) GLSM's.
Begin with an anomaly-free (0,2) GLSM describing a bundle ${\cal E}'$, say,
\begin{displaymath}
0 \: \longrightarrow \: {\cal E}' \: \longrightarrow \:
\oplus_a {\cal O}(n_a) \: \stackrel{F}{\longrightarrow} \:
\oplus_i {\cal O}(m_i) \: \longrightarrow \: 0,
\end{displaymath}
over a hypersurface in a weighted projective space
${\mathbb P}^d_{w_0, \cdots, w_d}[w_0 + \cdots + w_d]$.
Now, build a new (0,2) GLSM constructed from the one above by multiplying
all gauge charges by an integer $k > 0$.  The result is a bundle ${\cal E}$,
\begin{displaymath}
0 \: \longrightarrow \: {\cal E} \: \longrightarrow \:
\oplus_a {\cal O}(k n_a) \: \stackrel{F}{\longrightarrow} \:
\oplus_i {\cal O}(k m_i) \: \longrightarrow \: 0,
\end{displaymath}
over a hypersurface in a weighted projective stack
${\mathbb P}^d_{[k w_0, \cdots, k w_d]}[k(w_0 + \cdots + w_d)]$.
The bundle map $F$ and hypersurface polynomial are unchanged.
If one now goes to the Landau-Ginzburg point of this theory and computes
the massless spectrum, it is trivial to see that the spectrum will consist
of $k$ copies of the spectrum of the original theory, in agreement with the
prediction of the decomposition conjecture.

The analogue of the decomposition conjecture for nonbanded gerbes is not
currently known.  It is tempting to speculate that it should be the
statement that a heterotic string on $([X/G], {\cal E})$ is the same as a
heterotic strings on 
$( [(X \times \hat{K})/H],
{\cal E})$, where (as in the (2,2) case) $\hat{K}$ is the set of
irreducible representations of $K$, and we extend ${\cal E}$ trivially
from $[X/H]$ to $[(X \times \hat{K})/H]$.  However, on the (2,2) locus,
the special case that
${\cal E} = TX$ with its natural $G$-equivariant structure, 
${\cal E}$ reinterpreted as an $H$-equivariant bundle and extended
trivially over $\hat{K}$ does not in general\footnote{
Only if $K$ lies in the center of $G$ would the tangent bundle have
a trivial extension over $\hat{K}$.}
define the tangent bundle of
$[(X \times \hat{K})/H]$, and so this would not reduce correctly
to (2,2) decomposition.

\section{Class II:  Dualities}
\label{sect:het-gsomods}

\subsection{Basic proposal}

In the special case of a heterotic string in which a ${\mathbb Z}_2$
that acts nontrivially on the base, acts nontrivially on a rank 8 bundle,
that subgroup of the gauge group is locally duplicating the effect of one
of the ten-dimensional left-moving GSO projections.  If one starts with
a Spin$(32)/{\mathbb Z}_2$ string, then the dual looks locally like an
$E_8 \times E_8$ string.

In this section, we will describe\footnote{
We have not been able to locate this particular 
duality in the literature, but would
not be surprised if it has been discussed somewhere previously,
presumably in a different context.
The closest of which we are aware is old work on T-duality in
toroidally compactified heterotic strings, relating Spin$(32)/{\mathbb Z}_2$
strings and $E_8 \times E_8$ strings after the gauge group has been
Higgsed to a common subgroup, see for example \cite{ginsparghet}.
} a precise duality relating such
Spin$(32)/{\mathbb Z}_2$ compactifications to ordinary $E_8 \times E_8$
compactifications, and discuss some examples.

First, let us consider the easiest case.
If the ${\mathbb Z}_2$ gerbe is trivial, the result is automatic:
the worldsheet left-moving GSO projection is duplicated exactly,
not just locally.  When the gerbe is nontrivial, one must think a little
harder to find a precise duality.

We propose\footnote{
We would link to thank J.~Distler for suggesting this construction to us.
} a duality to heterotic $E_8 \times E_8$ strings as follows.
To set conventions, suppose our stack
$\mathfrak{X} = [X/\tilde{G}]$, where 
\begin{displaymath}
1 \: \longrightarrow \: {\mathbb Z}_2 \: \longrightarrow \:
\tilde{G} \: \longrightarrow \: G \: \longrightarrow \: 1
\end{displaymath}
and ${\mathbb Z}_2$ acts trivially on $X$.
Suppose furthermore that ${\cal E}$ is a bundle on $\mathfrak{X}$,
{\it i.e.} a $\tilde{G}$-equivariant bundle on $X$, whose
embedding into $E_8$ is via the standard worldsheet fermionic 
construction, in which left-moving fermions are in the fundamental 
representation of the structure group.
Suppose that the ${\mathbb Z}_2$ acts nontrivially on ${\cal E}$.

In general ${\cal E}$ will not 
admit a $G$-equivariant structure.
Nevertheless,
at least in the special case that the ${\mathbb Z}_2$ is central in
$\tilde{G}$, the bundles ${\cal E}^* \otimes {\cal E}$ and
$\wedge^{\rm even} {\cal E}$ will admit a $G$-equivariant structure,
and so can be defined on $[X/G]$.  Moreover, it was observed in
\cite{dist-greene} that, for the `typical' worldsheet
embeddings of $SU(n)$ in $E_8$ (including the present one),
massless spectra of heterotic compactifications
on smooth spaces can be defined solely in terms of sheaf cohomology
with coefficients in ${\cal E}^* \otimes {\cal E}$ and
$\wedge^{\rm even} {\cal E}$; other sheaf cohomology groups are related
by Serre duality.  There is a good reason for this.
In the heterotic compactifications discussed in \cite{dist-greene},
the $SU(n)$ gauge bundle is embedded into $E_8$ by first embedding
in ${\rm Spin}(2n) \subset {\rm Spin}(16)$, projecting to
${\rm Spin}(16)/{\mathbb Z}_2$ (as a result of the left GSO projection),
and then ${\rm Spin}(16)/{\mathbb Z}_2$ naturally embeds into $E_8$
\cite{adamse8}.
The only coefficient bundles that survive the left GSO projection
are ${\cal E}^* \otimes {\cal E}$ and $\wedge^{\rm even} {\cal E}$;
they suffice to define an $E_8$ bundle, and that is why they suffice
to define massless spectra.

Thus, we propose that a heterotic ${\rm Spin}(32)/{\mathbb Z}_2$ string
compactified on a ${\mathbb Z}_2$ gerbe $\mathfrak{X}$ as above,
with the ${\mathbb Z}_2$ central, acting by signs on a rank 8
bundle ${\cal E} \rightarrow \mathfrak{X}$, embedded in a typical fashion,
defines the same SCFT as a heterotic $E_8 \times E_8$ string compactified
on $[X/G]$ with $E_8$ gauge bundle determined by
${\cal E}^* \otimes {\cal E}$ and $\wedge^{\rm even} {\cal E}$
(which are defined on $[X/G]$, even if ${\cal E}$ itself is not).

In the special case that the ${\mathbb Z}_2$ gerbe is trivial, 
the dual $E_8 \times E_8$ string on $[X/G]$ is defined by
the bundle ${\cal E}$ -- in this special case, the $E_8$ bundle
determined by ${\cal E}^* \otimes {\cal E}$, $\wedge^{\rm even} {\cal E}$
is the same determined by the usual embedding of ${\cal E}$ into
$E_8$.  More generally, the $E_8$ bundle
need not have a description in terms of a similarly-embedded
$SU(n)$ gauge bundle; a direct construction might have to appeal to the
fibered WZW methods discussed in 
\cite{anom,gates1,gates2,gates3,gates4,gates5}.

So far we have discussed Spin$(32)/{\mathbb Z}_2$ compactifications on
a ${\mathbb Z}_2$ gerbe with rank 8 bundle.  Now let us briefly consider
an $E_8 \times E_8$ compactification on a ${\mathbb Z}_2$ gerbe with
rank 8 bundle.  Nearly the same analysis applies as in the
Spin$(32)/{\mathbb Z}_2$ case.
At the level of SCFT, before imposing the left GSO projections,
the same duality argument we have just given suggests the gerbe
theory should be dual to an $E_8$ bundle, as above.  The left GSO for
the corresponding bundle duplicates the gerbe ${\mathbb Z}_2$, and so
should act trivially on the theory.  The dual should be thus be
interpreted as class I, and so the result should have the
form of a disjoint union of two copies of an $E_8 \times E_8$ compactification.
As the details are largely duplicative of the Spin$(32)/{\mathbb Z}_2$
case just discussed, and for which we will see examples below,
we will not treat this case further.

We have discussed bundles with structure group $SU(n)$ embedded into
${\rm Spin}(32)/{\mathbb Z}_2$ and $E_8 \times E_8$ in the form of
the standard
worldsheet construction, but more general embeddings exist, and admit
worldsheet descriptions \cite{anom}.
One open question we leave for future work is to generalize the duality
discussed here to more general embeddings.

\subsection{Toroidal orbifold example}
\label{sect:class2-ex1}

Consider a Spin$(32)/{\mathbb Z}_2$ heterotic string compactified
on a ${\mathbb Z}_2$ gerbe over $[T^4/{\mathbb Z}_2]$, with a rank eight
bundle, defined as follows.  The ${\mathbb Z}_2$ gerbe is 
$[T^4/{\mathbb Z}_4]$, where the ${\mathbb Z}_4$ acts on the $T^4$ by
\begin{displaymath}
x \: \mapsto \: \exp\left( \frac{2\pi i (2k)k}{4} \right) x
\: = \: (-)^k x,
\end{displaymath}
so that there is a trivially-acting ${\mathbb Z}_2$ subgroup; only the
sectors $k=1, 3$ have twisted bosons.
(Mathematically, this is a nontrivial\footnote{
This gerbe is the obstruction to lifting the principal ${\mathbb Z}_2$ bundle
$T^4 \rightarrow [T^4/{\mathbb Z}_2]$ to a principal ${\mathbb Z}_4$ bundle on
$[T^4/{\mathbb Z}_2]$.  But a principal ${\mathbb Z}_k$ 
bundle on any space $X$ is the
same thing as a homomorphism $\pi_1(X) \rightarrow {\mathbb Z}_k$.
Therefore, we can study nontriviality of the gerbe as a question about
lifts of group homomorphisms.  The bundle $T^4 \rightarrow
[T^4/{\mathbb Z}_2]$ corresponds to a homomorphism
\begin{displaymath}
\phi: \: \pi_1\left( [T^4/{\mathbb Z}_2] \right) \: \longrightarrow \: 
{\mathbb Z}_2.
\end{displaymath}
(In particular, since $T^4 \rightarrow [T^4/{\mathbb Z}_2]$ is a
principal ${\mathbb Z}_2$ bundle, we have a long exact sequence with relevant
part
\begin{displaymath}
\pi_1(T^4) \: \longrightarrow \: \pi_1([T^4/{\mathbb Z}_2) \:
\stackrel{\phi}{\longrightarrow}
\: \pi_0\left( {\mathbb Z}_2 \right) \: \left( \cong \: {\mathbb Z}_2 \right)
\: \longrightarrow \: \pi_0(T^4),
\end{displaymath}
and as $T^4$ is connected, we see that $\phi$ is surjective.)
We want to understand whether $\phi$ lifts to a homomorphism
\begin{displaymath}
\psi: \: \pi_1\left( [T^4/{\mathbb Z}_2] \right) \: \longrightarrow \:
{\mathbb Z}_4.
\end{displaymath}
First note
\begin{displaymath}
\pi_1\left( [T^4/{\mathbb Z}_2] \right) \: = \: {\mathbb Z}_2 \rtimes
{\mathbb Z}^4,
\end{displaymath}
where the nontrivial element in ${\mathbb Z}_2$ acts as multiplication by
$-1$ on ${\mathbb Z}^4$.  The homomorphism $\phi$ is the projection to
${\mathbb Z}_2$.  The maximal 2-group quotient of
${\mathbb Z}_2 \rtimes {\mathbb Z}^4$ is ${\mathbb Z}_2 \times 
({\mathbb Z}_2)^4$, so any
homomorphism 
${\mathbb Z}_2 \rtimes {\mathbb Z}^8 \rightarrow {\mathbb Z}_4$ will
factor through ${\mathbb Z}_2 \times ({\mathbb Z}_2)^4$.  But in the map
${\mathbb Z}_4 \rightarrow {\mathbb Z}_2$, the generator of ${\mathbb Z}_4$ 
maps onto
the generator of ${\mathbb Z}_2$.  Since ${\mathbb Z}_2 \times 
({\mathbb Z}_2)^4$ does not
contain any element of order 4, there is no map 
${\mathbb Z}_2 \times ({\mathbb Z}_2)^4
\rightarrow {\mathbb Z}_4$ that lifts the projection onto the first factor.
Therefore, the ${\mathbb Z}_2$ gerbe is nontrivial.
More generally, if $[T^4/{\mathbb Z}_{2k}]$ is a ${\mathbb Z}_k$ gerbe over
${\mathbb Z}_2$, where the $Z_{2k}$ acts by first projecting to ${\mathbb Z}_2$,
then it is nontrivial.
} ${\mathbb Z}_2$ gerbe.)
The bundle is the rank eight bundle ${\cal O}^{\oplus 8}$,
on which the ${\mathbb Z}_4$ acts (effectively) by fourth roots of unity.

We will compute the spectrum, and discover not only that it is 
consistent, but in addition that it has the same form as the
spectrum of a perturbative $E_8 \times E_8$ compactification on a space,
as expected from the duality proposal.

We use $X^{3-4}$ to denote the bosons in the $T^4$, and
$\psi^{3-4}$ their right-moving superpartners.
We shall use $\lambda^{1-8}$ to denote the free left-moving fermions
and $\lambda^{9-16}$ to denote the left-moving fermions in the bundle
above.

Let us begin the spectrum computation in the untwisted sector.

First, consider (NS,NS) states.  Here, the left- and right-moving
vacuum energies are given by $E_{\rm left} = -1$, $E_{\rm right} =
-1/2$.  The ${\mathbb Z}_4$-invariant states have the form
\begin{center}
\begin{tabular}{cc}
State & Count \\ \hline
$\left( \lambda^{1-8}_{-1/2}, \overline{\lambda}^{1-8}_{-1/2} \right)^2
\otimes \left( \psi^{1-2}_{-1/2}, \overline{\psi}^{1-2}_{-1/2} \right)$
& spacetime vector, valued in adjoint of $so(16)$ \\
$\overline{\partial} X^{1-2}_{-1} \otimes
\left( \psi^{1-2}_{-1/2}, \overline{\psi}^{1-2}_{-1/2} \right)$
& gravity, tensor multiplet contribution \\
$\left( \lambda^{9-16}_{-1/2} \overline{\lambda}^{9-16}_{-1/2} \right)
\otimes \left( \psi^{1-2}_{-1/2}, \overline{\psi}^{1-2}_{-1/2} \right)$
& spacetime vector, valued in adjoint, ${\bf 1}$ of $su(8)$ \\
& (${\bf 1}$ from the trace) \\
$\overline{\partial} X^{3-4}_{-1}
\otimes \left( \psi^{3-4}_{-1/2}, \overline{\psi}^{3-4}_{-1/2} \right)$
& 16 spacetime scalars (toroidal moduli),\\
& forming 4 hypermultiplets \\
$\left( \left( \lambda^{9-16}_{-1/2} \right)^2,
\left( \overline{\lambda}^{9-16}_{-1/2} \right)^2 \right)
\otimes \left( \psi^{3-4}_{-1/2}, \overline{\psi}^{3-4}_{-1/2} \right)$
& 4 spacetime scalars, \\
& valued in $\wedge^2 {\bf 8} = {\bf 28}$,
$\wedge^2 {\bf
\overline{8}} = {\bf \overline{28}}$ of $su(8)$, \\
& forming 1 hypermultiplet in ${\bf 28}$, \\
& another in ${\bf \overline{28}}$
\end{tabular}
\end{center}

There are no (R,NS) states in the untwisted sector, since
$E_{\rm left} > 0$.

Next, consider the twisted sector $k=1$.

In the (NS,NS) sector, fields have the following boundary conditions:
\begin{eqnarray*}
X^{1-2}(\sigma + 2 \pi) & = & + X^{1-2}(\sigma), \\
X^{3-4}(\sigma + 2 \pi) & = & - X^{3-4}(\sigma), \\
\psi^{1-2}(\sigma + 2\pi) & = & - \psi^{1-2}(\sigma), \\
\psi^{3-4}(\sigma + 2 \pi) & = & + \psi^{3-4}(\sigma), \\
\lambda^{1-8}(\sigma + 2 \pi) & = & - \lambda^{1-8}(\sigma), \\
\lambda^{9-16}(\sigma + 2 \pi) & = & - \exp\left( \frac{2\pi i}{4} \right)
\lambda^{9-16}(\sigma).
\end{eqnarray*}
It is straightforward to compute $E_{\rm left} = -1/2$,
$E_{\rm right} = 0$.
The available field modes are
\begin{displaymath}
X^{3-4}_{-1/2}, \: \: \:
\lambda^{1-8}_{-1/2}, \overline{\lambda}^{1-8}_{-1/2}, \: \: \:
\lambda^{9-16}_{-1/4}, \overline{\lambda}^{9-16}_{-3/4}.
\end{displaymath}

There is a multiplicity of right-moving Fock vacua, arising from the
periodicity of $\psi^{3-4}$.  Briefly, the vacua $| \pm \mp \rangle$
are invariant, and $| \pm \pm \rangle$ get a sign under the action of
the generator of ${\mathbb Z}_4$.

In this sector, the ${\mathbb Z}_4$- and GSO-invariant states are
\begin{center}
\begin{tabular}{cc}
State & Count \\ \hline
$\overline{\partial} X^{3-4}_{-1/2} \otimes
| \pm \pm \rangle$ & 8 spacetime scalars \\
$\left( \lambda^{9-16}_{-1/4} \right)^2 \otimes | \pm \pm \rangle$
& 2 spacetime scalars valued in $\wedge^2 {\bf 8} = {\bf 28}$ of $su(8)$
\end{tabular}
\end{center}

There are no massless states in (R,NS) in this sector, as $E_{\rm left} =
+ 1/2$.

Copies of the states in the $k=1$ sector occur at each of the sixteen fixed
points, hence the total state count should be obtained by multiplying the
totals for this sector by sixteen.

Next, consider the twisted sector $k=2$.

In the (NS,NS) sector, fields have the following boundary conditions:
\begin{eqnarray*}
X^{1-4}(\sigma + 2\pi) & = & + X^{1-4}(\sigma), \\
\psi^{1-4}(\sigma + 2 \pi) & = & - \psi^{1-4}(\sigma), \\
\lambda^{1-8}(\sigma + 2 \pi) & = & - \lambda^{1-8}(\sigma), \\
\lambda^{9-16}(\sigma + 2 \pi) & = & + \lambda^{9-16}(\sigma).
\end{eqnarray*}
It is straightforward to compute $E_{\rm left} = 0$,
$E_{\rm right} = -1/2$.  The available field modes are
\begin{displaymath}
\psi^{\mu}_{-1/2}, \overline{\psi}^{\mu}_{-1/2}, \: \: \:
\lambda^{1-8}_{-1/2}, \overline{\lambda}^{1-8}_{-1/2}.
\end{displaymath}
There is a multiplicity of left Fock vacua, arising from $\lambda^{9-16}$.
Let $|m,n\rangle$ denote a vacuum with $m$ +'s and $n$ -'s (note $m+n=8$),
then under the action of the generator of ${\mathbb Z}_4$, it is straightforward
to check that
$|m=0,4,8\rangle$ are invariant, $|m=2,6\rangle$ get a sign flip,
and the others get other fourth roots of unity.

The ${\mathbb Z}_4$- and GSO-invariant states in this sector are of the form
\begin{center}
\begin{tabular}{cc}
State & Count \\ \hline
$|m=0,4,8\rangle \otimes \left( \psi^{1-2}_{-1/2},
\overline{\psi}^{1-2}_{-1/2} \right)$ & spacetime vectors, in the
${\bf 1}$, ${\bf 1}$, $\wedge^4 {\bf 8} = {\bf 70}$ of
$su(8)$ \\
$|m=2,6\rangle \otimes \left( \psi^{3-4}_{-1/2},
\overline{\psi}^{3-4}_{-1/2} \right)$ & 1 hypermultiplet in
$\wedge^2 {\bf 8} = {\bf 28}$, $\wedge^2 {\bf \overline{8}} =
{\bf \overline{28}}$ of $su(8)$
\end{tabular}
\end{center}

The (R,NS) sector in $k=2$ is closely related.  Here, fields
have the following boundary conditions:
\begin{eqnarray*}
X^{1-4}(\sigma + 2\pi) & = & + X^{1-4}(\sigma), \\
\psi^{1-4}(\sigma + 2 \pi) & = & - \psi^{1-4}(\sigma), \\
\lambda^{1-8}(\sigma + 2 \pi) & = & + \lambda^{1-8}(\sigma), \\
\lambda^{9-16}(\sigma + 2 \pi) & = & - \lambda^{9-16}(\sigma).
\end{eqnarray*}
Just as in the (NS,NS) sector, $E_{\rm left} = 0$ and
$E_{\rm right} = -1/2$.  Here, the left Fock vacua form a spinor of
the low-energy $so(16)$.

The ${\mathbb Z}_4$-invariant states in this sector are of the form
\begin{center}
\begin{tabular}{cc}
State & Count \\ \hline
$(\mbox{spinor}) \otimes  \left( \psi^{1-2}_{-1/2},
\overline{\psi}^{1-2}_{-1/2} \right)$ & spacetime vector, in chiral spinor of
$so(16)$
\end{tabular}
\end{center}

Finally, let us consider the $k=3$ sector.
There are no massless states in (R,NS), so we only consider (NS,NS).
Fields in this sector have the following boundary conditions:
\begin{eqnarray*}
X^{1-2}(\sigma + 2\pi) & = & + X^{1-2}(\sigma), \\
X^{3-4}(\sigma + 2\pi) & = & - X^{3-4}(\sigma), \\
\psi^{1-2}(\sigma + 2\pi) & = & - \psi^{1-2}(\sigma), \\
\psi^{3-4}(\sigma + 2 \pi) & = & + \psi^{3-4}(\sigma), \\
\lambda^{1-8}(\sigma + 2\pi) & = & - \lambda^{1-8}(\sigma), \\
\lambda^{9-16}(\sigma + 2\pi) & = & \exp\left( \frac{\pi i}{2} \right)
\lambda^{9-16}(\sigma).
\end{eqnarray*}
It is straightforward to compute $E_{\rm left} = -1/2$,
$E_{\rm right} = 0$.  The available field modes are
\begin{displaymath}
\overline{\partial} X^{3-4}_{-1/2}, \: \: \:
\lambda^{1-8}_{-1/2}, \overline{\lambda}^{1-8}_{-1/2}, \: \: \:
\lambda^{9-16}_{-3/4}, \overline{\lambda}^{9-16}_{-1/4}.
\end{displaymath}
Because $\psi^{3-4}$ is periodic, there is a multiplicity of right Fock
vacua.  The states $|+-\rangle$, $|-+\rangle$ are invariant under the
generator of ${\mathbb Z}_4$, whereas the states $|++\rangle$,
$|--\rangle$ get a sign flip.

Putting this together, we find ${\mathbb Z}_4$- and GSO-invariant 
massless states
of the form:
\begin{center}
\begin{tabular}{cc}
State & Count \\ \hline
$\left( \overline{\partial} X^{3-4}_{-1/2},
\overline{\partial} \overline{X}^{3-4}_{-1/2} \right) \otimes
| \pm \pm \rangle$ & 8 scalars \\
$\left( \overline{\lambda}^{9-16}_{-1/4} \right)^2 \otimes
| \pm \pm \rangle$ & 2 sets of scalars each in the $\wedge^2 {\bf
\overline{8}} = {\bf \overline{28}}$ of $su(8)$
\end{tabular}
\end{center}

Furthermore, copies of the states above occur at each fixed point,
hence the total number of states is obtained by multiplying the tally above
by sixteen.

Now, let us summarize our results so far.  We have found the following
states:
\begin{single_itemize}
\item 1 gravity multiplet,
\item 1 tensor multiplet,
\item vector multiplets transforming in the adjoint, chiral spinor of $so(16)$,
\item vector multiplets transforming in the adjoint, ${\bf 70}$,
${\bf 1}$, ${\bf 1}$, ${\bf 1}$ of $su(8)$,
\item 10 hypermultiplets in ${\bf 28}$ of $su(8)$  ($k=0,1,2$),
\item 10 hypermultiplets in ${\bf \overline{28}}$ of $su(8)$ ($k=0,2,3$),
\item 4 ($k=0$) plus 32 ($k=1$) plus 32 ($k=3$) singlet hypermultiplets.
\end{single_itemize}

We can describe this spectrum more compactly as follows.
First, the vectors transforming in the adjoint and chiral spinor
of $so(16)$ clearly combine to form a vector in the adjoint of $e_8$.
Second, under its $su(8)$ subalgebra,
the adjoint representation of $e_7$ decomposes as \cite{slansky}[table 52]
\begin{displaymath}
{\bf 133} \: = \: {\bf 63} \: + \: {\bf 70},
\end{displaymath}
so we see that the remaining non-singlet vectors combine to form
the adjoint of $e_7$.  In the same decomposition,
\begin{displaymath}
{\bf 56} \: = \: {\bf 28} \: + \: {\bf \overline{28}},
\end{displaymath}
so we see that the hypermultiplets in the ${\bf 28}$ and
${\bf \overline{28}}$ combine to form 10 hypermultiplets in the
${\bf 56}$ of $e_7$.

Putting this together, we find that the spectrum can be described as
follows:
\begin{single_itemize}
\item 1 gravity multiplet,
\item 1 tensor multiplet,
\item vector multiplets transforming in the adjoint of $E_7 \times E_8
\times U(1)^3$,
\item 10 hypermultiplets in the ${\bf 56}$ of $E_7$,
\item 68 hypermuliplets that are singlets under $E_7 \times E_8$.
\end{single_itemize}
The number of hypermultiplets is greater than the number of vector
multiplets by 244, which is a necessary condition for anomaly cancellation.

The duality proposal in this example predicts that the dual is defined
by a heterotic $E_8 \times E_8$ compactification on 
$[T^4/{\mathbb Z}_2]$, with $E_8$ bundle defined by 
${\cal E}^* \otimes {\cal E}$, 
$\wedge^{\rm even} {\cal E}$, for ${\cal E} = {\cal O}^8$ on $T^4$,
but such that ${\cal E}^* \otimes {\cal E}$ and 
$\wedge^{\rm even} {\cal E}$ are odd under the
action of the ${\mathbb Z}_2$ defining $[T^4/{\mathbb Z}_2]$.
We do not see how such an $E_8$ bundle on $[T^4/{\mathbb Z}_2]$ could
be obtained from embedding an $SU(n)$ bundle in the usual fashion, and indeed,
as remarked earlier, it need not be, the duals in general may only be
describable by fibered WZW models.  That said, the reader should note
that the spectrum computed above
is nearly the same as the massless spectrum of
an $E_8 \times E_8$ string compactified on a (2,2) $[T^4/{\mathbb Z}_2]$,
which in general terms is consistent with the
existence of a duality between the current
Spin$(32)/{\mathbb Z}_2$ gerbe compactification and an $E_8 \times
E_8$ compactification.  So, although we cannot check the details at this
time, certainly in broad brushstrokes this is consistent.

\subsection{Examples in Distler-Kachru models}

In table~\ref{table:DK-duality-exs}
we tabulate the combinatorial data for a number of anomaly-free
Distler-Kachru
(0,2) GLSM's of the pertinent form.  Each describes a bundle ${\cal E}$ over
a Calabi-Yau hypersurface in a weighted projective stack, 
\begin{displaymath}
{\mathbb P}^n_{[w_0, \cdots, w_n]}[w_0 + \cdots + w_n],
\end{displaymath}
a ${\mathbb Z}_2$ gerbe over a Calabi-Yau space,
where the (rank 8)
bundle is given as a kernel of the form
\begin{displaymath}
0 \: \longrightarrow \: {\cal E} \: \longrightarrow
\: \oplus_a {\cal O}(n_a) \: \longrightarrow \:
\oplus_i {\cal O}(m_i) \: \longrightarrow \: 0.
\end{displaymath}

\begin{table}
\centering
\begin{tabular}{cccc}
$w_0, \cdots, w_4$ &  $n_a$ & $m_i$ \\ \hline
$2,2,2,4$ &  $1^9$ & $9$ \\
$2,2,2,2,2$ &  $1^9, 9$ & $7, 11$ \\
$2,2,2,2,2$ &  $3^9, 19$ & $9, 21$ \\
$2,2,2,2,4$ &  $3^9, 27$ & $9, 29$ \\
$2,2,2,4,6$ &  $1^8, 5^2$ &  $9, 13$ \\
$2,2,2,2,6$ &  $1^9, 9$ & $3, 15$ \\
$2,2,2,4,4$ &  $1^9, 15$ & $5,19$ 
\end{tabular}
\caption{This table lists combinatorial data for anomaly-free (0,2) GLSM's
describing rank 8 bundles over ${\mathbb Z}_2$ gerbes on Calabi-Yau's.}
\label{table:DK-duality-exs}
\end{table}

For example, the first entry in table~\ref{table:DK-duality-exs}
describes a rank 8 bundle given as a kernel
\begin{displaymath}
0 \: \longrightarrow \: {\cal E} \: \longrightarrow \:
\oplus_1^9 {\cal O}(1) \: \longrightarrow \: {\cal O}(9) \:
\longrightarrow \: 0
\end{displaymath}
over the stack ${\mathbb P}^3_{[2,2,2,4]}[10]$, a ${\mathbb Z}_2$
gerbe over ${\mathbb P}^3_{[1,1,1,2]}[5]$.

We list a few rank 9 examples over ${\mathbb Z}_3$ gerbes in
section~\ref{sect:otherexs-good}.
These rank 8 examples are listed in this section because we are
enumerating rank 8 bundles over ${\mathbb Z}_2$ gerbes, and the
rank 9 examples are not candidates for the dualities discussed here.

Curiously, 
we were unable to find solutions of the combinatorial consistency conditions
for GLSM's for bundles of rank less than 8.  We do not know whether this
reflects a fundamental limitation of GLSM's, or merely the inadequacy
of our parameter space search.

Given a Distler-Kachru model with a phase describing a Landau-Ginzburg model
over an orbifold of a vector space, methods exist to compute the massless
spectrum in that Landau-Ginzburg phase \cite{kw,dk1}.
When these methods are applied to, for example, 
heterotic ${\rm Spin}(32)/{\mathbb Z}_2$ compactifications on 
typical examples from the table above,
we find a large number of single vectors and matter representations which
likely combine to form representations of a larger nonabelian gauge symmetry,
but unfortunately the corresponding worldsheet global symmetry does not seem
to be visible in the UV.  We conclude that in these examples, much of the
needed worldsheet global symmetry appears in the IR, where we have no
direct access.  This is atypical of Distler-Kachru models, where
spacetime gauge symmetries typically appear as worldsheet global symmetries
visible in the UV, but is not contradicted by any physics we know.
In any event, spectrum computations at Landau-Ginzburg points in these
theories have not proven insightful.

\section{Class III:  Twisted bundles}  \label{sect:type3:twisted}

The third fundamental class of examples we shall discuss
involve cases in which the trivially-acting part of the gauge
group acts nontrivially on the bundle, but is not one of the
special cases discussed in section~\ref{sect:het-gsomods} in which the
effect is merely to recreate part of the ten-dimensional
left-moving GSO projection.  

One reason for interest is that examples of this form have the potential
to define new heterotic string compactifications.  Other reasons also
exist, revolving around making sense of heterotic orbifolds with
invariant non-equivariant bundles.  We review such motivations
in subsection~\ref{sect:class3-motivations}.

In subsection~\ref{sect:otherexs-good}, 
we describe some (indirect) constructions of
(0,2) SCFT's of this form, via dimensional reduction of consistent
four-dimensional theories, and via anomaly-free (0,2) GLSM's.  

Unfortunately, although there seem to exist
consistent (0,2) SCFT's, they do not seem to yield consistent 
perturbative heterotic
string compactifications.
The essential problem is that any finite group that acts only on
left-movers, locally looks like a modification of the
ten-dimensional left-moving GSO projection, and as the consistent
ten-dimensional GSO projections are already known, if it is not one of
them, the results cannot be well-behaved.  We will give several examples
of six-dimensional compactifications of this form, in which the six
dimensional theory has anomalies and cannot be consistent.
We outline in detail some examples in which heterotic string
compactifications on these (0,2) SCFT's break down in 
subsections~\ref{sect:class3-caution1}, \ref{sect:class3-caution2},
and \ref{sect:class3-caution4}.

That said, it may sometimes be possible to restore these theories by
adding suitable phases to twisted sectors.  For example, the
ten-dimensional nonsupersymmetric
$SO(8) \times SO(24)$ string seems to be obtainable by
a procedure along these lines.  Specifically, in the worldsheet theory,
if one takes the Spin$(32)/{\mathbb Z}_2$ string and performs an
additional left-moving ${\mathbb Z}_2$ orbifold on 4 complex fermions,
the result satisfies level-matching but does not define a modular-invariant
theory; if one then adds phases to restore modular invariance, the result
is the nonsupersymmetric ten-dimensional $SO(8) \times SO(24)$ string.
(See {\it e.g.} \cite{dix-harv,klt-ns}, \cite{polv2}[section 11.3] 
for more information on this nonsupersymmetric
string.)
Unfortunately, we do not have a procedure
for finding such phases (or even checking whether they exist), and if they
do, the previous example suggests that the results will not be supersymmetric.
In addition, see {\it e.g.} \cite{afiu,afiuv} for a different set of ideas
which may be relevant, though we have not considered them carefully in this
context.

In subsection~\ref{sect:possible-anomcanc} 
we outline a few attempts to find a way to 
understand these issues in terms of some sort of
anomaly cancellation.

\subsection{Motivations}  \label{sect:class3-motivations}

One reason for interest
in this class of examples is that they potentially could describe
new (0,2) SCFT's.

Another reason to be interested in them is that they may give a way
of understanding heterotic compactifications on ordinary spaces but
with non-equivariant bundles.  In this section we will explain this motivation.

Let $X$ be a Calabi-Yau manifold, with stable holomorphic vector bundle
${\cal E} \rightarrow X$ satisfying anomaly cancellation, so that the
pair $(X, {\cal E})$ defines a consistent large-radius heterotic
Calabi-Yau compactification.

Now, suppose a finite group $G$ acts on $X$.  In order to construct
a $G$-orbifold of the heterotic string on $(X, {\cal E})$, we need
for the bundle ${\cal E}$ to admit a $G$-equivariantizable structure,
which means that for every $g \in G$, we need a lift
$\tilde{g}: {\cal E} \rightarrow {\cal E}$ such that
\begin{displaymath} 
\xymatrix{
{\cal E} \ar[r]^{\tilde{g}} \ar[d] & {\cal E} \ar[d] \\
X \ar[r]^{g} & X
}
\end{displaymath}
and also such that the lifts obey the group law:  $\tilde{g} \circ
\tilde{h} = \widetilde{gh}$.

We need such an equivariant structure on the bundle ${\cal E}$ for
the following two reasons:
\begin{itemize}
\item In the worldsheet theory, such an equivariant structure enables us to
define a group action on the worldsheet fermions/bosons describing the
bundle, such that summing over twisted sectors in the orbifold yields
an honest projection operator onto $G$-invariant states.
\item In the low-energy supergravity, if ${\cal E}$ does not have
an equivariant structure, then even if $G$ acts freely, on the quotient
$X/G$ the bundle ${\cal E}$ will descend to a `twisted' bundle, not an
honest bundle, whose transition functions $g_{\alpha \beta}$ obey
\begin{displaymath}
g_{\alpha \beta} g_{\beta \gamma} g_{\gamma \alpha} \: = \: h_{\alpha
\beta \gamma} I
\end{displaymath}
on triple overlaps, and whose gauge field $A$ obeys
\begin{displaymath}
A_{\beta} \: = \: g_{\alpha \beta} A_{\alpha} g_{\alpha \beta}^{-1}
\: + \: g_{\alpha \beta}^{-1} d g_{\alpha \beta} \: + \:
\Lambda_{\alpha \beta} I
\end{displaymath}
across intersections, for some affine translation $\Lambda_{\alpha \beta}$.
As ten-dimensional super-Yang-Mills only describes honest bundles and
ordinary gauge transformations, the structure above cannot be used to
define a consistent string compactification.
\end{itemize}

However, there is a workaround.  If the bundle ${\cal E}$ is
invariant (meaning, its characteristic classes are invariant under the
group action), but not equivariant, then we can find a larger group
$\tilde{G}$, an extension of $G$ by a trivially-acting subgroup, such that
${\cal E}$ does admit a $\tilde{G}$-equivariant structure, and then take a
$G'$ orbifold.  This is precisely an example of a heterotic string on
a gerbe, in this case a gerbe over $[X/G]$.

First, let us review some generalities on the construction of $G'$.
There is a `universal', `maximal' extension $\tilde{G}_{\rm max}$, 
which extends
$G$ by the group of all automorphisms of the total space of
${\cal E}$ that cover the action of the elements of $G$ on $X$.
It fits into a short exact sequence
\begin{displaymath}
1 \: \longrightarrow \: {\rm Aut}({\cal E}) \: \longrightarrow \:
\tilde{G}_{\rm max} \: \longrightarrow \: G \: \longrightarrow \: 1,
\end{displaymath}
where ${\rm Aut}({\cal E})$ is the group of global bundle automorphisms
of ${\cal E}$.
The group we want, $\tilde{G}$, will necessarily be a subgroup of
this universal extension of $\tilde{G}_{\rm max}$.

In general, the extension defining $\tilde{G}_{\rm max}$ will not be central,
but if ${\cal E}$ is stable or simple then ${\rm Aut}({\cal E}) = 
{\mathbb C}^{\times}$, and the extension is central.
The group $\tilde{G}_{\rm max}$ 
acts by definition on ${\cal E}$ and so defines an
equivariant structure.  Every other group for which one has an equivariant
structure will map to $\tilde{G}_{\rm max}$ 
and the equivariant structure will factor
through that map.

Now, clearly, $\tilde{G}_{\rm max}$ is not a finite group, and we only want to
consider cases in which the trivially-acting subgroup is finite.
If $G$ is finite and ${\cal E}$ is stable or simple, 
then $\tilde{G}_{\rm max}$ is
a central extension of $G$ by ${\mathbb C}^{\times}$ and, because
\begin{displaymath}
H^2(G,{\mathbb C}^*) \: = \: H^2(G,{\mathbb Q}/{\mathbb Z})
\end{displaymath}
for $G$ finite, the relevant $H^2$ is finite and so every extension
is induced from some central extension $G_{min}$ of $G$ by
a finite group of order 
bounded by the maximal order of elements in $G$.
In this fashion, we can construct a $\tilde{G}$.

So far, we have described how, given a bundle that is invariant but
not equivariant with respect to an orbifold group $G$, one can extend
$G$ to a larger finite group $\tilde{G}$, where the extension acts
trivially on the base.  Now, not any $\tilde{G}$ will be acceptable:
the orbifold by $\tilde{G}$ must, at minimum, satisfy
level-matching, and as discussed earlier, even more in order to define
a consistent heterotic string compactification.

For completeness, let us now consider some possible examples.

One example is described in the paper \cite{dopr}.
(See also 
\cite{Donagi:2000zf,Donagi:2000zs,Donagi:2000fw,Ovrut:2002jk,Ovrut:2003zj,
Braun:2004xv}.)
In that paper, the authors first construct an elliptically-fibered
Calabi-Yau threefold $Z$ with fundamental group ${\mathbb Z}_2 \times
{\mathbb Z}_2$, built as a freely-acting\footnote{
For further examples of Calabi-Yau threefolds with this property, see
{\it e.g.} \cite{cd}.  Examples include ${\mathbb P}^7[2,2,2,2]$ and
$({\mathbb P}^1)^4$ with a degree (2,2,2,2) hypersurface.  For both,
the restriction of an ambient hyperplane class to the Calabi-Yau defines
a line bundle which is invariant
but not equivariant.
}
${\mathbb Z}_2 \times {\mathbb Z}_2$ quotient of a simply-connected
Calabi-Yau threefold $X$:
\begin{displaymath}
Z \: = \: X / ( {\mathbb Z}_2 \times {\mathbb Z}_2 )
\end{displaymath}
together with a bundle $V$ on $X$ that is not quite equivariant
with respect to the ${\mathbb Z}_2 \times {\mathbb Z}_2$ action, and so
descends to a twisted bundle on $Z$.

Consider the gerbe presented as $[X/G]$, where
\begin{displaymath}
1 \: \longrightarrow \: {\mathbb Z}_2 \: \longrightarrow \:
G \: \longrightarrow \:
{\mathbb Z}_2 \times {\mathbb Z}_2 \: \longrightarrow \:
1,
\end{displaymath}
with the ${\mathbb Z}_2$ kernel acting trivially.
(Explicitly, the extension above is the Heisenberg extension,
and $G = D_4$ \cite{tonypriv}.)
The bundle $V$ above descends to a bundle on a gerbe.
Furthermore, the entire bundle is an eigenbundle under the nontrivial
element of the center of $D_4$, with eigenvalue $-1$ (since it must square
to the identity and can not itself be the identity) \cite{tonypriv}.

For completeness,
let us now work through the example of \cite{dopr} in more detail.
Their Calabi-Yau manifold $X$ is an elliptic fibration over a
rational elliptic surface, and in fact can be described as the fiber
product over ${\mathbb P}^1$ of two rational elliptic surfaces $B$, $B'$:
\begin{displaymath}
X \: = \: B \times_{ {\mathbb P}^1 } B'
\end{displaymath}
where $\pi: X \rightarrow B'$, $\pi': X \rightarrow B$,
$\beta': B' \rightarrow {\mathbb P}^1$, $\beta: B \rightarrow B$:
\begin{displaymath}
\xymatrix{
& X \ar[rd]^{\pi} \ar[ld]_{\pi'} & \\
B \ar[rd]_{\beta} & & B' \ar[ld]^{\beta'} \\
& {\mathbb P}^1 &
}
\end{displaymath}
$B$ and $B'$ are both chosen to admit an automorphism group containing
${\mathbb Z}_2 \times {\mathbb Z}_2$.
A stable rank four vector bundle $V \rightarrow X$ is constructed
as an extension
\begin{displaymath}
0 \: \longrightarrow \: V_1 \: \longrightarrow \: V \: \longrightarrow \:
V_2 \: \longrightarrow \: 0,
\end{displaymath}
where
\begin{displaymath}
V_i \: = \: \pi'^* W_i \otimes \pi^* L_i,
\end{displaymath}
for $W_i$ a pair of rank 2 vector bundles on $B$ and
$L_i$ a pair of line bundles on $B'$.

Briefly, \cite{dopr} first argues that each $V_i$ is
${\mathbb Z}_2\times {\mathbb Z}_2$-equivariant.  As a result, the group
of extensions ${\rm Ext}^1(V_2,V_1)$ decomposes into subspaces associated
with characters of ${\mathbb Z}_2 \times {\mathbb Z}_2$.  
By picking an extension
in a subspace associated with the trivial representation, we get a bundle
$V$ which is at least ${\mathbb Z}_2 \times {\mathbb Z}_2$-invariant,
though not necessarily ${\mathbb Z}_2 \times {\mathbb Z}_2$-equivariant.

Again, for this example to be physically meaningful, the
orbifold group would have to, at minimum, satisfy level-matching.
As our purpose in this section was merely to outline one of the motivations
for considering heterotic compactifications on gerbes, and we will argue
later that these examples are, in most cases, not physically useful,
we will end our discussion here.

\subsection{Constructions of consistent CFT's}
\label{sect:otherexs-good}

In this section, we will describe some constructions of what seem to be
consistent (0,2) SCFT's describing heterotic strings on gerbes with
fractional gauge bundles.  For reasons described elsewhere, these
cannot be consistently used in supersymmetric heterotic string
compactifications, but nevertheless they do seem to be examples of consistent
(0,2) SCFT's.

Our first example was discussed in \cite{anton1}.
Specifically, in
\cite{anton1}[section 3.1], an ${\cal N}=2$ gauge theory in four
dimensions with hypermultiplets transforming in the $R$ representation
of the gauge group,
was reduced along a Riemann surface $C$ to a two-dimensional
$(0,4)$ theory, a heterotic nonlinear sigma model whose target is
the Hitchin moduli space ${\cal M}_H(G,C)$ and with a twisted
gauge bundle ${\cal R}$, defined by the representation $R$ in which
the hypermultiplets transform.  The four-dimensional theory was
partially topologically twisted along a $U(1)_R$ (and only exists
for superconformal field theories for which that $U(1)_R$ is nonanomalous).

In this example, the
gauge bundle is twisted, in the sense that the transition functions
only close to a cocycle on triple overlaps.
Now, ordinarily heterotic strings cannot couple to such twisted bundles,
only D-branes can couple to such twisted bundles, as described in
section~\ref{sect:twisting}.
Despite that fact, it was claimed in \cite{anton1}[section 3.1] that the $(0,4)$
theory nevertheless consistently couples to a twisted gauge bundle.
In order to make that possible, the nonlinear sigma model was restricted
to maps such that the pullback of the twisted bundle, is an honest bundle.

Such nonlinear sigma models, with a restriction on nonperturbative
sectors, are equivalent to sigma models on gerbes, as reviewed in
{\it e.g.} section~\ref{sect:generalreview}, and so this is an
example of a heterotic string compactification on a gerbe with a
non-pullback bundle.  

More generally, more of
the
analysis of \cite{anton1} can also be rephrased in this language,
following a discussion in \cite{summ}[section 12.3], which discussed
how gerbes could be used
to slightly simplify the analysis of the physical realization of
geometric Langlands.  Briefly,
Hitchin moduli spaces arising from $G$ gauge theories are defined by
modding out adjoint actions, under which the center $Z(G)$ is trivial
and so formally one can replace them with moduli stacks which are
$Z(G)$-gerbes.  After reduction to two dimensions, one obtains sigma
models on gerbes, which physics sees \cite{summ} as a sigma model on a
disjoint union of spaces, matching results of \cite{edanton}.

In any event, after performing the dimensional reduction from
a four-dimensional ${\cal N}=2$ theory to two dimensions, one gets
\cite{anton1}[section 3.1] a heterotic sigma model on the
Hitchin moduli space ${\cal M}_H(G,C)$, with a twisted bundle over that
moduli space, twisted by an element of $H^2(Z(G))$.  Since the Hitchin
moduli space is defined by modding out the adjoint action of $G$,
the center is trivial, and so one could naturally replace the
Hitchin moduli space with a moduli stack which is a $Z(G)$ gerbe,
just as in \cite{summ}[section 12.3].
A heterotic sigma model on such a stack would appear to be a sigma
model on the moduli space but with a restriction on allowed maps,
exactly as described in \cite{anton1}[section 3.1].

As these two-dimensional (0,2) theories are constructed by
dimensional reduction of a consistent four-dimensional theory,
it is difficult to believe that they are not consistent.

Other naively-consistent examples can be constructed in
(0,2) GLSM's.
For example, consider the two examples:
\begin{itemize}
\item The rank 9 bundle
\begin{displaymath}
0 \: \longrightarrow \: {\cal E} \: \longrightarrow \:
\bigoplus_1^9 {\cal O}(1) \oplus {\cal O}(10) \: \longrightarrow \:
{\cal O}(19) \: \longrightarrow \: 0
\end{displaymath}
over ${\mathbb P}^4_{[3,3,3,3,6]}[18]$, a ${\mathbb Z}_3$ gerbe over
${\mathbb P}^4_{[1,1,1,1,2]}[6]$,
\item The rank 9 bundle
\begin{displaymath}
0 \: \longrightarrow \: {\cal E} \: \longrightarrow \:
\bigoplus_1^9 {\cal O}(1) \oplus {\cal O}(13) \: \longrightarrow \:
{\cal O}(22) \: \longrightarrow \: 0
\end{displaymath}
over ${\mathbb P}^3_{[3,3,6,9]}[21]$, a ${\mathbb Z}_3$ gerbe over
${\mathbb P}^3_{[1,1,2,3]}[7]$.
\end{itemize}
It is straightforward to check, just at the level of combinatorics,
that they satisfy the usual conditions for a GLSM to be anomaly-free.
However, the usual danger with GLSM's is that we do not have perfect
control over the RG flow -- although we have described them in terms
of data associated to twisted bundles, along the RG flow they might
pick up `phases' (as suggested earlier), for example.

In the next subsections, we shall show explicitly that examples of
this form do not yield consistent supersymmetric heterotic string
compactifications, unfortunately.

\subsection{Cautionary example}  \label{sect:class3-caution1}

Let ${\cal E}$ be a rank 4 bundle on a Calabi-Yau $X$, defining a consistent
(0,2) SCFT.  Now, consider a ${\mathbb Z}_2$ orbifold 
in which the orbifold group
acts trivially on $X$, but by a sign flip on ${\cal E}$ (so that all of
${\cal E}$ is an eigenbundle of weight $-1$).

This example can be shown to satisfy level-matching in the sense
of \cite{freedvafa}, as well as the conditions in
appendix~\ref{app:spectra:fockconstraints}.
However, in principle this theory is nevertheless deeply suspicious.  Since the 
${\mathbb Z}_2$
acts trivially on right-moving fields, and only on left-moving fields,
we could just as well think of this as a compactification of a 
ten-dimensional theory in which the left-moving GSO projection has been
altered.  Since the resulting new GSO does not coincide with either
the existing Spin$(32)/{\mathbb Z}_2$ or $E_8 \times E_8$ strings, this
theory must be inconsistent.  (Indeed, this is the starting point for
one construction of the ten-dimensional nonsupersymmetric
$SO(8) \times SO(24)$ string \cite{dix-harv,klt-ns,polv2}, 
though this orbifold must be 
supplemented by further phases.)

Another argument for inconsistency arises from considering massless spectra.
Specifically, if we take $X$ to be a K3 surface, and consider a compactification
of a ten-dimensional $E_8 \times E_8$ string, in which the gauge
bundle is embedded in one $E_8$, then the six-dimensional
spectrum is anomalous.  We summarize the details below, following the
methods outlined in appendix~\ref{app:spectra}.
(The integer $n$ is the dimension of $X$; we will quickly
specialize to $n=2$, but will remain general for as long as possible.)

Following the appendix, there are two components in the inertia stack,
which are identical:
\begin{displaymath}
I_{\mathfrak{X}} \: = \: \mathfrak{X} \amalg \mathfrak{X}.
\end{displaymath}
Below we list results for both states and left and right $U(1)_R$
charges.

First, consider the untwisted sector.

In the (R,R) sector, the vacuum energy $E_{{\rm Id}} = 0$.
The massless charged states are
\begin{itemize}
\item $H^m(X, \wedge^{\rm even} {\cal E})$, charge $({\rm even}-2, m-n/2)$,
giving spacetime states valued 
in a spinor of $so(8)$.
\end{itemize}

In the (NS,R) sector, the vacuum energy $E_{{\rm Id}} = -1$.
The massless charged states are
\begin{itemize}
\item $H^m(X, {\cal E}^* \otimes {\cal E})$, charge $(0,m-n/2)$,
spacetime gauge neutral,
\item $H^m(X, \wedge^2 {\cal E})$, charge $(2,m-n/2)$,
spacetime gauge neutral,
\item $H^m(X, {\cal O})$, charge $(0,m-n/2)$,
in the adjoint representation of $so(8)$,
\item $H^m(X, \wedge^2 {\cal E}^*)$, charge $(-2,m-n/2)$,
spacetime gauge neutral.
\end{itemize}

Now, consider the twisted sector.
Here, all of ${\cal E}$ is an eigenbundle with eigenvalue $-1$.

In the (R,R) sector, $E = -1/2$.
There are no massless charged states
in this sector.

In the (NS,R) sector, $E = -1/2$.
Again, there are no massless charged states
in this sector.

States above are listed with charges $(q_-,q_+)$.
The $q_+$ charge distinguishes chiral multiplets from vector multiplets;
the $q_-$ charge is the charge of the $u(1)$ that combines with
$so(8)$ to build $so(10)$.

For a compactification to four dimensions, ($n=3$,)
states with $q_+ = -1/2$ would be spacetime fermions in chiral multiplets
(and $q_+ = +1/2$ their antichiral partners); states with $q_+ = +3/2$
would be spacetime fermions in vector multiplets (and $q_+ = -3/2$ their
partners).

For a compactification to six dimensions, ($n=2$,) which is the
pertinent case,
states with $q_+ = \pm 1$ are spacetime fermions in vector multiplets;
states with $q_+ = 0$ are spacetime fermions in hypermultiplets.

Since we have a rank 4 bundle, in principle the $E_8$ should be broken to
${\rm Spin}(10)$, which in the worldsheet theory will be assembled
from representations of $so(8) \times u(1)$ (the $so(8)$ rotating the
remaining free left-moving fermions in the first $E_8$, and the $u(1)$
being an overall phase rotation on the bundle fermions, which on the (2,2)
locus would become the left R symmetry). 
Under the $so(8)\times u(1)$ subalgebra, representations of $so(10)$
decompose as follows:
\begin{eqnarray*}
{\bf 45} & = & {\bf 8}_{-2} \oplus {\bf 28}_0 \oplus {\bf 1}_0 \oplus
{\bf 8}_2, \\
{\bf 16} & = & {\bf 8}_{-1} \oplus {\bf 8}_{+1}, \\
{\bf 10} & = & {\bf 1}_{-2} \oplus {\bf 8}_0 \oplus {\bf 1}_{2}, \\
{\bf 1} & = & {\bf 1}_0,
\end{eqnarray*}
where the subscript indicates the $q_-$ charge.

We arrange the (untwisted sector) states into $so(10)$
representations, with the following results:
\begin{itemize}
\item The adjoint of $so(10)$ arises from $H^*(X, {\cal O})$.
Contributing terms are:
\begin{itemize}
\item $H^*(X, {\cal O})$ in (R,R), transforming as ${\bf 8}_{-2}$,
\item $H^*(X, \wedge^4 {\cal E} \cong {\cal O})$ in (R,R),
transforming as ${\bf 8}_{+2}$,
\item $H^*(X, {\rm Tr} \, {\cal E}^* \otimes {\cal E} \cong {\cal O})$
in (NS,R), transforming as ${\bf 1}_0$,
\item $H^*(X, {\cal O})$ in (NS,R), transforming as ${\bf 28}_0$.
\end{itemize}
\item Copies of ${\bf 10}$ of $so(10)$ arise from
$H^*(X, \wedge^2 {\cal E})$.  Contributing terms are:
\begin{itemize}
\item $H^*(X, \wedge^2 {\cal E})$ in (R,R), transforming as
${\bf 8}_0$,
\item $H^*(X, \wedge^2 {\cal E})$ in (NS,R), transforming as
${\bf 1}_2$,
\item $H^*(X, \wedge^2 {\cal E}^* \cong \wedge^2 {\cal E})$ in (NS,R),
transforming as ${\bf 1}_{-2}$.
\end{itemize}
\item Gauge singlets, arising as $H^*(X, {\rm End} \, {\cal E})$
(where we use End to denote the traceless endomorphisms), arising in the
(NS,R) sector.
\end{itemize}

In addition, there is one vector in the adjoint representation of the
second $E_8$, which is always present in computations of the form
of appendix~\ref{app:spectra}.

In any event, altogether in this six-dimensional theory we have
\begin{single_itemize}
\item $h^0(X, {\cal O})=1$ vector multiplets in the adjoint of
${\rm Spin}(10)$,
\item One vector multiplet in the adjoint of $E_8$,
\item $h^1(X, \wedge^2 {\cal E}) = 36$ half-hypermultiplets\footnote{
The dimension of this sheaf cohomology group can be determined from index
theory, and applies to any stable irreducible rank 4 bundle ${\cal E}$ on a K3
surface.
} in the
${\bf 10}$ of ${\rm Spin}(10)$,
\item 20 singlet hypermultiplets for $K3$ moduli,
\item $h^1({\rm End}\, {\cal E}) = 162$ singlet half-hypermultiplets for
bundle moduli\footnote{
It is a standard result that the moduli in an irreducible
rank $r$ vector bundle ${\cal E}$ on K3 with 
$c_1({\cal E}) = 0$, $c_2({\cal E}) = c_2(T K3)$ is
encoded in $24r + 1 - r^2$ hypermultiplets, 
or $2(24r+1-r^2$ half-hypermultiplets.  Here, $r=4$. 
},    
\end{single_itemize}
so that we find
\begin{eqnarray*}
n_V & = & 45 \: + \: 248 \: = \: 293, \\
n_H & = & (1/2)\left( (36)(10) \: + \: 162 \right) \: + \: 20 \: = \: 281, \\
n_H - n_V & = & -12 \: \neq \: 244,
\end{eqnarray*}
and so we see that this cannot satisfy anomaly cancellation,
mechanically verifying our previous observation that this theory cannot
be consistent.

More generally, any heterotic compactification on a gerbe, in which the
bundle is twisted, will be of this same general type, unless
the bundle has rank 8 and the trivially-acting group is ${\mathbb Z}_2$.
Locally each theory will look like a compactification
of a ten-dimensional theory with an altered GSO projection,
and except for the case that the GSO projection switches between
Spin$(32)/{\mathbb Z}_2$ and $E_8 \times E_8$, the resulting theory cannot
be consistent.

For purposes of comparison, and to help illuminate the methods encoded
in appendix~\ref{app:spectra}, let us also outline the results
in a closely related consistent compactification.  If we did not orbifold,
if we took a compactification of an $E_8 \times
E_8$ heterotic string on a smooth large-radius $K3$ with a rank 4
vector bundle, then from a similar computation we would find
\begin{single_itemize}
\item $h^0(X, {\cal O})=1$ vector multiplets in the adjoint of
${\rm Spin}(10)$,
\item One vector multiplet in the adjoint of $E_8$,
\item $h^1(X, {\cal E}) = 16$ half-hypermultiplets in the
${\bf 16}$ of ${\rm Spin}(10)$,
\item $h^1(X, \Lambda^2 {\cal E}) = 36$ half-hypermultiplets in the
${\bf 10}$ of ${\rm Spin}(10)$,
\item $h^1(X, \Lambda^3 {\cal E} = {\cal E}^*) = 16$
half-hypermultiplets in the ${\bf 16}$ of ${\rm Spin}(10)$,
\item 20 singlet hypermultiplets for $K3$ moduli,
\item $h^1({\rm End}\, {\cal E}) = 162$ singlet half-hypermultiplets for
bundle moduli,
\end{single_itemize}
where the representations of ${\rm Spin}(10)$ are 
constructed in the same fashion.
Altogether, we find that
\begin{eqnarray*}
n_V & = & 45 \: + \: 248 \: = \: 293, \\
n_H & = & (1/2)\left( (16)(16) \: + \: (36)(10) \: + \: (16)(16)
\: + \: 162 \right) \: + \: 20 \: = \: 537,
\\
n_H - n_V & = & 244,
\end{eqnarray*}
consistent with anomaly cancellation, in that standard compactification.
Unfortunately, our gerbe example is not so well-behaved.

\subsection{Second cautionary example}  \label{sect:class3-caution2}

For completeness, we give here a second cautionary example,
here involving a heterotic Spin$(32)/{\mathbb Z}_2$ compactification on
a nontrivial toroidal orbifold.
This will involve a rank 10 bundle over
a ${\mathbb Z}_2$ gerbe on $[T^4/{\mathbb Z}_2]$, 
and although level matching holds, the spectrum
is anomalous in six dimensions.

The ${\mathbb Z}_2$ gerbe is defined by $[T^4/{\mathbb Z}_4]$, where
the ${\mathbb Z}_4$ acts on the $T^4$ by
\begin{displaymath}
x \: \mapsto \: \exp\left( \frac{2\pi i (2k)k}{4} \right) x
\: = \: (-)^k x,
\end{displaymath}
so that there is a trivially-acting ${\mathbb Z}_2$ subgroup.
(This is the same ${\mathbb Z}_2$ gerbe discussed in 
a different context in section~\ref{sect:class2-ex1}.)
The gauge bundle is a rank 10 bundle,
where the generator of ${\mathbb Z}_4$ acts on an ${\cal O}^{\oplus 2}$ factor
by multiplication by $\exp(2 \pi i (2/4) ) = -1$, and on the
${\cal O}^{\oplus 8}$ factor by $\exp(2 \pi i /4)$.
It is straightforward to check that this satisfies level-matching,
in the sense of \cite{freedvafa}, as well as the conditions in
appendix~\ref{app:spectra:fockconstraints}.

Let us now outline the massless spectrum.

In the untwisted sector, there are massless states in the (NS,NS) sector.
It is straightforward to compute $E_{\rm left} = -1$,
$E_{\rm right} = -1/2$, and one has ${\mathbb Z}_4$-invariant states of the
form
\begin{center}
\begin{tabular}{cc}
State & Count \\ \hline
$\left( \lambda^{1-6}_{-1/2}, \overline{\lambda}^{1-6}_{-1/2} \right)^2
\otimes \left( \psi^{1-2}_{-1/2}, \overline{\psi}^{1-2}_{-1/2}
\right)$ & spacetime vector, \\
& valued in adjoint of $so(12)$ \\
$\overline{\partial} X^{1-2}_{-1} \otimes
\left( \psi^{1-2}_{-1/2}, \overline{\psi}^{1-2}_{-1/2}
\right)$ & gravity, tensor multiplet contributions \\
$\left( \lambda^{7-14}_{-1/2} \overline{\lambda}^{7-14}_{-1/2} \right)
\otimes \left( \psi^{1-2}_{-1/2}, \overline{\psi}^{1-2}_{-1/2}
\right)$ & spacetime vector, \\
& valued in adjoint, ${\bf 1}$ (trace) of
$su(8)$ \\
$\left( \lambda^{15-16}_{-1/2}, \overline{\lambda}^{15-16}_{-1/2} \right)^2
\otimes \left( \psi^{1-2}_{-1/2}, \overline{\psi}^{1-2}_{-1/2}
\right)$ & spacetime vector, \\
& valued in adjoint of $so(4)$ \\
$\left( \lambda^{1-6}_{-1/2}, \overline{\lambda}^{1-6}_{-1/2} \right)
\left( \lambda^{15-16}_{-1/2}, \overline{\lambda}^{15-16}_{-1/2} \right)
\otimes \left( \psi^{3-4}_{-1/2}, \overline{\psi}^{3-4}_{-1/2}
\right)$ & 4 sets of scalars, \\
& valued in $({\bf 12},{\bf 4})$ of $so(12) \times so(4)$\\
$\overline{\partial} X^{3-4}_{-1}
\otimes \left( \psi^{3-4}_{-1/2}, \overline{\psi}^{3-4}_{-1/2}
\right)$ & 16 scalars (toroidal moduli) \\
$\left( \left( \lambda^{7-14}_{-1/2} \right)^2,
\left( \overline{\lambda}^{7-14}_{-1/2} \right)^2 \right)
\otimes \left( \psi^{3-4}_{-1/2}, \overline{\psi}^{3-4}_{-1/2}
\right)$ & 4 sets of scalars, \\
& valued in $\wedge^2 {\bf 8} = {\bf 28}$,
$\wedge^2 {\bf \overline{8}} = {\bf \overline{28}}$ of $su(8)$
\end{tabular}
\end{center}

There are no massless states in the untwisted (R,NS) sector,
and in fact also no massless states in the $k=1$ or $k=3$ sectors.

All of the remaining massless states are in the $k=2$ sector.
In the (NS,NS) sector, fields have the following boundary conditions:
\begin{eqnarray*}
X^{1-2}(\sigma + 2 \pi) & = & + X^{1-2}(\sigma), \\
X^{3-4}(\sigma + 2 \pi) & = & + X^{3-4}(\sigma), \\
\psi^{1-2}(\sigma + 2 \pi) & = & - \psi^{1-2}(\sigma), \\
\psi^{3-4}(\sigma + 2 \pi) & = & - \psi^{3-4}(\sigma), \\
\lambda^{1-6}(\sigma + 2 \pi) & = & - \lambda^{1-6}(\sigma), \\
\lambda^{7-14}(\sigma + 2 \pi) & = & - \exp\left( 2 \pi i
\frac{2}{4} \right) \lambda^{7-14}(\sigma) \: = \:
+ \lambda^{7-14}(\sigma), \\
\lambda^{15-16}(\sigma + 2 \pi) & = & - \lambda^{15-16}(\sigma).
\end{eqnarray*}
It is straightforward to compute that $E_{\rm left} = 0$,
$E_{\rm right} = -1/2$.  There is a multiplicity of left vacua,
arising from $\lambda^{7-14}$.  Let $|m\rangle$ denote a vacuum
with $m$ +'s and $8-m$ -'s, {\it i.e.} annihilated by $m$ $\lambda$'s and
$8-m$ $\overline{\lambda}$'s, then under the action of the generator of
${\mathbb Z}_4$, it is straightforward to check that $|m=0,4,8\rangle$
are invariant, $|m=2,6\rangle$ get a sign flip, and the
others are multiplied by various fourth roots of unity.

The ${\mathbb Z}_4$-invariant states in this sector are of the form
\begin{center}
\begin{tabular}{cc}
State & Count \\ \hline
$|m=0,4,8 \rangle \otimes
 \left( \psi^{1-2}_{-1/2}, \overline{\psi}^{1-2}_{-1/2}
\right)$ & spacetime vector, valued in ${\bf 1}$, ${\bf 1}$, $\wedge^4 {\bf 8}
= {\bf 70}$ of $su(8)$
\\
$|m=6,2\rangle \otimes
\left( \psi^{3-4}_{-1/2}, \overline{\psi}^{3-4}_{-1/2}
\right)$ & 4 sets of scalars, in $\wedge^2 {\bf 8} = {\bf 28}$,
$\wedge^2 {\bf \overline{8}} = {\bf \overline{28}}$ of $su(8)$
\end{tabular}
\end{center}

In the $k=2$ (R,NS) sector, fields have the following boundary
conditions:
\begin{eqnarray*}
X^{1-2}(\sigma + 2 \pi) & = & + X^{1-2}(\sigma), \\
X^{3-4}(\sigma + 2 \pi) & = & + X^{3-4}(\sigma), \\
\psi^{1-2}(\sigma + 2 \pi) & = & - \psi^{1-2}(\sigma), \\
\psi^{3-4}(\sigma + 2 \pi) & = & - \psi^{3-4}(\sigma), \\
\lambda^{1-6}(\sigma + 2 \pi) & = & + \lambda^{1-6}(\sigma), \\
\lambda^{7-14}(\sigma + 2 \pi) & = & + \exp\left( 2 \pi i
\frac{2}{4} \right) \lambda^{7-14}(\sigma) \: = \:
- \lambda^{7-14}(\sigma), \\
\lambda^{15-16}(\sigma + 2 \pi) & = & + \lambda^{15-16}(\sigma).
\end{eqnarray*}
It is straightforward to compute that $E_{\rm left} = 0$,
$E_{\rm right} = -1/2$.  There is a multiplicity of left vacua,
as $\lambda^{1-6}$ and $\lambda^{15-16}$ are periodic.
In particular, $|\pm \mp \rangle_{15-16}$
are invariant under ${\mathbb Z}_4$, whereas
$|\pm \pm \rangle_{15-16}$ get a sign flip.
Therefore, the ${\mathbb Z}_4$-invariant massless states are of the form
\begin{center}
\begin{tabular}{cc}
State & Count \\ \hline
$| \pm \cdots \pm \rangle_{1-6}
| \pm \mp \rangle_{15-16}
\otimes
 \left( \psi^{1-2}_{-1/2}, \overline{\psi}^{1-2}_{-1/2}
\right)$  & spacetime vector \\
& in $({\bf 32},{\bf 2})$ of $so(12)\times so(4)$ \\
$| \pm \cdots \pm\rangle_{1-6} |\pm \pm \rangle_{15-16}
\otimes
\left( \psi^{3-4}_{-1/2}, \overline{\psi}^{3-4}_{-1/2}
\right)$ &  4 sets of scalars \\
& in $({\bf 32}',{\bf 2}')$ of $so(12)\times so(4)$
\end{tabular}
\end{center}

We can rearrange the spacetime vectors more sensibly as follows.
The $so(12)\times so(4) \cong so(12) \times su(2) \times su(2)$
is enhanced to an $e_7 \times su(2)$, using the fact that the
adjoint representation of $e_7$ decomposes under $so(12) \times su(2)$
as  
\cite{slansky}[table 52]
\begin{displaymath}
{\bf 133} \: = \: ({\bf 66},{\bf 1}) \oplus
({\bf 32},{\bf 2}) \oplus ({\bf 1},{\bf 3}).
\end{displaymath}
The ${\bf 66}$ is the adjoint representation of $so(12)$,
which arises in $k=0$, as does the ${\bf 3}$ of $su(2)$
(half of the adjoint representation of $so(4)$), and the
$({\bf 32},{\bf 2})$ arises in the sector $k=2$.
Similarly, the $su(8)$ is enhanced to $e_7$.  The adjoint representation
of $e_7$ decomposes under $su(8)$ as
\cite{slansky}[table 52]
\begin{displaymath}
{\bf 133} \: = \: {\bf 63} \oplus {\bf 70}.
\end{displaymath}
The ${\bf 63}$ arises in the $k=0$ sector, and the ${\bf 70}$ in $k=2$.
In addition, there are three remaining vector multiplets, in the
$k=0$ and $k=2$ sectors.
Therefore, the complete gauge (algebra) symmetry in this compactification is
$e_7 \times e_7 \times su(2) \times u(1)^3$.

The matter fields align themselves with the gauge algebra above.
In the $k=0$ and $k=2$ sectors, the hypermultiplets valued in ${\bf 28}$,
${\bf \overline{28}}$ of $su(8)$ form hypermultiplets in the ${\bf 56}$
of $e_7$, using the fact that under the $su(8)$ subalgebra
\cite{slansky}[table 52],
\begin{displaymath}
{\bf 56} \: = \: {\bf 28} \oplus {\bf \overline{28}}.
\end{displaymath}
Similarly, since under the $so(12) \times su(2)$ subalgebra
\cite{slansky}[table 52],
\begin{displaymath}
{\bf 56} \: = \: ({\bf 32}',{\bf 1}) \oplus ({\bf 12},{\bf 2}),
\end{displaymath}
the $k=0$ hypermultiplet valued in $({\bf 12},{\bf 4})$ of $so(12)\times so(4)$
and the $k=2$ hypermultiplet valued in $({\bf 32}',{\bf 2}')$ form
a hypermultiplet valued in $({\bf 56},{\bf 2})$ of $e_7 \times su(2)$.

Let us summarize our results so far.  We have found the following states:
\begin{single_itemize}
\item 1 gravity multiplet,
\item 1 tensor multiplet,
\item 1 vector multiplet in the adjoint representation of
$e_7 \times e_7 \times su(2) \times u(1)^3$,
\item 2 hypermultiplets in the $({\bf 56},{\bf 1},{\bf 1})$ of
$e_7 \times e_7 \times su(2)$,
\item 1 hypermultiplet in the $({\bf 1},{\bf 56},{\bf 2})$ of
$e_7 \times e_7 \times su(2)$,
\item 4 singlet hypermultiplets.
\end{single_itemize}
It is straightforward to compute that there are 272 vector multiplets and
228 hypermultiplets.  Since the difference is not 244, this six-dimensional
theory is anomalous.

\subsection{Third cautionary example}
\label{sect:class3-caution4}

Now consider an $E_8 \times E_8$ string on
a ${\mathbb Z}_3$ gerbe over a different
$[T^4/{\mathbb Z}_2]$, constructed as $[T^4/{\mathbb Z}_6]$.  
Let the generator $g$ of
${\mathbb Z}_6$ act on the $T^4$ with coordinates $(X^3, X^4)$ as
\begin{displaymath}
g: \: \left( X^3, X^4 \right) \: \mapsto \:
\left( \exp(+4 \pi i/3), \exp(-4 \pi i/3) \right).
\end{displaymath}
Define a rank 2 bundle over this stack by taking ${\cal O}^{\oplus 2}$ over
$T^4$, and let $g$ act with eigenvalues
\begin{displaymath}
\left( \exp(-2 \pi i/3), \exp(-4 \pi i/3) \right).
\end{displaymath}
It is straightforward to check that this satisfies anomaly cancellation in the 
sense of \cite{freedvafa}, and also the constraints in
appendix~\ref{app:spectra:fockconstraints}.

In an $E_8 \times E_8$ compactification, we can describe this as the
following action on fields:
\begin{eqnarray*}
g \cdot X^{1-2} & = & + X^{1-2}, \\
g \cdot X^3 & = & \exp(+4 \pi i/3) X^3, \\
g \cdot X^4 & = & \exp(-4 \pi i/3) X^4, \\
g \cdot \psi^{1-2} & = & + \psi^{1-2}, \\
g \cdot \psi^{3} & = & \exp(+ 4 \pi i/3) \psi^3, \\
g \cdot \psi^4 & = & \exp(-4 \pi i/3) \psi^4, \\
g \cdot \lambda^{1-6} & = & + \lambda^{1-6}, \\
g \cdot \lambda^7 & = & \exp(-2 \pi i/3) \lambda^7, \\
g \cdot \lambda^8 & = & \exp(-4 \pi i/3) \lambda^8.
\end{eqnarray*}

Let us now outline the massless spectrum.

In the untwisted sector, there are massless states in the (NS,NS) sector.
It is straightforward to compute that $E_{\rm left} = -1$,
$E_{\rm right} = -1/2$, and one has ${\mathbb Z}_6$-invariant states of the form
\begin{center}
\begin{tabular}{cc}
State & Count \\ \hline
$\overline{\partial} X^{1-2}_{-1} \otimes \left( \psi^{1-2}_{-1/2},
\overline{\psi}^{1-2}_{-1/2} \right)$ &
gravity, tensor multiplet contributions\\
$\left( \lambda^{1-6}_{-1/2}, \overline{\lambda}^{1-6}_{-1/2} \right)^2
\otimes \left( \psi^{1-2}_{-1/2},\overline{\psi}^{1-2}_{-1/2} \right)$ &
vector in adjoint of $so(12)$ \\
$\left( \lambda^7_{-1/2} \lambda^8_{-1/2},
\overline{\lambda}^7_{-1/2} \overline{\lambda}^8_{-1/2} \right) \otimes
\left( \psi^{1-2}_{-1/2},\overline{\psi}^{1-2}_{-1/2} \right)$ &
vectors in adjoint of $U(1)^2$ \\
$\left( \lambda^7_{-1/2} \overline{\lambda}^7_{-1/2},
\lambda^8_{-1/2} \overline{\lambda}^8_{-1/2} \right) \otimes
\left( \psi^{1-2}_{-1/2},\overline{\psi}^{1-2}_{-1/2} \right)$ &
vectors in adjoint of $U(1)^2$ \\
$\lambda^8_{-1/2}\left( \lambda^{1-6}_{-1/2}, \overline{\lambda}^{1-6}_{-1/2}
\right) \otimes \left( \psi^3_{-1/2}, \overline{\psi}^4_{-1/2} \right)$ &
half-hypermultiplet in ${\bf 12}$ of $so(12)$ \\
$\overline{\lambda}^8_{-1/2} \left( \lambda^{1-6}_{-1/2}, 
\overline{\lambda}^{1-6}_{-1/2}\right)  \otimes \left(
\overline{\psi}^3_{-1/2}, \psi^4_{-1/2} \right)$ &
half-hypermultiplet in ${\bf 12}$ of $so(12)$ \\
$\lambda^7_{-1/2} \left( \lambda^{1-6}_{-1/2}, \overline{\lambda}^{1-6}_{-1/2}
\right) \otimes \left(
\overline{\psi}^3_{-1/2}, \psi^4_{-1/2} \right)$ &
half-hypermultiplet in ${\bf 12}$ of $so(12)$ \\
$\overline{\lambda}^7_{-1/2}\left(
\lambda^{1-6}_{-1/2}, \overline{\lambda}^{1-6}_{-1/2}
\right) \otimes \left(
\psi^3_{-1/2}, \overline{\psi}^4_{-1/2} \right)$ &
half-hypermultiplet in ${\bf 12}$ of $so(12)$ \\
$\left( \overline{\partial} X^3, \overline{\partial} \overline{X}^4 \right)
\otimes \left( \overline{\psi}^3_{-1/2}, \psi^4_{-1/2} \right)$ &
1 singlet hypermultiplet \\
$\left( \overline{\partial} \overline{X}^3, \overline{\partial} X^4 \right)
\otimes \left( \psi^3_{-1/2}, \overline{\psi}^4_{-1/2} \right)$ &
1 singlet hypermultiplet \\
$\lambda^7_{-1/2} \overline{\lambda}^8_{-1/2} \otimes \left(
 \psi^3_{-1/2}, \overline{\psi}^4_{-1/2} \right)$ &
1/2 singlet hypermultiplet \\
$\overline{\lambda}^7_{-1/2} \lambda^8_{-1/2} \otimes \left(
\overline{\psi}^3_{-1/2}, \psi^4_{-1/2} \right)$ &
1/2 singlet hypermultiplet
\end{tabular}
\end{center}

There are no massless states in the untwisted (R,NS) sector,
and no massless states in $k=1$, $k=2$ sectors.
The $k=3$, $4$, $5$ sectors are copies of the $k=0$, $1$, $2$ sectors,
respectively.  Thus, altogether, the spectrum is two copies of the
states above.

Note that since there are no (R,NS) states, the nonabelian gauge symmetry
is only $so(12)$; it is not enhanced to $e_7$.  Also, since the spectrum
is two copies of the states above, the spectrum contains two gravitons, and
hence would be a likely candidate for decomposition.

Unfortunately, the spectrum is also anomalous.  The gauge symmetry is $so(12)
\times e_8 \times u(1)^4$ (including the second $E_8$, which until now
has been suppressed),
so the total number of vector multiplets is 318, and the number of
hypermultiplets is 27.  Clearly $n_H - n_V \neq 244$, so this model
is anomalous in six dimensions.

\subsection{Potential refinements of anomaly cancellation}
\label{sect:possible-anomcanc}

So far we have described some consistent (0,2) SCFT's of the class III form,
and also illustrated in detail how theories of this form cannot be
consistently used in supersymmetric heterotic string compactifications.
This begs the question of whether there exists a criterion, perhaps
a generalization of anomaly cancellation, that can be used to distinguish
theories of this form.  In this section, we will examine one 
such possibility.

In appendix~\ref{app:chern-reps} we describe a modified notion of Chern
classes and characters, labelled $c^{\rm rep}$ and ${\rm ch}^{\rm rep}$,
that contain extra information in twisted sectors.
It is tempting to speculate that one might be able to use these
to obtain
additional finite-group anomaly constraints on theories by demanding
matching ${\rm ch}_2^{\rm rep}$'s.
Let us check this by studying GLSM's, for which anomaly cancellation 
conditions are more or less well understood.  We will argue that 
although ${\rm ch}^{\rm rep}$'s play a vital
role in index theory, confusingly they do not seem to define any new
anomaly-cancellation conditions.

Consider a (0,2) theory over the hypersurface
$\mathfrak{X} = {\mathbb P}^n_{[k,k,\cdots,k]}[d]$, with gauge bundle
${\cal E}$:
\begin{displaymath}
0 \: \longrightarrow \: {\cal E} \: \longrightarrow \: 
\oplus_a {\cal O}(n_a) \: \longrightarrow \: {\cal O}(m)
\: \longrightarrow \: 0.
\end{displaymath}

It is straightforward to compute that
\begin{displaymath}
c_1^{\rm rep}(T\mathfrak{X})|_{\alpha} \: = \:
(n+1) \frac{k}{k} J \: - \: \frac{d}{k} \alpha^{-d} J,
\end{displaymath}
\begin{eqnarray*}
{\rm ch}_2^{\rm rep}(T\mathfrak{X})|_{\alpha} & = &
{\rm ch}_2^{\rm rep}( \oplus_{n+1} {\cal O}(k) )|_{\alpha} \: - \:
{\rm ch}_2^{\rm rep}( {\cal O}(d) )|_{\alpha}, \\
& = &
\frac{1}{2} (n+1) \left( \frac{k}{k} J \right)^2  
\: - \: \frac{1}{2} \left( \frac{d}{k} J \right)^2 \alpha^{-d},
\end{eqnarray*}
and for the bundle ${\cal E}$,
\begin{displaymath}
c_1^{\rm rep}({\cal E})|_{\alpha} \: = \:
\sum_a \frac{n_a}{k} J \alpha^{-n_a} \: - \: \frac{m}{k} J \alpha^{-m},
\end{displaymath}
\begin{eqnarray*}
{\rm ch}_2^{\rm rep}({\cal E})|_{\alpha} & = &
{\rm ch}_2^{\rm rep}( \oplus_a {\cal O}(n_a) )|_{\alpha} \: - \:
{\rm ch}_2^{\rm rep}( {\cal O}(m) )|_{\alpha}, \\
& = &
\frac{1}{2} \sum_a \left( \frac{n_a}{k} J \right)^2 \alpha^{-n_a}
\: - \:
\frac{1}{2} \left( \frac{m}{k} J \right)^2 \alpha^{-m}.
\end{eqnarray*}

By contrast, anomaly cancellation in the GLSM is merely the statement that
\begin{displaymath}
\sum_a n_a^2 \: - \: m^2 \: = \: (n+1) k^2 \: - \: d^2,
\end{displaymath}
a much weaker statement than demanding ${\rm ch}_2^{\rm rep}({\cal E}) = 
{\rm ch}_2^{\rm rep}(T \mathfrak{X})$ in each sector $\alpha$.
Anomaly cancellation in the GLSM is well-understood -- in the present case,
this is just the gauge anomaly in a $U(1)$ gauge theory, which is under 
extremely good control.  Demanding matching ${\rm ch}^{\rm rep}$'s
gives a stronger condition -- some theories that would satisfy 
GLSM anomaly cancellation, would not satisfy the constraint of matching
${\rm ch}^{\rm rep}$'s.

For this reason, we do not believe that one should demand 
matching ${\rm ch}_2^{\rm rep}$'s.  This is a somewhat puzzling
conclusion, as these are not only the most natural notion of Chern classes
on stacks, but they are also vital in index theory, which ordinarily would
be a route to {\it deriving} their utility.
(On the other hand, we briefly remark on a possible application of
$c_1^{\rm rep}$ in appendix~\ref{app:spectra:fockconstraints}.)

\section{Combinations}  \label{sect:combos}

So far we have discussed three fundamental classes of examples of
heterotic string compactifications on gerbes.

Those three classes do not exhaust all possibilities; rather,
one should think of them as `building blocks' that can be used to
assemble more complicated possibilities.

For one example, consider a string on a ${\mathbb Z}_4$ gerbe, of which
a ${\mathbb Z}_2$ subgroup acts on a rank 8 bundle, but the ${\mathbb Z}_2$
coset leaves the bundle invariant.  A version of the decomposition
conjecture should apply here, relating this (0,2) SCFT to a disjoint
union of two (0,2) SCFT's, each of which would involve a heterotic
string on a ${\mathbb Z}_2$ gerbe with a nontrivial action on the
gauge bundle.  Those individual SCFT's might be dual to a different
string compactification (class II), or might not define a consistent
heterotic string compactification (class III).

It is straightforward to assemble more complicated possibilities,
following similar patterns.

\section{Conclusions}

In this paper we have examined general aspects of heterotic string
compactifications on generalized spaces known as stacks, focusing on the
particularly interesting special case of stacks that are gerbes.

Briefly, we have described how
heterotic string compactifications on gerbes are built from
three basic classes:
\begin{itemize}
\item In the special case that the gauge bundle on the gerbe is a pullback
from an underlying space, the heterotic theory on the gerbe is
equivalent to a heterotic theory on a disjoint union of spaces,
the same sort of decomposition as type II strings on gerbes
\cite{summ}.
\item In the special case that the gauge bundle on the gerbe is twisted
in such a way as to locally 
duplicate a different ten-dimensional GSO projection,
the gerbe compactification seems to be dual to
a compactification of the corresponding
different heterotic string.
\item In other cases in which the gauge bundle is different from a pullback
from the base, although at least sometimes one can define consistent
(0,2) SCFT's, there do not seem to be any viable perturbative heterotic string
compactifications.
\end{itemize}

There are several open questions that would be interesting to pursue.
For example,
\begin{itemize}
\item We have not identified a complete set of sufficient conditions for
a stack $\mathfrak{X}$ with bundle ${\cal E} \rightarrow \mathfrak{X}$ to
define a consistent heterotic string compactification.  We have identified
a number of necessary conditions, such as anomaly cancellation on the
cover and level-matching in orbifolds, we have derived additional 
necessary conditions from well-definedness of Fock vacua,
but we have also observed that these conditions
do not suffice in general.  We have speculated on some enhancements of
anomaly cancellation (involving the ${\rm ch}^{\rm rep}$'s that can be
defined on stacks), but do not at this time have any definitive statements
to make.
\item We have discussed a heterotic analogue of the decomposition conjecture
for banded gerbes, with bundle a pullback from the base.  We do not at this
time have an analogue for nonbanded gerbes.
\end{itemize}
These questions are left for future work.

\section{Acknowledgements}

Various parts of this work have been in progress for approximately five
years, and so we have a number
of people to thank, including M.~Ando (for discussions of
elliptic genera with twisted bundles), P.~Clarke (for assistance
constructing Distler-Kachru examples),
K.~Dienes (for discussions of free fermion models and sufficiency
of level matching), J.~Distler,
D.~Freed (for discussions of uses and analogues of
${\rm ch}^{\rm rep}$'s), J.~Gray, S.~Hellerman, I.~Melnikov,
E.~Scheidegger, 
K.~Wendland (for discussions of the nonsupersymmetric $SO(8) \times SO(24)$
string), and especially T.~Pantev for very many useful discussions and
collaborations revolving around stacks. 

L.~Anderson was supported by the Fundamental Laws Initiative of the Center
for the Fundamental
Laws of Nature, Harvard University.
B.~Ovrut was supported in part by the DOE under contract
DE-AC02-76-ER-03071 and the NSF under grant 1001296.
Over the course of this work, E.~Sharpe was
partially supported by NSF grants DMS-0705381, PHY-0755614,
and PHY-1068725.

\appendix

\section{Massless spectra of heterotic strings on stacks}
\label{app:spectra}

\subsection{Basic definitions}

In this section, we will describe the computation of the massless spectrum
of a perturbative heterotic $E_8 \times E_8$ string compactified on a
smooth Deligne-Mumford stack $\mathfrak{X}$ with suitable gauge bundle
${\cal E} \rightarrow \mathfrak{X}$, following (in spirit when
not detail) \cite{dist-greene} and \cite{manion-toappear}.  
Not only will this be useful for 
computations, but the existence of such a computational method is a good
consistency check for the existence of heterotic string compactifications
on stacks.

Let $\mathfrak{X}$ be a smooth Deligne-Mumford stack of complex 
dimension\footnote{
For simplicity, as we wish to work in light-cone gauge, we will assume
that the complex dimension is bounded by 4.
} $n \leq 4$, 
and ${\cal E}$ a holomorphic
vector bundle over $\mathfrak{X}$ of rank $r$, 
satisfying suitable anomaly-cancellation conditions.
We will embed the bundle in one of the $E_8$'s of the ten-dimensional
heterotic string, so we will assume that $r < 8$.
As in \cite{dist-greene}, all our computations will be in a right-moving
R sector (hence, spacetime fermions), 
but spacetime supersymmetry can be used to derive the NS sector (spacetime
bosons) in principle.

Let $I_{\mathfrak{X}}$ denote the inertia stack associated to $\mathfrak{X}$.
Roughly speaking, the inertia stack is a geometric mechanism for
encoding twisted sectors; it has multiple components, each of which
corresponds to a twisted sector in a standard global orbifold.  For example, 
if $\mathfrak{X} = [{\mathbb C}^2/{\mathbb Z}_2]$, where the 
${\mathbb Z}_2$ acts by sign flips, then
\begin{displaymath}
I_{\mathfrak{X}} \: = \: [{\mathbb C}^2/{\mathbb Z}_2]
\amalg [{\rm point}/{\mathbb Z}_2].
\end{displaymath}
For another example, if $\mathfrak{X} \: = \: [ {\mathbb C}/{\mathbb Z}_3]$,
where the ${\mathbb Z}_3$ acts by multiplying by phases,
then
\begin{displaymath}
I_{\mathfrak{X}} \: = \: [ {\mathbb C}/{\mathbb Z}_3]
\amalg [{\rm point}/{\mathbb Z}_3] \amalg
[{\rm point}/{\mathbb Z}_3].
\end{displaymath}
For yet another example, if $\mathfrak{X} \: = \: [{\mathbb C}/{\mathbb Z}_2]$,
where the ${\mathbb Z}_2$ acts trivially (so that all of ${\mathbb C}$ is
fixed), then
\begin{displaymath}
I_{\mathfrak{X}} \: = \: [{\mathbb C}/{\mathbb Z}_2]
\amalg [{\mathbb C}/{\mathbb Z}_2].
\end{displaymath}
(See {\it e.g.} 
\cite{vistoli,gomez,lmb,bx,metzler1,noohi1,noohi2,noohi3,heinloth2,bss1} 
for more information on the inertia stack.)
In general,
points in the inertia stack are pairs $(x,\alpha)$, where $x$ is a point
of $\mathfrak{X}$, 
and $\alpha$ is an automorphism of $x$, which for an orbifold $[Y/G]$ by
$G$ a finite group, would define the twisted sectors.
In the $[{\mathbb C}^3/{\mathbb Z}_3]$ example, 
if $g$ generates ${\mathbb Z}_3$, then the two copies of
$[{\rm point}/{\mathbb Z}_3]$ correspond to $\alpha = g, g^2$.
The inertia stack $I_{\mathfrak{X}}$ always contains a copy of $\mathfrak{X}$
as one component, corresponding to $\alpha = {\rm Id}$.

Let us describe how to compute the spectrum on each component $\alpha$ of
$I_{\mathfrak{X}}$.  
(We will use $\alpha$ to denote both a component of $I_{\mathfrak{X}}$ and
the automorphism defining that component.)

First, let $q: I_{\mathfrak{X}} \rightarrow \mathfrak{X}$ 
denote the natural projection onto a single component,
and for $\alpha \neq {\rm Id}$,
decompose the pullback bundles into eigenbundles\footnote{
Since $\alpha$ leaves the points invariant, this component of the inertia
stack must have a $\langle \alpha \rangle$ gerbe structure, and bundles on
such gerbes have an eigenbundle decomposition as given here.
} of $\langle \alpha
\rangle$:
\begin{eqnarray*}
q^* T\mathfrak{X}|_{\alpha} & = & \oplus_n T_n^{\alpha}, \\
q^* {\cal E} |_{\alpha} & = & \oplus_n {\cal E}_n^{\alpha}.
\end{eqnarray*}
Define $t_{\alpha}$ to be the order of the corresponding automorphism,
and take $T_n^{\alpha}$ and ${\cal E}_n^{\alpha}$
to be associated with character
\begin{displaymath}
\exp(2 \pi i n / t_{\alpha}).
\end{displaymath}
By this we mean that the (R-sector) worldsheet fermions corresponding to
$T_n^{\alpha}$ and ${\cal E}_n^{\alpha}$ have boundary
conditions of the form
\begin{displaymath}
\psi(\sigma + 2 \pi) \: = \: \exp(2 \pi i n / t_{\alpha})
\psi(\sigma).
\end{displaymath}
We will denote fermions couplings to $T_n^{\alpha}$
(respectively, ${\cal E}_n^{\alpha}$) by $\psi_{+,n}$
(respectively, $\lambda_{-,n}$).

Let us pause to briefly discuss some concrete examples, to illuminate these
abstract definitions.
For global orbifolds by finite groups, it should hopefully be clear
that the description above is an abstraction of the standard prescription
for distinguishing various worldsheet fermions with different boundary
conditions.  Let us turn to an example which does not have such a 
realization, but which is relevant to (0,2) GLSMs.  
Take $\mathfrak{X} = {\mathbb P}^4_{[1,1,1,2,2]}$, with bundle
\begin{displaymath}
0 \: \longrightarrow \: {\cal E} \: \longrightarrow \:
\oplus_a {\cal O}(n_a) \: \stackrel{F_a}{\longrightarrow} \: {\cal O}(m)
\: \longrightarrow \: 0
\end{displaymath}
where $\det {\cal E}^* \cong K_{\mathfrak{X}}$: 
\begin{displaymath}
\sum n_a \: - \: m \: = \: 7,
\end{displaymath}
and second Chern classes match:
\begin{displaymath}
\sum n_a^2 \: - \: m^2 \: = \: 11.
\end{displaymath}
This is not Calabi-Yau, so it would not be directly useful for a string
compactification, but can help illuminate some general aspects.
This stack has a ${\mathbb P}^1$ of ${\mathbb Z}_2$ orbifolds, so the
inertia stack has the form
\begin{displaymath}
I_{\mathfrak{X}} \: = \: \mathfrak{X} \amalg 
{\mathbb P}^1_{[2,2]}.
\end{displaymath}
On the nontrivial component ${\mathbb P}^1_{[2,2]}$, call it $\alpha$,
we can work out the decomposition of the
gauge bundle.  Suppose, for example, that $m$ is odd.
For any given $a$, if $n_a$ is even, then $F_a$ is odd, so $F_a = 0$;
if $n_a$ is even on the other hand, there is no constraint on $F_a$.
In this case, we can decompose
\begin{displaymath}
q^* {\cal E} |_{\alpha} \: = \: {\cal E}_+ \oplus {\cal E}_-,
\end{displaymath}
where ${\cal E}_+$ is invariant, ${\cal E}_-$ anti-invariant under
${\mathbb Z}_2$, and specifically
\begin{eqnarray*}
{\cal E}_+ & = & \oplus {\cal O}(n_a \, {\rm even}),
\\
{\cal E}_- & = & {\rm ker}\left( \oplus {\cal O}(n_a \, {\rm odd}) \:
\longrightarrow \: {\cal O}(m) \right).
\end{eqnarray*}
A closely related decomposition exists for $m$ even.

Now that we have illuminated the definitions, 
let us return to our description of the general procedure for
spectrum computation.
At this point, the computation of spectra becomes more or less identical
to that in an ordinary global orbifold by a finite group, if we think of
$\alpha$ as denoting a twisted sector.
We will walk through the details, as there are a few important subtleties
for general cases not usually discussed in the literature, especially
regarding Fock vacua, but the
rest of the computation is nearly standard, once one masters the 
description.

\subsection{Vacuum energies}

We need to compute left- and right-moving zero point energies in each
twisted sector.
Recall that a complex worldsheet fermion $\psi$ with boundary conditions
\begin{displaymath}
\psi(\sigma + 2 \pi) \: = \: \exp(i(\pi + \theta)) \psi(\sigma), \: \: \:
- \pi \leq \theta \leq \pi
\end{displaymath}
contributes
\begin{displaymath}
- \frac{1}{24} \: + \: \frac{1}{8}\left( \frac{\theta}{\pi} \right)^2
\end{displaymath}
to the vacuum energy, and a complex boson with the same boundary conditions
contributes with the opposite sign.

Let $\theta^{T,\alpha}_n$ denote the $\theta$ corresponding to
worldsheet fermions associated with $T_n^{\alpha}$, and
$\theta^{{\cal E},\alpha}_n$ denote the $\theta$ corresponding to
worldsheet fermions associated with ${\cal E}_n^{\alpha}$.
For the moment, we will assume that we are in an (R,R) sector
(meaning, left-moving fermions in the first $E_8$ and right-moving
fermions in an R sector, second $E_8$ will be held fixed in an NS sector).
In an (NS,R) sector (left-moving fermions in the first $E_8$ in an NS sector
instead), we would modify the $\theta$'s for left-moving worldsheet fermions
to take into account an extra sign in boundary conditions.

Then, in an (R,R) sector,
the left-moving vacuum energy is
\begin{eqnarray*}
E_{{\rm (R,R)}, {\rm Id}} & = & 
8\left(- \frac{1}{24} \right) \: + \: 8\left(+\frac{1}{12}\right)
\: + \: 4\left( - \frac{1}{12} \right), \\
& = & 0,
\end{eqnarray*}
in the untwisted sector ($\alpha = {\rm Id}$) and in twisted sectors,
\begin{eqnarray*}
E_{{\rm (R,R)},\alpha} & = &
8\left(- \frac{1}{24} \right) \: + \:
\sum_n ({\rm rk}\, {\cal E}^{\alpha}_n)\left(
- \frac{1}{24} \: + \: \frac{1}{8}\left( 
\frac{ \theta^{{\cal E},\alpha}_n }{\pi} \right)^2 \right)
\: + \: (8-r)\left( + \frac{1}{12} \right)
\\
& &
\: + \: \sum_n ( {\rm rk}\, T^{\alpha}_n )\left(
+ \frac{1}{24} \: - \: \frac{1}{8} \left(
\frac{ \theta^{T,\alpha}_n }{\pi} \right)^2 \right)
\: + \: (4-n) \left( - \frac{1}{12} \right),
\\
& = &
 \frac{n-r}{8} \: + \:
\frac{1}{8} \sum_n ({\rm rk}\, {\cal E}^{\alpha}_n)
\left( \frac{ \theta^{{\cal E},\alpha}_n }{\pi} \right)^2 
\: - \: \frac{1}{8} \sum_n ( {\rm rk}\, T^{\alpha}_n )\left(
\frac{ \theta^{T,\alpha}_n }{\pi} \right)^2.
\end{eqnarray*}
In all cases
the right-moving vacuum energy vanishes, since the right-moving
bosons and fermions make equal and opposite contributions.

Vacuum energies in (NS,R) sectors (meaning, left-moving fermions of the
first $E_8$ in an NS sector) can be computed similarly.
For completeness, we list them below:
in an untwisted sector,
\begin{eqnarray*}
E_{{\rm (NS,R)}, {\rm Id}} & = & 8\left(- \frac{1}{24} \right) \: + \:
8\left( - \frac{1}{24} \right) \: + \:  4\left( - \frac{1}{12} \right), \\
& = & -1,
\end{eqnarray*}
and in a twisted sector,
\begin{eqnarray*}
E_{{\rm (NS,R)}, \alpha} & = &
8\left(- \frac{1}{24} \right) \: + \:
\sum_n ({\rm rk}\, {\cal E}^{\alpha}_n)\left(
- \frac{1}{24} \: + \: \frac{1}{8}\left( 
\frac{ \tilde{\theta}^{{\cal E},\alpha}_n }{\pi} \right)^2 \right)
\: + \: (8-r)\left( - \frac{1}{24} \right)
\\
& &
\: + \: \sum_n ( {\rm rk}\, T^{\alpha}_n )\left(
+ \frac{1}{24} \: - \: \frac{1}{8} \left(
\frac{ \theta^{T,\alpha}_n }{\pi} \right)^2 \right)
\: + \: (4-n) \left( - \frac{1}{12} \right),
\\
& = &
-1 \: + \: \frac{n}{8} 
\: + \: \frac{1}{8} \sum_n ({\rm rk}\, {\cal E}^{\alpha}_n)\left(
\frac{ \tilde{\theta}^{{\cal E},\alpha}_n }{\pi} \right)^2 
\: - \: \frac{1}{8} \sum_n ( {\rm rk}\, T^{\alpha}_n )\left(
\frac{ \theta^{T,\alpha}_n }{\pi} \right)^2,
\end{eqnarray*}
where $\tilde{\theta}$ denotes $\theta$'s as modified to include a sign
in the boundary conditions.
Vacuum energies in (NS,R) sectors (meaning, left-moving fermions of the
first $E_8$ in an NS sector) can be computed similarly.

\subsection{Fock vacua}

The fractional charges of the Fock vacua can and should be understood
in terms of coupling to nontrivial bundles.  Recall (see {\it e.g.} \cite{kw})
that a complex left-moving
fermion $\lambda$ with boundary conditions
\begin{displaymath}
\lambda(\sigma + 2 \pi) \: = \: e^{-i \theta} \lambda(\sigma)
\end{displaymath}
contributes fractional fermion number
\begin{displaymath}
\frac{\theta}{2\pi} \: - \: \left[ \frac{\theta}{2 \pi} \right]
\: - \: \frac{1}{2}
\end{displaymath}
and a complex right-moving fermion $\psi$ with the same boundary conditions
contributes fractional fermion number 
\begin{displaymath}
- \left( 
\frac{\theta}{2\pi} \: - \: \left[ \frac{\theta}{2 \pi} \right]
\: - \: \frac{1}{2}
\right)
\end{displaymath}
In the present case, in the sector defined by automorphism $\alpha$,
we have complex left-moving fermions $\lambda_{-,n}$ coupling to bundle
${\cal E}^{\alpha}_n$, with boundary conditions
\begin{displaymath}
\lambda_{-,n}(\sigma + 2 \pi) \: = \: \exp\left( 2 \pi i n / t_{\alpha} \right)
\lambda_{-,n}(\sigma)
\end{displaymath}
and complex right-moving fermions $\psi_{+,n}$ coupling to bundle
$T^{\alpha}_n$, with boundary conditions
\begin{displaymath}
\psi_{+,n}(\sigma + 2 \pi) \: = \: \exp\left( 2 \pi i n/t_{\alpha}\right)
\psi_{+,n}(\sigma)
\end{displaymath}
Putting this together, we see that from each set of $\lambda_{-,n}$, the
Fock vacuum couples to
\begin{equation}  \label{eq:left-Fock-nonzero}
\left( \det {\cal E}^{\alpha}_n \right)^{- \frac{n}{t_{\alpha}} \: - \: \left[
- \frac{n}{t_{\alpha}} \right] \: - \: \frac{1}{2} }
\end{equation}
and from each set of $\psi_{+,n}$, the Fock vacuum couples
to
\begin{equation}  \label{eq:right-Fock-nonzero}
\left( \det T^{\alpha}_n \right)^{\frac{n}{t_{\alpha}} \: + \: \left[ 
- \frac{n}{t_{\alpha}} \right] \: + \: \frac{1}{2} }
\end{equation}

Since the $\alpha$-sector has components which are $t_{\alpha}$ gerbes,
$t_{\alpha}$-th roots of bundles might exist, though not necessarily.
(See appendix~\ref{app:canonical-roots} for examples of bundles on
${\mathbb Z}_n$-gerbes which do and do not admit $n$th roots.)
Existence of these roots is a necessary condition for the existence
of the physical theories.  When multiple roots exist, as will happen
if the components are not simply-connected, the roots must be specified
as part of the data defining the sigma model.

When there are periodic fermions,
there are multiple Fock vacua, each with different (fractional) charges.  
The different Fock vacua are defined by which subset of the fermi zero
modes annihilate.  In our case, we will work in conventions in which
our Fock vacuum $| 0 \rangle$ has the properties
\begin{displaymath}
\lambda_{-,0}^{a} | 0 \rangle \: = \: 0 \: = \:
\psi_{+,0}^{\overline{\imath}} | 0 \rangle .
\end{displaymath}

As before, reflecting the fact that the $\lambda$'s and $\psi$'s couple to
nontrivial bundles, this Fock vacuum is itself a section of a line bundle.
From those periodic fermions, the Fock vacuum behaves as a section of
a square root of the determinant of the periodic modes, specifically,
\begin{equation}   \label{eq:squareroot}
\sqrt{ K_{\alpha} \otimes \det {\cal E}^{\alpha}_0 },
\end{equation}
(square root chosen with periodic boundary conditions), where
\begin{displaymath}
K_{\alpha} \: = \: \det (T^{\alpha}_0)^*
\end{displaymath}
{\it i.e.} the canonical bundle of the $\alpha$ component of 
$I_{\mathfrak{X}}$.
(Note that in an (NS,R) sector, the `invariant' subbundle ${\cal E}_0$
is defined to be invariant under the combination of spacetime group action
and spin state boundary condition, and hence will be different from the
${\cal E}_0$ in an (R,R) sector.)  If the square root above does not
exist, then the orbifold is not well-defined, which we shall come back to
after we derive the expression above.

We can derive the result above for periodic fermions as follows.
Different choices of Fock vacua act as sections of different line bundles,
related by fermions acting as raising and lowering operators.
Just as in fractional charges, the square root and bundles above are constrained
by the fact that the set of Fock vacua must be consistent with those
raising and lowering operations.
For example, the `opposite' Fock vacuum $| 0 \rangle^{\rm op}$ is defined by
applying raising operators maximally:
\begin{displaymath}
| 0 \rangle^{\rm op} \: = \: \lambda_{-,0}^{\overline{a}_1} \cdots 
\lambda_{-,0}^{\overline{a}_r}
\psi_{+,0}^{i_1} \cdots
\psi_{+,0}^{i_d} | 0 \rangle ,
\end{displaymath}
(where $r$ is the rank of ${\cal E}_0^{\alpha}$ and
$d$ the rank of $T_0^{\alpha}$),
so if our Fock vacuum $| 0 \rangle$ couples to a line bundle
${\cal L}$, then the opposite or dual Fock vacuum above must couple to
\begin{displaymath}
\left( \det {\cal E}_0^{\alpha *} \right) \otimes \left( \det T_0^{\alpha }
\right) \otimes {\cal L},
\end{displaymath}
which, by symmetry, should also be the same as ${\cal L}^*$.  In other
words,
\begin{displaymath}
\left( \det {\cal E}_0^{\alpha *} \right) \otimes \left( \det T_0^{\alpha }
\right) \otimes {\cal L}
\: \cong \: {\cal L}^*
\end{displaymath}
or more simply
\begin{displaymath}
{\cal L}^2 \: \cong \:
\left( \det {\cal E}_0^{\alpha} \right) \otimes \left( \det 
T_0^{\alpha *} \right) \: = \:
K_{\alpha} \otimes \det {\cal E}_0^{\alpha},
\end{displaymath}
from which our claim is derived.
In particular, taking ${\cal L} = {\cal O}$ will not, in general,
be consistent.

In passing, note that the set of all Fock vacua in sector $\alpha$
form a vector bundle
\begin{displaymath}
\left( \wedge^{\bullet} {\cal E}_0^{\alpha *}
\right) \otimes \left(
\wedge^{\bullet} T_0^{\alpha} \right) \otimes
\sqrt{ K_{\alpha} \otimes \det {\cal E}^{\alpha}_0 }
\otimes \otimes_{n > 0} 
\left( \left( \det {\cal E}^{\alpha}_n \right) \left(
\det T^{\alpha}_n \right)^{-1}\right)^{- \frac{n}{t_{\alpha}} \: - \: \left[
- \frac{n}{t_{\alpha}} \right] \: - \: \frac{1}{2} }
\end{displaymath}
over $I_{\mathfrak{X}}|_{\alpha}$, taking into account contributions
from all boundary conditions.

The phenomenon of Fock vacua coupling to nontrivial bundles
has also been noted
in this context in \cite{manion-toappear}, \cite{ando-s}[section 2.1].  
However, aside from those two sources, we are not aware of many discussions
of Fock vacua as sections of line bundles over target
spaces\footnote{
Fock vacua have been much more commonly described in terms of sections of
bundles over CFT moduli spaces, see {\it e.g.} \cite{distler-trieste,bcov}, 
but descriptions as
sections of bundles over target spaces are much more rare.
} in the literature, so it is
perhaps useful to elaborate on this point.
As we shall see
in the present case
and also in \cite{manion-toappear}, it plays a crucial role in closing the
spectrum under Serre duality of the sheaf cohomology groups, a basic
symmetry of the spectra discussed in \cite{dist-greene}.
The same behavior also arises elsewhere.  For example, in open string
theories, the Fock vacuum also transforms as a section of a line bundle,
a square root of the canonical bundle of the D-brane worldvolume $B$
(assumed Spin),
if the D-brane worldvolume is not Calabi-Yau.  This can be understood
simply from the matter representations:  a spinor in the worldvolume
theory can be represented mathematically in the form
\cite{lawson-m}
\begin{displaymath}
\left( \wedge^{\bullet} TB \right) \otimes \sqrt{K_B}.
\end{displaymath}
In terms of the worldsheet RNS formalism,
perturbative modes realize the $TB$ factors,
and the $\sqrt{K_B}$ is implemented
by the Fock vacuum itself.  This phenomenon is also reminiscent of factors
arising from the Freed-Witten anomaly \cite{ks-ext,s-branes}, 
though we shall not pursue
that direction here.

\subsection{Consistency conditions derived from existence of Fock vacua}
\label{app:spectra:fockconstraints}

In some cases, the $t_{\alpha}$th roots~(\ref{eq:left-Fock-nonzero}),
(\ref{eq:right-Fock-nonzero}) or the
square root~(\ref{eq:squareroot}) might not\footnote{
Since the $\alpha$-sector has components which are $t_{\alpha}$ gerbes,
$t_{\alpha}$-th roots of bundles might exist, though not necessarily.
(See appendix~\ref{app:canonical-roots} for examples of bundles on
${\mathbb Z}_n$-gerbes which do and do not admit $n$th roots.)
} exist
as honest equivariant line bundles.  In such a case, the heterotic string
on the stack is not well-defined.  In an ordinary orbifold, this
is the case that the Fock vacua (and hence perturbative states built
from them) form a merely projective representation of the orbifold group,
instead of an honest representation, and the projection operator built
implicitly in the structure of the string one-loop partition function no
longer functions.  This condition represents a new (to our knowledge)
consistency condition, so let us take a few paragraphs to elaborate on this
point.

At least morally, this condition is a generalization to stacks of the
old requirement that ``$c_1 \equiv 0$ mod 2'' for bundles embedded
in $E_8$ in the standard fashion.  That constraint could be understood
in two ways:
\begin{itemize}
\item In low-energy supergravity, this is ultimately the statement that the
$U(n)$ bundle can be lifted to ${\rm Spin}(16)$, realized by the
left GSO projection, whose embeddeding into $E_8$ then factors through
Spin$(16)/{\mathbb Z}_2$, 
\item On the worldsheet, this is the statement that the Fock vacua are
well-defined in a left R sector.  The Fock vacua couple to a square root
of the gauge bundle; that square root will exist if and only if
``$c_1 \equiv 0$ mod 2.''
\end{itemize}
(For another recent discussion of constraints of this form,
see for example \cite{enstx}.)

In toroidal orbifolds, this constraint is very mild, but illustrates
an important point:  not only the bundle must admit a square root, but also
the equivariant structure.  For a typical toroidal orbifold, the bundle
factors are all trivial, only the equivariant structures are nontrivial.
In typical such orbifolds, $K_{\alpha}$ is the trivial line bundle
with trivial connection, but although $\det {\cal E}^{\alpha}_0$ is
a trivial bundle, the equivariant structure may be nontrivial.
In left R sector, ${\cal E}^{\alpha}_0$ describes couples to fermions that
are both periodic and invariant under the orbifold group, so the
equivariant structure is trivial.
In a left NS sector, on the other hand, ${\cal E}^{\alpha}_0$ describes
periodic fermions, which are anti-invariant under the orbifold group.
In a left NS sector,
if the rank of ${\cal E}^{\alpha}_0$ is even, the induced equivariant
structure on $\det {\cal E}^{\alpha}_0$ is trivial; if the rank of
${\cal E}^{\alpha}_0$ is odd, then the induced equivariant structure
is nontrivial, and does not admit a square root, hence there is an
obstruction to the existence of the orbifold in this case.

We can build an example of a toroidal orbifold in which this condition
appears nontrivially as follows.  Consider an $E_8 \times E_8$ string on
a $[T^4/{\mathbb Z}_6]$ orbifold,
in which the generator $g$ of
${\mathbb Z}_6$ act on $T^4$ by multiplication by $-1$.
Define a rank 4
bundle over this stack by taking ${\cal O}^{\oplus 4}$ over $T^4$, and
let $g$ act with eigenvalues
\begin{displaymath}
\left(\exp(6 \pi i/6) = -1, \exp(4 \pi i/6) = \exp(2 \pi i/3),
\exp(2 \pi i/3), \exp(- 2 \pi i/6) \right).
\end{displaymath}
It is straightforward to check that this satisfies level-matching, in the
sense of \cite{freedvafa}.
In the $g$-twisted left NS sector, there will be one periodic fermion,
which is problematic as above.

It is tempting to speculate that a necessary condition for
the existence of the square
root~(\ref{eq:squareroot}) can be written in the form
\begin{displaymath}
c_1^{\rm rep}({\cal E}) \: \equiv \:
c_1^{\rm rep}(T \mathfrak{X}) \mbox{ mod } 2
\end{displaymath}
applying the Chern-rep's discussed in sections~\ref{sect:possible-anomcanc}
and appendix~\ref{app:chern-reps}.  
We will leave such an interpretation to future
work.

\subsection{Spectrum result and Serre duality}

Finally, we are ready to associate sheaf cohomology groups to
elements of the spectrum.  A general element of the spectrum will have
the form
\begin{displaymath}
\lambda_-^{a_1} \cdots \lambda_-^{a_m} \psi_+^{\overline{\imath}_1} \cdots
\psi_+^{\overline{\imath}_k} | 0 \rangle ,
\end{displaymath}
where each $\lambda$ and $\psi$ has some unspecified moding, such that the
sum of the modings equals the vacuum energy computed earlier.
Canonical commutation relations descend to statements of the form
\begin{displaymath}
\{ \lambda_{p}^a, \lambda_{-p}^{\overline{b}} \} \: \propto \: h^{a 
\overline{b}}, \: \: \:
\{ \psi_p^i, \psi^{\overline{\jmath}}_{-p} \} \: \propto \: g^{i
\overline{\jmath}} ,
\end{displaymath}
where $p$ is a moding.  So long as the modings are all negative, both
holomorphic and antiholomorphic-indexed fermions can appear in states.
For zero modes, our Fock vacuum conventions are such that only
$\lambda_{-,0}^{\overline{a}}$ and $\psi_{+,0}^i$ contribute.

In any event, it should now be clear, following \cite{dist-greene},
that on component $\alpha$, states of the form\footnote{
We have omitted modings for reasons of notational sanity.
}
\begin{displaymath}
\prod_n \left( 
\lambda_{-,n}^{a_1} \cdots \lambda_{-,n}^{a_{m_n}} 
\lambda_{-,n}^{\overline{b}_1} \cdots 
\lambda_{-,n}^{\overline{b}_{p_n}}
\psi_{+,n}^{j_1} \cdots \psi_{+,n}^{j_{\ell_n}}
\psi_{+,n}^{\overline{\imath}_1} \cdots
\psi_{+,n}^{\overline{\imath}_{k_n}} \right) | 0 \rangle ,
\end{displaymath}
(where the fermion modings add up to the vacuum energy in the
$\alpha$ sector)
are counted by the sheaf cohomology group
\begin{equation}   \label{eq:countstates1}
H^{k_0}\left( I_{\mathfrak{X}}|_{\alpha},
\left( \wedge^{m_0} {\cal E}_0^{\alpha *} \right)
\otimes_{n>0}\left( \wedge^{m_n} {\cal E}_n^{\alpha } 
\otimes \wedge^{p_n} {\cal E}_n^{\alpha *}
\otimes \wedge^{\ell_n} T_n^{\alpha}
\otimes \wedge^{k_n} T_n^{\alpha *} \right) \otimes
{\cal F}
\right) ,
\end{equation}
where 
\begin{displaymath}
{\cal F}^{\alpha} \: = \: \sqrt{ K_{\alpha} \otimes \det {\cal E}^{\alpha}_0 }
\otimes_{n>0} \left( \left( \det {\cal E}^{\alpha}_n \right) \left(
\det T^{\alpha}_n \right)^{-1}\right)^{- \frac{n}{t_{\alpha}} \: - \: \left[
- \frac{n}{t_{\alpha}} \right] \: - \: \frac{1}{2} } 
\end{displaymath}
(reflecting the Fock vacuum).
Strictly speaking, not all states need be of the form above -- for
example, one might also be able to multiply in bosonic
$\partial \phi$ modes.  As their inclusion is standard and their treatment
should now be clear, for reasons of brevity we shall move on.

For example, if $I_{\mathfrak{X}}|_{\alpha} = [ {\rm point}/{\mathbb Z}_2 ]$,
then this becomes
\begin{displaymath}
H^{k_0}\left( {\rm point}, 
\left( \wedge^{m_0} {\cal E}_0^{\alpha *} \right)
\otimes_{n>0}\left( \wedge^{m_n} {\cal E}_n^{\alpha *} 
\otimes \wedge^{p_n} {\cal E}_n^{\alpha *}
\otimes \wedge^{\ell_n} T_n^{\alpha}
\otimes \wedge^{k_n} T_n^{\alpha *} \right) \otimes
{\cal F}^{\alpha}
\right)^{\mathbb{Z}_2}.
\end{displaymath}
(Taking group invariants is encoded implicitly in taking sheaf cohomology
on the quotient stack.)
This group vanishes if $k_0 \neq 0$, and when $k_0 = 0$, is the dimension of
the ${\mathbb Z}_2$-invariant part of the vector space fibers.

Finally, in a physical computation, one must impose the left- and right-
GSO projections.  For states of the form above, this will amount to a 
chirality constraint on $k_0$ and $m_0$.  As the procedure is standard,
we will say no more.

One of the central observations of the heterotic spectrum computation on
smooth manifolds in
\cite{dist-greene} is that it is closed under Serre duality.
The same is true here.  First, for any component of the inertia stack
indexed by an automorphism $\alpha$, there is another (not necessarily
distinct) component indexed by $\alpha^{-1}$, which is isomorphic:
\begin{displaymath}
I_{\mathfrak{X}}|_{\alpha} \: 
\cong \: I_{\mathfrak{X}} |_{\alpha^{-1}}.
\end{displaymath}
Eigenbundle decompositions are closely related:
\begin{eqnarray*}
T_n^{\alpha^{-1}} & \cong & T_{-n}^{\alpha}, \: \: \:
T_0^{\alpha^{-1}} \: \cong \: T_0^{\alpha}, \\
{\cal E}_n^{\alpha^{-1}} & \cong & {\cal E}_{-n}^{\alpha}, \: \: \:
{\cal E}^{\alpha^{-1}}_0 \: \cong \: {\cal E}^{\alpha}_0,
\end{eqnarray*}
(in conventions where $-n$ denotes the component associated to the
character of the inverse).
Let us now consider the following factor in the Fock vacuum bundle,
\begin{displaymath}
{\cal F}^{\alpha}_+ \: = \: \otimes_{n>0} \left( \left( \det {\cal E}^{\alpha}_n \right) \left(
\det T^{\alpha}_n \right)^{-1}\right)^{- \frac{n}{t_{\alpha}} \: - \: \left[
- \frac{n}{t_{\alpha}} \right] \: - \: \frac{1}{2} } 
\end{displaymath}
(where the tensor product runs over all nontrivial representations of
${\mathbb Z}_{t^{\alpha}}$).  Using relations such as 
${\cal E}^{\alpha^{-1}}_n \cong {\cal E}^{\alpha}_{-n}$,
we see that each factor in ${\cal F}_+^{\alpha^{-1}}$ is equivalent to a factor
in ${\cal F}_+^{\alpha}$, but with an exponent of the opposite sign,
hence
\begin{equation}  \label{eq:pos-Fock-duality}
{\cal F}_+^{\alpha^{-1}} \: \cong \: \left( {\cal F}_+^{\alpha} \right)^* .
\end{equation}
As the combinatorics in these exponents is slightly complicated, let us
consider some special cases to explicitly confirm this prediction.  
When $t_{\alpha} = 2$,
\begin{eqnarray*}
{\cal F}_+^{\alpha} & = & \left( \left( \det {\cal E}_1^{\alpha} \right)
\left( \det T_1^{\alpha} \right)^{-1} \right)^{- \frac{1}{2} - \left[ 
- \frac{1}{2}\right] - \frac{1}{2} }, \\
& = & \left( \left( \det {\cal E}_1^{\alpha} \right)
\left( \det T_1^{\alpha} \right)^{-1} \right)^{0} \: \cong \: {\cal O}
\: \cong \: \left( {\cal F}_+^{\alpha^{-1}} \right)^* .
\end{eqnarray*}
When $t_{\alpha}=3$,
\begin{eqnarray*}
{\cal F}_+^{\alpha} & = & \left( \left( \det {\cal E}_1^{\alpha} \right)
\left( \det T_1^{\alpha} \right)^{-1} \right)^{-\frac{1}{3} - \left[ 
- \frac{1}{3} \right] - \frac{1}{2}} \otimes
\left( \left( \det {\cal E}_2^{\alpha} \right) \left( \det T_2^{\alpha} 
\right)^{-1} \right)^{- \frac{2}{3} - \left[ - \frac{2}{3} \right] - 
\frac{1}{2} }, \\
& = & \left( \left( \det {\cal E}_1^{\alpha} \right)
\left( \det T_1^{\alpha} \right)^{-1} \right)^{+1/6} \otimes
\left( \left( \det {\cal E}_2^{\alpha} \right) \left( \det T_2^{\alpha} 
\right)^{-1} \right)^{-1/6}  ,
\end{eqnarray*}
and
\begin{eqnarray*}
{\cal F}_+^{\alpha^{-1}} & = & \left( \left( \det {\cal E}_1^{\alpha^{-1}}
\right) \left( \det T_1^{\alpha^{-1}} \right)^{-1} \right)^{+1/6} \otimes
\left( \left( \det {\cal E}_2^{\alpha^{-1}} \right)
\left( \det T_2^{\alpha^{-1}} \right)^{-1} \right)^{-1/6}, \\
& = &
\left( \left( \det {\cal E}_2^{\alpha}
\right) \left( \det T_2^{\alpha} \right)^{-1} \right)^{+1/6} \otimes
\left( \left( \det {\cal E}_1^{\alpha} \right)
\left( \det T_1^{\alpha} \right)^{-1} \right)^{-1/6}, \\
& = &
\left( {\cal F}_+^{\alpha} \right)^* .
\end{eqnarray*}
In this fashion we confirm equation~(\ref{eq:pos-Fock-duality})
explicitly.

Vacuum energies are invariant:  if a fermion boundary condition in sector
$\alpha$ is determined by $\theta$, then in $\alpha^{-1}$ it is determined
by $- \theta$, but vacuum energies only depend upon $(\theta)^2$,
and so are invariant.
Contributions to the spectrum from sector $\alpha$ are matched by
Serre duals in sector $\alpha^{-1}$.  In terms of global quotients
by finite groups, this means the untwisted sector closes into itself
under Serre duality, but twisted sectors are exchanged.
For example, 
the Serre duals to~(\ref{eq:countstates1}) are given by
\begin{eqnarray*}
\lefteqn{
H^{{\rm dim}-k_0} \Bigl(  
I_{\mathfrak{X}}|_{\alpha},
\left( 
\wedge^{m_0} {\cal E}_0^{\alpha } \right)
\otimes_{n>0}\left( \wedge^{m_n} {\cal E}_n^{\alpha *} 
\otimes \wedge^{p_n} {\cal E}_n^{\alpha }
\otimes \wedge^{\ell_n} T_n^{\alpha *}
\otimes \wedge^{k_n} T_n^{\alpha } \right)
}
\\
& & \hspace*{3.25in}  \left. 
\otimes ({\cal F}^{\alpha}_+)^* \otimes
\sqrt{ K_{\alpha}^* \otimes \det {\cal E}^{\alpha *}_0 }
\otimes K_{\alpha}
\right)^*
 \\
& = &
H^{{\rm dim}-k_0}\Bigl(  
I_{\mathfrak{X}}|_{\alpha^{-1}}, 
\left( \wedge^{{\rm rk} - m_0} {\cal E}_0^{\alpha^{-1} } \right)
\otimes_{n>0}\left( \wedge^{m_n} {\cal E}_n^{\alpha^{-1} *} 
\otimes \wedge^{p_n} {\cal E}_n^{\alpha^{-1} }
\otimes \wedge^{\ell_n} T_n^{\alpha^{-1} *}
\otimes \wedge^{k_n} T_n^{\alpha^{-1} } \right) 
\\
& & \hspace*{3.5in} \left.
\otimes {\cal F}^{\alpha^{-1}}_+ \otimes
\sqrt{ K_{\alpha^{-1}} \otimes \det {\cal E}^{\alpha^{-1} }_0 }
\right)^*,
\end{eqnarray*}
which is of the same form as
equation~(\ref{eq:countstates1}), as desired.
Note that the Fock vacuum contribution is essential for the spectrum
to close under Serre duality in this fashion: otherwise, Serre duality
would generate a factor of $K_{\alpha}$ in the coefficients, unmatched by
anything else, and which is nontrivial if the $\alpha$ component is not
Calabi-Yau\footnote{
To make it clear that this condition is nontrivial,
here is an example of a global orbifold in which a twisted sector has
support on a non-Calabi-Yau subvariety.  Let $X$ be a branched double cover
of ${\mathbb P}^n$, branched over a degree $2n+2$ locus.  Now, orbifold by
the globally-acting ${\mathbb Z}_2$ that exchanges the sheets of the cover.
This leaves invariant the degree $2n+2$ branch locus, which is not 
Calabi-Yau.
}.
Our computations so far have focused on the (R,R) sector, but one
should note that identical considerations hold in the (NS,R) sector as
well.

In the special case that the stack $\mathfrak{X}$ is a smooth
Calabi-Yau manifold $X$, these computational methods reduce to those
of \cite{dist-greene}.  In this case, the inertia stack $I_{\mathfrak{X}}$
has no nontrivial components:  $I_{\mathfrak{X}} = X$.  Furthermore,
we typically take $\det {\cal E}$ to be trivial, so the Fock vacuum is
a section of a trivial line bundle.

In the special case that the stack $\mathfrak{X}$ is a toroidal orbifold,
again these methods reduce to known results.  In this case, all of the
bundles involved are trivial, so sheaf cohomology is nontrivial only in
degree zero, and sheaf cohomology on a stack just takes group invariants
of the coefficients.

A less trivial example is discussed in section~\ref{sect:class3-caution1}.
Further examples and computational techniques will appear in
\cite{manion-toappear}.

Just as in \cite{dist-greene}, in principle the number of generations can
be computed as an index based on the spectrum.  We shall not work through
details here, but appendix~\ref{app:chern-reps} 
contains general results on index
theory computations on stacks.

\subsection{A/2 model spectra}

In this appendix we have focused on physical heterotic string spectra.
It is possible to apply the same methods to the A/2 model
to formulate a mathematical
theory of sheaf cohomology of orbifolds, and this has been done in
\cite{manion-toappear}.

Briefly, the A/2 model is a heterotic analogue of the A model topological
field theory.  If $X$ is a smooth space and ${\cal E} \rightarrow X$ a
holomorphic vector bundle, then the A/2 model is well-defined if
both\footnote{
The second condition arises from the need to make the path integral
measure a scalar, ultimately.  On stacks, one might wonder whether one
should impose an analogous condition in each individual twisted sector,
something of the form
\begin{displaymath}
\det {\cal E}_0^{\alpha *} \: \cong \: K_{\alpha}.
\end{displaymath}
Reference \cite{manion-toappear} does not impose a stronger condition of
this sort.  One reason is that there is no analogue of such a condition in
GLSM's (whereas the original condition $\det {\cal E}^* \cong K_{\mathfrak{X}}$
on the 
entire stack
does manifest in GLSM's).  In terms of making sense of path integral
measures, in twisted sectors one must insert twist fields to get nonzero
results, and which would modify any such constraint one wished to impose
on individual twisted sectors.
}
\begin{displaymath}
{\rm ch}_2({\cal E}) \: = \: {\rm ch}_2(TX)
\mbox{ and }
\det {\cal E}^* \: \cong \: K_X.
\end{displaymath}
See {\it e.g.} \cite{katz-s,ade,s-b,jock-ilarion,dgks1,dgks2,mss,tan1,tan2} 
for more information on the A/2 and B/2 models.
As this is no longer a physical theory, constraints on the dimension of
$X$ and rank of ${\cal E}$ are dropped.
When $X$ is smooth, the massless spectrum consists of sheaf
cohomology groups of the form
\begin{displaymath}
H^{\bullet} ( X, \wedge^{\bullet} {\cal E}^* )
\end{displaymath}
When $X$ is a stack $\mathfrak{X}$, reference
\cite{manion-toappear} applies methods similar to
those in this appendix (modulo restricting to (R,R) sector states and
omitting the GSO projections) to define a generalization, which broadly speaking
adds in various sheaf cohomology groups associated to twisted sectors
(nontrivial components of the inertia stack).

\section{Line bundles on gerbes over projective spaces}
\label{app:linebundles}

For any stack $\mathfrak{X}$ presented as $\mathfrak{X} = [X/G]$ for
some space $X$ and group $G$, a vector bundle (sheaf) on $\mathfrak{X}$ is
the same as a $G$-equivariant vector bundle (sheaf) on $X$.
Suppose that
$G$ is an extension
\begin{displaymath}
1 \: \longrightarrow \: K \: \longrightarrow \: G \: \longrightarrow \:
H \: \longrightarrow \: 1,
\end{displaymath}
where $K$ acts trivially on $X$, and $G/K \cong H$ acts effectively.
In this case, $\mathfrak{X} = [X/G]$ is a $K$-gerbe.
A vector bundle on $\mathfrak{X}$ is a $G$-equivariant vector bundle on $X$,
and as such, the $K$ action is defined by a representation of $K$ on the
fibers of that vector bundle. 

In this section, we will discuss in greater detail the special case of
line bundles on gerbes over projective spaces.

\subsection{Generalities}

Let us first review some basic properties of line bundles on
gerbes over projective spaces, and then we will outline their
sheaf cohomology.

First, let us consider some simple explicit examples.
The total space of the
line bundle ${\cal O}(-m)$ over the projective space ${\mathbb P}^n$
can be described\footnote{For $m>0$.  The total spaces of
line bundles of positive
degree over projective spaces do not seem to admit a GLSM description,
even though they are toric varieties -- they can be described as
GIT quotients of open subsets of ${\mathbb C}^{n+2}$ by ${\mathbb C}^{\times}$,
but not as a GIT quotient of the full complex vector space, and they
naturally compactify to ${\mathbb P}^{n+1}_{[1,\cdots,1,m]}$.
We would like to thank D.~Skinner for asking a question that
made this manifest.}
by a gauged linear sigma model with fields of
$U(1)$ charges
\begin{center}
\begin{tabular}{cccc}
$x_1$ & $\cdots$ & $x_{n+1}$ & $p$ \\ \hline
$1$ & $\cdots$ & $1$ & $-m$
\end{tabular}
\end{center}

Now, a ${\mathbb Z}_k$ gerbe over ${\mathbb P}^n$ can be described by
a gauged linear sigma model in which the $n+1$ fields/homogeneous
coordinates have weight $k$ instead of weight $1$, as discussed
in {\it e.g.} \cite{glsm}.  Then, for example, the GLSM with fields and $U(1)$
charges
\begin{center}
\begin{tabular}{cccc}
$x_1$ & $\cdots$ & $x_{n+1}$ & $p$ \\ \hline
$k$ & $\cdots$ & $k$ & $-k$
\end{tabular}
\end{center}
is surely going to be the pullback of ${\cal O}(-1) \rightarrow {\mathbb P}^n$
to the gerbe.

However, how does one interpret GLSM's defined by, for example:
\begin{center}
\begin{tabular}{cccc}
$x_1$ & $\cdots$ & $x_{n+1}$ & $p$ \\ \hline
$k$ & $\cdots$ & $k$ & $-1$
\end{tabular}
\end{center}
This is the total space of what is sometimes referred to as the
``${\cal O}(1/k)$'' line bundle over the ${\mathbb Z}_k$ gerbe
${\mathbb P}^n_{[k,\cdots,k]}$.
It is an example of a line bundle on the gerbe that is not a pullback
of a line bundle on the base space -- the gerbe has more bundles than the
base space.  More to the point, it can only be understood as the total
space of a line bundle on a gerbe -- so a physicist who was very careful
in a study of GLSM's would eventually be forced to discover gerbes in order
to make sense of this example.

In addition to being a line bundle over the stack,
the total space of the ${\cal O}(1/k)$ line bundle is also a fibered orbifold
over the projective space ${\mathbb P}^n$ -- it is a type of fiber bundle over
${\mathbb P}^n$, in which the fibers are the orbifolds
$[ {\mathbb C}/{\mathbb Z}_k ]$.  For this reason, these
are sometimes known as `orbibundles;' see {\it e.g.}
\cite{qureshi-szendroi} for references to the literature under this name.
(This same structure has also
been discussed in connection with interpreting hybrid Landau-Ginzburg
models, see {\it e.g.} \cite{alg22}.)

Not all ${\mathbb Z}_k$ gerbes on projective spaces are of the form of weighted
projective stacks.  A more general class was discussed in
{\it e.g.} \cite{glsm}[section 3.3], and, roughly, are given by
${\mathbb C}^{\times}$ quotients of principal ${\mathbb C}^{\times}$
bundles over ${\mathbb P}^n$.  Specifically, consider a GLSM with
fields $x_i$, $z$, and two ${\mathbb C}^{\times}$ actions, as follows:
\begin{center}
\begin{tabular}{c|cc}
& $x_i$ & $z$ \\ \hline
$\lambda$ & $1$ & $-n$ \\
$\mu$ & $0$ & $k$ 
\end{tabular}
\end{center}
The first ${\mathbb C}^{\times}$, $\lambda$, defines the total space of a 
line bundle on ${\mathbb P}^n$ 
of degree $-n$.  The second ${\mathbb C}^{\times}$, $\mu$, 
quotients out the fibers, leaving a ${\mathbb Z}_k$ kernel.  The result
is a ${\mathbb Z}_k$ gerbe over ${\mathbb P}^n$, of characteristic
class $-n$~mod~$k$.  The weighted projective stacks we have been discussing
correspond to an alternative presentation in the special case that $n=1$.
One can define line bundles over these gerbes in the obvious fashion.

The notation ${\cal O}(1/k)$, while initially catchy, is unfortunately
ambiguous -- for example, it does not distinguish a twisted bundle of $c_1=k$
over the gerbe from the pullback from ${\mathbb P}^n$ of an ordinary
line bundle of $c_1=1$.  Let us introduce a more precise notation. 

We will use 
``${\cal O}_{\Lambda}(m)$'' to
denote a line bundle defined by
a superfield of charge $m$.  For bundles on, say,
ordinary projective spaces, the $k=1$ case,
a superfield of charge $m$ couples
to the line bundle ${\cal O}(m)$.

To understand the meaning of this notation, let us first consider a
${\mathbb Z}_2$ gerbe over ${\mathbb P}^n$ defined by
the weighted projective stack ${\mathbb P}^n_{[2,2,\cdots,2]}$.
Let $G{\mathbb P}^n$ denote the gerbe, and $\pi:
G{\mathbb P}^n \rightarrow {\mathbb P}^n$ the natural projection from the
gerbe onto the underlying projective space.

Now,
coherent sheaves on the gerbe decompose into twisted sheaves on the
underlying space (see section~\ref{sect:twisting} or \cite{summ}).
Formally, if $\alpha \in H^2({\mathbb P}^n,{\mathbb Z}_2)$ is the characteristic
class of the gerbe, then
\begin{displaymath}
\mbox{Coh}(G {\mathbb P}^n) \: = \:
\mbox{Coh}({\mathbb P}^n, 1(\alpha)) \cup
\mbox{Coh}({\mathbb P}^n, \chi(\alpha)),
\end{displaymath}
where $\mbox{Coh}(X, \lambda)$ denotes coherent sheaves on $X$ twisted
by a 2-cocycle $\lambda$.  In the notation above, $1$ and
$\chi$ are the two irreducible representations of ${\mathbb Z}_2$,
so $1(\alpha)$ is the vanishing 2-cocycle and $\chi(\alpha)$ is
a cocycle that does not vanish identically.
Note that {\it both} cocycles are cohomologous to the identity --
both components of $\mbox{Coh}(G{\mathbb P}^n)$ are isomorphic to
ordinary coherent sheaves $\mbox{Coh}({\mathbb P}^n)$.
(This resolves a potential contradiction, in that the rank of a bundle
twisted by a cohomologically nontrivial cocycle, must be divisible by the
order of the cocycle, and so here would need to be divisible by $k$ --
truly twisted line bundles do not exist.)

In this language, we can immediately read off that
\begin{displaymath}
{\cal O}_{\Lambda}(k) \: = \: \left\{ \begin{array}{cl}
\mbox{Coh}({\mathbb P}^n, 1(\alpha)) & k \mbox{ even}, \\
\mbox{Coh}({\mathbb P}^n, \chi(\alpha)) & k \mbox{ odd}.
\end{array} \right.
\end{displaymath}
In other words, if $k$ is even, then ${\cal O}_{\Lambda}(k)$ is a pullback
to the gerbe from a line bundle on the base.  For other values of $k$,
the bundle is twisted by an action of the ${\mathbb Z}_2$.

Now, the projection map $\pi: G {\mathbb P}^n \rightarrow {\mathbb P}^n$ defines
a functor
\begin{displaymath}
\pi^*: \: \mbox{Coh}( {\mathbb P}^n ) \:\stackrel{\sim}{\longrightarrow} \:
\mbox{Coh}({\mathbb P}^n, 1(\alpha) ).
\end{displaymath}
In addition, there is another functor
\begin{displaymath}
\pi_1^* \: \equiv \: \pi^* \otimes {\cal O}_{\Lambda}(1): \:
\mbox{Coh}({\mathbb P}^n) \: \stackrel{\sim}{\longrightarrow} \:
\mbox{Coh}({\mathbb P}^n, \chi(\alpha) ).
\end{displaymath}
(In fact, there is an analogue of $\pi_1^*$ for every ${\cal O}_{\Lambda}(
\mbox{odd})$.)

To determine $\pi^* {\cal O}(m)$ in terms of ${\cal O}_{\Lambda}$'s,
consider the commutative diagram
\begin{displaymath}
\xymatrix{
\frac{ {\mathbb C}^{n+1} - 0 }{ {\mathbb C}^{\times} }
\ar[r] \ar[d] &
\frac{ {\mathbb C}^{n+1} - 0 }{ {\mathbb C}^{\times} }
\ar[d] \\
G {\mathbb P}^n \ar[r] & {\mathbb P}^n
}
\end{displaymath}
The line bundle ${\cal O}(k)$, defined by weights $1, \cdots, 1, k$,
pulls back to weights $2, \cdots, 2, 2k$,
from which we deduce that
\begin{displaymath}
\pi^* {\cal O}(k) \: = \: {\cal O}_{\Lambda}(2k),
\end{displaymath}
which implies
\begin{displaymath}
\pi_1^* {\cal O}(k) \: = \: {\cal O}_{\Lambda}(2k+1).
\end{displaymath}

Note that although $\pi^*$ preserves tensor products,
$\pi_1^*$ does {\it not} preserve tensor products:
\begin{eqnarray*}
\pi_1^* \left( {\cal O}(k) \otimes {\cal O}(m) \right) & \cong &
\pi_1^* {\cal O}(k+m), \\
& \cong & {\cal O}_{\Lambda}(2k+2m+1), \\
& \not\cong & {\cal O}_{\Lambda}(2k+2m+2) \: \cong \:
\left( \pi_1^* {\cal O}(k) \right) \otimes
\left( \pi_1^* {\cal O}(m) \right).
\end{eqnarray*}
Indeed, this is an immediate consequence of the definition of
$\pi_1^*$.
In addition, for the same reason, $\pi_1^*$ does not commute with
duality of bundles
\begin{displaymath}
\pi_1^* \left( {\cal L}^{\vee} \right) \: \not\cong \:
\left( \pi_1^* {\cal L} \right)^{\vee}.
\end{displaymath}

Now, for any finite gerbe over any space, the tangent bundle of the gerbe
is just the pullback (under $\pi$) of the tangent bundle to the space.
One way to see this is to work locally on the atlas, which is just a finite
cover, and so the tangent bundle should be the same.
We can see this explicitly in the present case as follows.
For the ${\mathbb Z}_2$ gerbe $G {\mathbb P}^n = {\mathbb P}^n_{[2,\cdots,2]}$,
the tangent bundle seen by the gauged linear sigma model is
\begin{displaymath}
0 \: \longrightarrow \: {\cal O}_{\Lambda} \: \longrightarrow \:
{\cal O}_{\Lambda}(2)^{n+1} \: \longrightarrow \:
T G {\mathbb P}^n \: \longrightarrow \: 0.
\end{displaymath}
Using the isomorphisms above, we see this short exact sequence is the
same as
\begin{displaymath}
0 \: \longrightarrow \: \pi^* {\cal O} \: \longrightarrow \:
\pi^* {\cal O}(1)^{n+1} \: \longrightarrow \: T G {\mathbb P}^n \: 
\longrightarrow
\: 0,
\end{displaymath}
which is just $\pi^*$ of the Euler sequence for the tangent bundle
\begin{displaymath}
0 \: \longrightarrow \: {\cal O} \: \longrightarrow \:
{\cal O}(1)^{n+1} \: \longrightarrow \: T {\mathbb P}^n \:
\longrightarrow \: 0.
\end{displaymath}

For ${\mathbb Z}_k$ gerbes over ${\mathbb P}^n$ built as the weighted
projective stack ${\mathbb P}^n_{[k,\cdots,k]}$, there is a closely
analogous story.  Here, coherent sheaves on $G {\mathbb P}^n$ decompose as
\begin{displaymath}
\mbox{Coh}(G{\mathbb P}^n) \: = \: \cup_{\chi} \mbox{Coh}({\mathbb P}^n,
\chi(\alpha)),
\end{displaymath}
where the union is over irreducible representations of ${\mathbb Z}_k$,
and there are $k$ different pullbacks, first the canonical
\begin{displaymath}
\pi^*: \: \mbox{Coh}({\mathbb P}^n) \: \stackrel{\sim}{\longrightarrow} \:
\mbox{Coh}({\mathbb P}^n, 1(\alpha)),
\end{displaymath}
followed by $\pi_i^*(-) \equiv \pi^*(-) \otimes {\cal O}_{\Lambda}(i)$.
Identifying $\pi_0^*$ with $\pi^*$, we have the general relation
\begin{displaymath}
\pi_i^* {\cal O}(m) \: = \: {\cal O}_{\Lambda}(km+i).
\end{displaymath}
An argument nearly identical to the one above shows that the
tangent bundle $T G {\mathbb P}^n$ seen by a gauged linear sigma model
is given by $\pi^* T {\mathbb P}^n$, exactly as must be true on general
grounds.

\subsection{Sheaf cohomology}

On a global quotient stack $\mathfrak{X} = [X/G]$, for $G$ finite,
given a vector bundle
${\cal E} \rightarrow \mathfrak{X}$, (equivalently, a $G$-equivariant
bundle on $X$,)
\begin{displaymath}
H^{\bullet}(\mathfrak{X}, {\cal E}) \: = \:
H^{\bullet}(X, {\cal E})^G.
\end{displaymath}
In our discussion of massless spectra of heterotic strings on stacks,
this is ultimately the reason why in orbifolds one gets $G$-invariants.

Now, nontrivial gerbes over projective spaces have a global quotient
description as some $[X/G]$ for $G$ nonfinite, and the simple
description of sheaf cohomology above in terms of $G$-invariants is only
valid for $G$ finite, so for general cases a different approach
is required.  For example,
let $\mathfrak{X} = {\mathbb P}^n_{[k,\cdots,k]}$, 
and ${\cal O}_{\mathfrak{X}}(m)$ as above, then
\begin{displaymath}
H^i(\mathfrak{X}, {\cal O}_{\mathfrak{X}}(m)) \: = \: \left\{
\begin{array}{cl}
0 & k \nmid  m, \\
H^i({\mathbb P}^n, {\cal O}_{ {\mathbb P}^n }(m/k)) & k \mid m.
\end{array} \right.
\end{displaymath}

For $m \geq 0$, we can check this as follows.
First,
\begin{displaymath}
H^i\left(\mathfrak{X}, {\cal O}_{\mathfrak{X}}(m) \right) \: = \:
H^i_{ {\mathbb C}^{\times} }\left( {\mathbb C}^{n+1}-\{0\}, {\cal O} \right),
\end{displaymath}
where the ${\cal O}$ coefficients have weight $m$ under the 
${\mathbb C}^{\times}$.
In principle, there is a spectral sequence converging to the right-hand
side, with level-two terms
\begin{displaymath}
H^p\left( {\mathbb C}^{\times}, H^q\left( {\mathbb C}^{n+1} - \{0\}, 
{\cal O}\right)
\right),
\end{displaymath}
but $H^q({\mathbb C}^{n+1}-\{0\}, {\cal O}) = 0$ for $q \neq 0, n$, and
\begin{displaymath}
H^0\left( {\mathbb C}^{n+1}-\{0\}, {\cal O} \right) \: = \:
{\mathbb C}[x_0, \cdots, x_n].
\end{displaymath}
(The degree $n$ cohomology is also nonzero and infinite-dimensional,
but it will not contribute any invariants for $m \geq 0$, only for
$m < 0$, so we omit it
from this discussion.)
For $\lambda \in {\mathbb C}^{\times}$, the representation
\begin{displaymath}
\rho: \: {\mathbb C}^{\times} \: \longrightarrow \: {\rm GL}(
{\mathbb C}[x_0,\cdots, x_n])
\end{displaymath}
is defined by
\begin{displaymath}
\rho_{\lambda}(f(x)) \: = \: \lambda^{-m}f(\lambda^k x_0, \cdots,
\lambda^k x_n).
\end{displaymath}
The group $H^p({\mathbb C}^{\times}, ( {\mathbb C}[x_0,\cdots,x_n],\rho))$
is zero unless $p=0$, since it is a reductive group, and for $p=0$
is given by the invariants.

Next, let us compute the invariants.
Decompose
\begin{displaymath}
f \: = \: f_0 \: + \: \cdots \: + \: f_N,
\end{displaymath}
where $f_d$ denotes a homogeneous polynomial of degree $d$.
Under the ${\mathbb C}^{\times}$ action,
\begin{displaymath}
\rho_{\lambda}(f) \: = \: \lambda^{-m} f_0 \: + \:
\lambda^{-m + k}f_1 \: + \: \cdots \: + \:
\lambda^{-m + kN} f_N.
\end{displaymath}
Thus, ${\mathbb C}^{\times}$ invariants only exist in the case that $k$
divides $m$, and in that case, are counted by degree $m/k$ polynomials
in $n+1$ variables.  

Now, let us compare to the original claim.
It is a standard result that for $\ell > 0$,
\begin{displaymath}
H^i({\mathbb P}^n, {\cal O}_{{\mathbb P}^n}(\ell)) \: = \:
\left\{ \begin{array}{cl}
0 & i \neq 0, \\
{\rm Sym}^{\ell} \mathbb{C}^{n+1} & i=0.
\end{array} \right.
\end{displaymath}
In other words, the only nonzero cohomology is in degree zero, and in that
degree, it is counted by homogeneous polynomials of degree $\ell$ in
$n+1$ variables.
The desired result follows.

\section{Chern classes on the inertia stack}
\label{app:chern-reps}

As we are manipulating bundles on stacks, it is worth spending a little
time reviewing corresponding Chern classes.  It is possible
to define Chern classes on a stack itself; for example, Chern classes
of a vector bundle ${\cal E}$ on a quotient stack $[X/G]$ are simply
$G$-equivariant Chern classes of ${\cal E}$ on $X$.
However, these Chern classes do not always behave well under
mathematical manipulations, and in any event a different notion of
Chern classes and Chern characters, 
denoted $c^{\rm rep}$ and ${\rm ch}^{\rm rep}$, exists and is relevant
for index theory.  These alternative notions of Chern classes do not
live in the cohomology of the
original stack, but rather of the inertia stack, which encodes twisted
sectors of string orbifolds.  (See appendix~\ref{app:spectra} 
for more information
on the inertia stack.)  

In this section, we will illustrate how to compute such Chern
classes and characters (denoted $c^{\rm rep}$ and ${\rm ch}^{\rm rep}$)
and describe their appearance in index theory in some examples.
It is tempting to wonder whether one could derive extra anomaly constraints
on orbifolds
from these stack Chern classes 
over nontrivial components of the inertia stack, but we argue that
does not seem to happen in heterotic compactifications in
section~\ref{sect:possible-anomcanc} (though see
section~\ref{app:spectra:fockconstraints} for a possible application
of $c_1^{\rm rep}$).

For any stack $\mathfrak{X}$, let $V$ be a vector bundle over
$\mathfrak{X}$, and $I_{\mathfrak{X}}$ the inertia stack of $\mathfrak{X}$.
Let $q: I_{\mathfrak{X}} \rightarrow \mathfrak{X}$ denote the natural
projection operator onto one component.

We define Chern classes of $V$ as follows.
First, pullback $V$ to $I_{\mathfrak{X}}$ along $q$.
Then, on each component $\alpha$ of $I_{\mathfrak{X}}$, 
$q^* V$ will decompose into
eigenbundles of the action of the stabilizer for that component:
\begin{displaymath}
q^* V|_{\alpha} \: = \: \oplus_{\chi} V_{\alpha, \chi}.
\end{displaymath}
(When $\alpha$ is the identity, our conventions are that there is only
one component, associated to the trivial character.)
Define ${\rm ch}^{\rm rep}(V)$ over a component $\alpha$ to be
\begin{displaymath}
{\rm ch}^{\rm rep}(V)|_{\alpha} \: \equiv \:
\bigoplus_{\chi} {\rm ch}(V_{\alpha, \chi}) \otimes \chi,
\end{displaymath}
where $\chi$ is the eigenvalue
of that component of $q^* V$ under the stabilizer, and ${\rm ch}$ denotes
the naive notion of Chern classes, living in equivariant cohomology
pertinent to the stack itself.
(These seem to be the same as the Chern classes in ``delocalized cohomology''
described in {\it e.g.} \cite{at-seg,bbmp,baum-fete},
though our starting point is different.)

Intuitively, the idea is that on any component of the inertia stack
determined by some generic automorphism, the bundle should decompose
into eigenbundles, and $\chi$ is the eigenvalue associated with the
action of that automorphism on the bundle.
Slightly more generally, one can define a ``diagonalization map''
\begin{displaymath}
d: \: K^0(I_{\mathfrak{X}}) \otimes {\mathbb C} \: \longrightarrow \:
K^0(I_{\mathfrak{X}}) \otimes {\mathbb C},
\end{displaymath}
which on a component $\alpha$ maps a sheaf ${\cal F}$ to
its isotypic decomposition, weighted by characters:
\begin{displaymath}
d( [{\cal F}] ) |_{\alpha}
 \: = \: \sum_{\chi} {\cal F}_{\alpha, \chi} \otimes \chi.
\end{displaymath}
In this language,
\begin{displaymath}
{\rm ch}^{\rm rep}(V) \: = \: {\rm ch}( d( q^* V) ).
\end{displaymath}

To clarify these ideas, let us work through some examples.

First, we shall consider a vector bundle on a trivial gerbe.
Consider a vector bundle $V \rightarrow \mathfrak{X} \equiv
X \times B {\mathbb Z}_k$,
so $V = p_1^* E \otimes p_2^* \zeta$ for some bundle $E \rightarrow X$
and representation $\zeta \in {\mathbb Z}_k^{\vee}$.

The inertia stack $I_{\mathfrak{X}}$ is given by
\begin{displaymath}
I_{\mathfrak{X}} \: = \: 
\coprod_{g \in {\mathbb Z}_k} X \times B{\mathbb Z}_k \times \{ g \}.
\end{displaymath}
There is a forgetful map $q: I_{\mathfrak{X}} 
\rightarrow X \times B {\mathbb Z}_k$.

Consider
\begin{displaymath}
q^* V \: = \: \oplus_{\chi \in {\mathbb Z}_k^{\vee} } V_{\chi},
\end{displaymath}
where $V_{\chi}$ is the $\chi$ eigenspace for the $g$ action on
$q^* V$:
\begin{displaymath}
q^* V|_{X \times B {\mathbb Z}_k \times \{ g \} } \: = \: V,
\end{displaymath}
\begin{displaymath}
V_{\chi}|_{X \times B{\mathbb Z}_k \times \{ g \} } \: = \:
\left\{ \begin{array}{cl}
V & \mbox{if } \chi(g) = \zeta(g), \\
0 & \rm{else}.
\end{array} \right.
\end{displaymath}

Now, we want to compute ${\rm ch}^{\rm rep}(V) \in 
H^{\bullet}(I_{\mathfrak{X}}, 
{\mathbb C})$.
\begin{displaymath}
V \: \mapsto \: q^* V \: = \: \oplus_{\chi} V_{\chi} \: \mapsto \:
\oplus_{\chi} V_{\chi} \otimes \chi,
\end{displaymath}
where $V_{\chi} \otimes \chi \in K^0(I_{\mathfrak{X}}) \otimes {\mathbb C}$.
(We think of $V_{\chi} \in K^0(I_{\mathfrak{X}})$, and $\chi \in {\mathbb C}$.)

Then,
\begin{displaymath}
{\rm ch}^{\rm rep}(V) \: = \: {\rm ch}\left( \oplus_{\chi} V_{\chi}
\otimes \chi \right) \in H^{\bullet}(I_{\mathfrak{X}},{\mathbb C}) \: = \: 
\oplus_g H^{\bullet}(X),
\end{displaymath}

\begin{displaymath}
V_{\chi} \otimes \chi |_{X \times B {\mathbb Z}_k \times \{ g \} } \: = \:
\left\{ \begin{array}{cl}
V \otimes \chi & \mbox{if } \chi(g) = \zeta(g), \\
0 & {\rm else}.  
\end{array} \right.
\end{displaymath}

Putting this together, we find
\begin{displaymath}
{\rm ch}^{\rm rep}(V) \: = \: \left( {\rm ch}^{\rm rep}(V)|_{(g)} \right)_{
g \in {\mathbb Z}_k},
\end{displaymath}
where
\begin{displaymath}
{\rm ch}^{\rm rep}(V)|_{(g)} \: = \: \oplus_{\chi \: {\rm s.t.} \:
\chi(g) = \zeta(g) } {\rm ch}(V) \otimes \chi.
\end{displaymath}
Similarly,
\begin{displaymath}
{\rm ch}^{\rm rep}(T\mathfrak{X})|_{(g)} \: = \: \oplus_{\chi \: {\rm s.t.} \:
\chi(g)=1 } {\rm ch}(T\mathfrak{X}) \otimes \chi.
\end{displaymath}

For $g=1$,
\begin{displaymath}
{\rm ch}^{\rm rep}(V)|_{(1)} \: = \: \oplus_{\chi} {\rm ch}(V) \otimes
\chi,
\end{displaymath}
and similarly for ${\rm ch}^{\rm rep}(TX)^{(1)}$.

Now, suppose $k$ is prime.  Then $\chi(g)=1$ implies $\chi = 1$.
Thus,
\begin{displaymath}
{\rm ch}^{\rm rep}(V)|_{(g)} \: = \: {\rm ch}(V) \otimes \zeta(g),
\end{displaymath}
\begin{displaymath}
{\rm ch}^{\rm rep}(T\mathfrak{X})|_{(g)} \: = \: {\rm ch}(T\mathfrak{X}) 
\otimes 1,
\end{displaymath}
for all $g$.

Next, let us consider a line bundle on a nontrivial gerbe.
Consider the prototypical example of a ${\mathbb Z}_k$ gerbe on
${\mathbb P}^n$:  $\mathfrak{X} = {\mathbb P}^n_{[k,k,\cdots,k]}$.
Let ${\cal O}_{\mathfrak{X}}(m)$ denote the holomorphic line bundle defined by
${\mathbb C}^{\times}$ weight $-m$.  In other words, if $m$ is divisible by
$k$, then ${\cal O}_{\mathfrak{X}}(m)$ is the pullback of 
${\cal O}_{ {\mathbb P}^n}(m/k)$ under the projection map from the
gerbe $\mathfrak{X}$ to the underlying space ${\mathbb P}^n$.

The components of the inertia stack are labelled by $k$th roots of unity
(not characters, but group elements).  The Chern classes
ch$^{\rm rep}$ have $k$ components, each component in a cohomology class
(with complex coefficients) on the stack.  If we let $\alpha$ denote
a $k$th root of unity, then on that component of the inertia stack,
\begin{displaymath}
c_1^{\rm rep}({\cal O}_{\mathfrak{X}}(m))|_{\alpha} \: = \: 
\frac{m}{k} \alpha^{-m} J,
\end{displaymath}
where $J$ is the pullback to the gerbe of the hyperplane class,
and the total Chern character is
\begin{displaymath}
{\rm ch}^{\rm rep}( {\cal O}_{\mathfrak{X}}(m))|_{\alpha} \: = \: 
\alpha^{-m} \exp\left( \frac{m}{k} J \right).
\end{displaymath}
To derive this, remember that for a line bundle $L$ over the stack 
$\mathfrak{X}$,
if $\pi: I_{\mathfrak{X}} \rightarrow \mathfrak{X}$ 
denotes the projection from the inertia stack
to $\mathfrak{X}$, then the Chern characters are
\begin{displaymath}
{\rm ch}^{\rm rep}(L)|_{\mathfrak{X} \times \{\alpha\} }
 \: = \: \pi^* \left. {\rm ch}\left(L\right)\right|_{\mathfrak{X} 
\times \{\alpha\} }
\otimes \chi,
\end{displaymath}
where $\chi$ is the eigenvalue of the stabilizer $\alpha$ on
$\pi^* L|_{\mathfrak{X} \times \{\alpha\} }$.  Here, $\chi = \alpha^{-m}$.

More generally, over all components, we write
\begin{displaymath}
c_1^{\rm rep}({\cal O}_{\mathfrak{X}}(m)) \: = \: \left( \frac{m}{k} J, \cdots, 
\frac{m}{k} \alpha^{-m} J, \cdots \right).
\end{displaymath}
Multiplication of components of ch$^{\rm rep}$ multiplies not only the
cohomology classes, but also the coefficients.  For example,
\begin{displaymath}
\left( c_1^{\rm rep}({\cal O}(m))|_{\mathfrak{X} \times \{\alpha\} }
\right)^2 \: = \: 
\left( \frac{m}{k} J \right)^2 \alpha^{-2m}.
\end{displaymath}
Now, for a line bundle $L$ on an ordinary space,
\begin{displaymath}
{\rm ch}_2(L) \: = \: (1/2) c_1^2(L),
\end{displaymath}
but here, by contrast,
\begin{eqnarray*}
{\rm ch}_2^{\rm rep}({\cal O}(m))|_{\mathfrak{X} \times\{\alpha\}} & = & 
\frac{1}{2}
\left( \frac{m}{k} J \right)^2 \alpha^{-m}, \\
& = & \alpha^{+m} \frac{1}{2}
\left( c_1^{\rm rep}({\cal O}(m))|_{\mathfrak{X} \times \{\alpha\} }
\right)^2,
\end{eqnarray*}
so that the usual relation between Chern classes and Chern characters
is modified on a stack.  (In fact, if we were computing Chern classes of
a bundle that split as several different eigenbundles, the relation would
be much more complicated than just an additional complex phase.)

As a consistency check, let us compute the index of this line bundle,
using Hirzebruch-Riemann-Roch.
For any bundle ${\cal E} \rightarrow \mathfrak{X}$, the 
Hirzebruch-Riemann-Roch index theorem says
\begin{displaymath}
\chi({\cal E}) \: = \: \int_{I_{\mathfrak{X}}} {\rm ch}^{\rm rep}({\cal E})
{\rm Td}(\mathfrak{X})
\end{displaymath}
where
\begin{displaymath}
\chi({\cal E}) \: = \: \sum_i (-)^i h^i(\mathfrak{X}, {\cal E}),
\end{displaymath}
and 
\begin{displaymath}
{\rm Td}(\mathfrak{X}) \: = \: \alpha_{\mathfrak{X}}^{-1} {\rm Td}( T 
I_{\mathfrak{X}} ),
\end{displaymath}
where
\begin{displaymath}
\alpha_{\mathfrak{X}} \: = \: {\rm ch}( d( \lambda_q) ), \: \: \:
\lambda_q \: = \: \sum_k (-)^k \wedge^k N_q^*,
\end{displaymath}
for $N_q$ the normal bundle.
(As $\lambda_q$ is not a pullback from $\mathfrak{X}$, but rather is
defined intrinsically on $I_{\mathfrak{X}}$, ${\rm ch}^{\rm rep}(\lambda_q)$
is not well-defined, so instead the pertinent Chern character is defined
via the diagonalization map $d$.)

In the present case, since each component of the inertia stack 
$I_{\mathfrak{X}}$
is isomorphic
to the original stack $\mathfrak{X}$, 
the normal bundle $N_q$ vanishes, and
each component of ${\rm ch}(d(\lambda_q))$ is  $1$.
Furthermore, as $\mathfrak{X}$ 
is essentially a $k$-fold quotient of ${\mathbb P}^n$,
\begin{displaymath}
\int_{\mathfrak{X}} \: = \: \frac{1}{k} \int_{{\mathbb P}^n}.
\end{displaymath}
Plugging into the index formula,
\begin{eqnarray*}
\int_{I_{\mathfrak{X}}} {\rm ch}^{\rm rep}({\cal O}_X(m)) 
{\rm Td}(TI_{\mathfrak{X}}) 
& = & \sum_{\alpha} \int_{\mathfrak{X}} \alpha^{-m} 
{\rm ch}({\cal O}_{\mathfrak{X}}(m)) {\rm Td}(T\mathfrak{X}),
\\
& = & \sum_{\alpha} \alpha^{-m} \int_{\mathfrak{X}} 
\sum_i {\rm ch}_i({\cal O}_{\mathfrak{X}}(m)) 
{\rm Td}_{n-i}(T\mathfrak{X}).
\end{eqnarray*}
Now, since $\alpha$ is a $k$th root of unity, the sum
\begin{displaymath}
\sum_{\alpha} \alpha^{-m}
\end{displaymath}
will vanish unless $m$ is divisible by $k$.  Thus, if $m$ is not divisible
by $k$, we find that $\chi({\cal O}_{\mathfrak{X}}(m))$ vanishes.
Next, suppose that $m=n k$ for some integer $n$.
Then,
\begin{eqnarray*}
\int_{I_{\mathfrak{X}}} {\rm ch}^{\rm rep}({\cal O}_{\mathfrak{X}}(m)) 
{\rm Td}(I_{\mathfrak{X}}) 
& = & \sum_{\alpha} \int_{\mathfrak{X}} \alpha^{-m} 
{\rm ch}({\cal O}_{\mathfrak{X}}(m)) {\rm Td}(T\mathfrak{X}),
\\
& = & \sum_{\alpha} \int_{\mathfrak{X}} \pi^* 
{\rm ch}({\cal O}_{{\mathbb P}^n}(n))
{\rm Td}(T {\mathbb P}^n), \\
& = & \sum_{\alpha} \frac{1}{k} \int_{ {\mathbb P}^n }
 {\rm ch}({\cal O}_{{\mathbb P}^n}(n))
{\rm Td}(T {\mathbb P}^n), \\
& = &
 \int_{ {\mathbb P}^n }
 {\rm ch}({\cal O}_{{\mathbb P}^n}(n))
{\rm Td}(T {\mathbb P}^n), \\
& = & \chi\left({\mathbb P}^n, {\cal O}_{ {\mathbb P}^n }(n) \right).
\end{eqnarray*}

Now, let us compare to expectations.  In the present case,
if $m$ is not divisible by $k$, then all the sheaf
cohomology groups of ${\cal O}_{\mathfrak{X}}(m)$ should vanish,
so the Euler class
$\chi({\cal O}_{\mathfrak{X}}(m))$ should vanish, exactly as we have
computed.  If $m$ is divisible by $k$,
then $\chi({\cal O}_{\mathfrak{X}}(m)) = 
\chi({\cal O}_{ {\mathbb P}^n }(m/k))$, again matching the result of
the computation.

Another example\footnote{
We would like to thank T.~Pantev for explaining this example to us.
} will be handy to understand.

Take $\mathfrak{X} = [T^4/{\mathbb Z}_2]$,
where the ${\mathbb Z}_2$ acts by sign flips
(and so has 16 fixed points).
Let us compute
\begin{displaymath}
\chi\left( {\cal O}_{\mathfrak{X}}[0] \right), \: \: \:
\chi\left( {\cal O}_{\mathfrak{X}}[1/2] \right),
\end{displaymath}
where ${\cal O}_{\mathfrak{X}}[0]$ denotes the structure sheaf with trivial
${\mathbb Z}_2$-equivariant structure, and
${\cal O}_{\mathfrak{X}}[1/2]$ denotes the
structure sheaf with nontrivial equivariant structure.
For this $\mathfrak{X}$, $I_{\mathfrak{X}}$ has 17 components:
one copy of $\mathfrak{X}$, and 16 copies of
$[{\rm pt}/{\mathbb Z}_2]$.  From the definition
\begin{displaymath}
{\rm ch}^{\rm rep}(L)|_{\alpha} \: = \: \pi^* {\rm ch}(L) |_{\alpha}
\otimes \chi,
\end{displaymath}
where $\alpha$ is a component of $I_{\mathfrak{X}}$ and $\chi$ the eigenvalue of
$\alpha$'s stabilizer on $\pi^* L$, it is straightforward to compute that
\begin{eqnarray*}
{\rm ch}^{\rm rep}( {\cal O}[0] ) & = &
(1, \vec{0}, 0; 1, \cdots 1), \\
{\rm ch}^{\rm rep}( {\cal O}[1/2] ) & = &
(1, \vec{0}, 0; -1, \cdots -1),
\end{eqnarray*}
where the leading three entries are for the $\mathfrak{X}$ component, 
corresponding to elements of $H^0(\mathfrak{X}) = {\mathbb C}$, 
$H^2(\mathfrak{X}) = {\mathbb C}^6$, $H^4(\mathfrak{X}) = {\mathbb C}$, 
respectively,
and the remaining
sixteen entries are each for a copy of $[{\rm pt}/{\mathbb Z}_2]$.

The normal bundle $N$ is $0$ for the trivial component $[T^4/{\mathbb Z}_2]$ of
$I_{\mathfrak{X}}$,
and is ${\mathbb C}^2$ with ${\mathbb Z}_2$ acting by sign flips for the
other components of $I_{\mathfrak{X}}$.  From that, we read off that
\begin{eqnarray*}
{\rm ch}(d(\wedge^0 N)) & = & (1, \vec{0}, 0; 1, \cdots, 1), \\
{\rm ch}(d(N)) & = & (0, \vec{0}, 0; -2, \cdots, -2), \\
{\rm ch}(d(\wedge^2 N)) & = & (0, \vec{0}, 0; 1, \cdots, 1), \\
{\rm ch}(d(\wedge^k N)) & = & 0 \: \mbox{  for } k > 2.
\end{eqnarray*}
From this we find
\begin{displaymath}
\alpha_{\mathfrak{X}} \: = \:
{\rm ch}(d(\lambda_q)) \: = \: {\rm ch}(d( \sum_i (-)^i \wedge^i N^*)
\: = \: (1, \vec{0}, 0; 4, \cdots, 4).
\end{displaymath}
In addition,
\begin{displaymath}
{\rm ch}^{\rm rep}({\rm Td}(TI_{\mathfrak{X}})) \: = \:
(1, \vec{0}, 0; 1, \cdots, 1),
\end{displaymath}
hence
\begin{displaymath}
{\rm Td}(\mathfrak{X}) \: = \: 
\alpha_{\mathfrak{X}}^{-1} {\rm Td}(T I_{\mathfrak{X}})
\: = \:
(1, \vec{0}, 0; 1/4, \cdots, 1/4).
\end{displaymath}

Putting this together, we find
\begin{eqnarray*}
\chi\left( {\cal O}_{\mathfrak{X}}[0] \right) & = &
\int_{I_{\mathfrak{X}}} {\rm ch}^{\rm rep}( {\cal O}_{\mathfrak{X}}[0])
{\rm Td}(\mathfrak{X}) \\
& = &
\int_{ [T^4/{\mathbb Z}_2] } (1)  (1)  \: + \:
16 \int_{ [{\rm pt}/{\mathbb Z}_2] } (1) (1/4),\\
& = & 0 \: + \: 4 \int_{ [ {\rm pt}/{\mathbb Z}_2] } 1, \\
& = & 4 \left(\frac{1}{2}\right) \: = \: 2, \\
\chi\left( {\cal O}_{\mathfrak{X}}[1/2] \right) & = & 
\int_{I_{\mathfrak{X}}} {\rm ch}^{\rm rep}( {\cal O}_{\mathfrak{X}}[1/2])
{\rm Td}(\mathfrak{X}) \\
& = &
\int_{ [T^4/{\mathbb Z}_2] } (1) (1)  \: + \:
16 \int_{ [{\rm pt}/{\mathbb Z}_2] } (-1) (1/4), \\
& = & 0 \: - \: 4 \int_{ [{\rm pt}/{\mathbb Z}_2] } 1, \\
& = & -4 \left( \frac{1}{2}\right) \: = \: -2.
\end{eqnarray*}

Let $Y$ denote a minimal resolution of $T^4/{\mathbb Z}_2$.
Applying 
the McKay correspondence \cite{bkr}, it can be shown \cite{tonypriv} that
the bundle ${\cal O}_{\mathfrak{X}}[0]$ maps to ${\cal O}_Y$,
and ${\cal O}_{\mathfrak{X}}[1/2]$ maps to ${\cal O}_Y( - (1/2) \sum E_a)$
where the $E_a$ are the exceptional divisors.
Furthermore, it can be shown that on $Y$, $\chi( {\cal O}_Y) = +2$
and $\chi( {\cal O}_Y( - (1/2) \sum E_a )) = -2$, matching the
Euler characteristics above.

So far we have discussed the index of the operator $\overline{\partial}$.
We are not aware of rigorous results concerning the Dirac index, which would
be of direct relevance for physics.  That said, it is very natural to
conjecture that, by analogy with smooth manifolds, the Dirac index is
computed by a closely analogous expression, except that
${\rm Td}(TI_{\mathfrak{X}})$ is
replaced by
\begin{displaymath}
{\rm Td}(TI_{\mathfrak{X}}) \exp\left( - \frac{1}{2} 
c_1^{\rm rep}(TI_{\mathfrak{X}}) \right),
\end{displaymath}
following the usual pattern that
\begin{displaymath}
\hat{A}(M) = {\rm Td}(M) \exp( - (1/2) c_1(M) )
\end{displaymath}
for a smooth manifold $M$.

See also {\it e.g.} \cite{edidin1,reid-icecream,tt1} and references therein
for more information on index theorems on
stacks.

\section{Roots of canonical bundles}
\label{app:canonical-roots}

On a ${\mathbb Z}_k$ gerbe, sometimes there exist $k$th roots of the
canonical bundle, and sometimes not, depending upon the gerbe.  Let us work
through some examples.

First, consider a nontrivial ${\mathbb Z}_k$ gerbe over ${\mathbb P}^1$.
In particular, let us consider the gerbe defined by the quotient
\begin{displaymath}
\frac{
 {\mathbb C}^2 - 0
}{
{\mathbb C}^{\times}
},
\end{displaymath}
where the ${\mathbb C}^{\times}$ acts with weight $k$.
We will show that the pullback of any line bundle on ${\mathbb P}^1$ to this
gerbe does admit a $k$th root.

A line bundle over this gerbe will have a total space of the form
\begin{displaymath}
\frac{
( {\mathbb C}^2 - 0 ) \times {\mathbb C}
}{
{\mathbb C}^{\times}
},
\end{displaymath}
where ${\mathbb C}^{\times}$ acts on $([x,y],z)$ as
\begin{displaymath}
([x,y],z) \: \mapsto \: ( [ \lambda^k x, \lambda^k y], \lambda^n z),
\end{displaymath}
and $n$ classifies the line bundle.
The pullback of ${\cal O}(m)$ on ${\mathbb P}^1$ to the gerbe has
$n = km$, so a line bundle on the gerbe with $n=m$ has the property that
its $k$th tensor power with itself is the pullback of ${\cal O}(m)$.

Thus, on this ${\mathbb Z}_k$ gerbe, $k$th roots of pullbacks of any line
bundle on the base space do exist.

Next, let us consider the trivial ${\mathbb Z}_k$ gerbe over ${\mathbb P}^1$.
Here, the total space any line bundle over this gerbe can be described as
\begin{displaymath}
\frac{
( {\rm Tot} \, L ) \times {\mathbb C}^{\times}
}{
{\mathbb C}^{\times}
},
\end{displaymath}
where $L$ is a line bundle on ${\mathbb P}^1$, and the 
${\mathbb C}^{\times}$ acts
only on ${\mathbb C}^{\times}$.  Here, there is clearly no way to construct
a $k$th root of $L$ (unless $L$ already had a $k$th root on ${\mathbb P}^1$).

\end{document}